\documentclass[12pt,preprint]{aastex}
\usepackage{fullpage}
\usepackage{epsf,amsfonts,amsmath,amssymb}
\usepackage{graphicx}
\usepackage{times}
\usepackage{natbib}
\usepackage{ulem}
\usepackage{epsfig}
\usepackage{mathrsfs}
\usepackage{footmisc}
\usepackage{float}
\usepackage{rotating}
\usepackage{pdflscape}
\usepackage[usenames,dvipsnames]{color}

\def\sigR{{\sigma_{\ln\!R}}}

\definecolor{blue-violet}{rgb}{0.54, 0.17, 0.89}

\newcommand{\fracb}[2]{\left(\frac{#1}{#2}\right)}

\begin{document}
\title{Properties of GRB Lightcurves from Magnetic Reconnection}
\author{Paz Beniamini\altaffilmark{1} and Jonathan Granot\altaffilmark{2}}
\altaffiltext{1}{Racah Institute for Physics, The Hebrew University, Jerusalem, 91904, Israel}
\altaffiltext{2}{Department of Natural Sciences, The Open University of Israel, 1 University Road, P.O. Box 808, Ra'anana 4353701, Israel}
\begin{abstract}
The energy dissipation mechanism within Gamma-Ray Burst (GRB) outflows, driving their extremely luminous 
prompt $\gamma$-ray emission is still uncertain. The leading candidates are internal shocks and magnetic reconnection. 
While the emission from internal shocks has been extensively studied, that from reconnection still has few quantitative predictions. 
We study the expected prompt-GRB emission from magnetic reconnection and compare its temporal and spectral properties 
to observations. The main difference from internal shocks is that for reconnection one expects relativistic bulk motions with 
Lorentz factors $\Gamma'\gtrsim\;$a few in the jet's bulk frame. We consider such motions of the emitting material in two 
anti-parallel directions (e.g. of the reconnecting magnetic-field lines) within an ultra-relativistic (with $\Gamma\gg1$) thin
spherical reconnection layer. The emission's relativistic beaming  in the jet's frame greatly affects the light-curves. 
For emission at radii $R_0<R<R_0+\Delta{}R$ (with $\Gamma=\rm{const}$) the observed pulse width is 
$\Delta{}T\sim(R_0/2c\Gamma^2)\max(1/\Gamma',\,\Delta{}R/R_0)$, i.e. up to $\sim\Gamma'$ times shorter than for 
isotropic emission in the jet's frame. We consider two possible magnetic reconnection modes: a quasi steady-state 
with continuous plasma flow into and out of the reconnection layer, and sporadic reconnection in relativistic turbulence 
that produces relativistic plasmoids. Both of these modes can account for many observed prompt-GRB properties: 
variability, pulse asymmetry, the very rapid declines at their end and pulse evolutions that are either hard to soft  
(for $\Gamma'\lesssim2$) or intensity tracking (for $\Gamma'>2$). However, only the relativistic turbulence mode 
can naturally account also for the following correlations: luminosity-variability, peak luminosity$\,$--$\,$peak frequency 
and pulse width energy dependence / spectral lags.
\end{abstract}

\section{Introduction}
\label{Int}
Gamma Ray Bursts (GRBs) are the most extreme explosions in nature. They are powered by ultra-relativistic jets 
(with Lorentz factors $\Gamma\gtrsim100$) and have huge isotropic equivalent luminosities ($L_{iso}\approx10^{53}\;$erg/s). 
A good part of the outflow energy is  thought to be dissipated at large distances ($R\sim10^{13}-10^{17}\;$cm) from the central engine, where it efficiently radiates,
predominantly in $\gamma$-rays that escape the emission region and eventually reach us.
One of the major open questions in GRB research concerns the outflow composition. In particular, whether the energy is
carried out from the central source to the emission region predominantly as kinetic energy -- a baryonic jet \citep{Shemi(1990)},
or as Poynting flux -- a highly magnetized (or Poynting flux dominated) jet \citep{Usov(1992),Thompson(1994),Meszaros(1997),Lyutikov(2003),Granot(2015)}.
Recently, \cite{Beniamini(2014)} have shown that the question of outflow composition is essentially tied to another major open question in GRB research - the nature of the prompt emission mechanism. In particular,
one zone emission models in magnetic jets naturally produce efficient synchrotron radiation and predict fluxes in the X-ray and optical bands in excess of the 
observed ones during the prompt. However, the magnetization might still be high up until the emission region, and then 
sharply drop within the emission region itself \citep{Drenkhahn(2002),Lyutikov(2003),Sironi(2015)}, if the dissipation that causes the emission is driven by efficient magnetic reconnection.

Magnetic jets can be powered by the rotational energy (ultimately arising from gravitational energy) 
of a magnetized torus around the black hole \citep{Meszaros(1997)} or of a highly magnetized millisecond neutron star \citep{Usov(1992),Spruit(1999)}.
Initially highly-magnetized jets are favored on energetic grounds, as modeling of GRB 
central engines that rely on accretion disks suggest that their power is significantly larger than that of thermally driven outflows powered by neutrino-anti neutrino annihilation \citep[see e.g. ][]{Kawanaka(2013)}.
Moreover, a high value of the magnetization parameter $\sigma$ (the magnetic to particle enthalpy density or energy flux ratio) near the source
can help prevent excessive baryon loading that may prevent the jet from reaching sufficiently high Lorentz factors far 
from the source, at the emission region. Such initially highly-magnetized jets may naturally lead to magnetic reconnection.
In a striped wind magnetic field configuration \citep{Coroniti(1990)}, with either periodic or stochastic flipping of the magnetic field direction near the source,
reconnection at large distances from the source has a natural preferred direction. Moreover, for 
large $\sigma$-values just before the dissipation region, reconnection will lead to local relativistic bulk motion of the outgoing particles away from 
the reconnection sites (in the bulk frame of the jet). Therefore, reconnection models may be able to overcome the difficulties 
present in an isotropic arrangement of the magnetic fields regarding overproduction of optical and X-ray radiation (as compared with
observations) due to synchrotron emission.

As mentioned above, synchrotron is likely the dominant source of emission in magnetic jets.
Many studies have considered the implications of synchrotron radiation being the source of the prompt emission in GRBs \citep{Katz(1994),Rees(1994),Sari(1996),Sari(1998),Kumar(2008),Daigne(2011),Beniamini(2013)}.
One of the greatest challenges for reconciling this model with the observations is related to the low energy spectral slope.
The synchrotron fast cooling (which is the likely cooling regime in order to achieve the required large radiative efficiencies) 
photon index below the peak, $dN/dE\propto E^{\alpha_B}$ with $\alpha_B=-1.5$, is inconsistent with the average observed slope of 
$\alpha_B=-1$. Although it may be possible to achieve softer spectra, $\alpha_B<-1.5$ (for instance by
having the peak frequency fluctuate on a shorter time-scale than that for which the spectra are obtained, and in effect smearing 
the transition from below to above the synchrotron peak), it is not at all trivial to increase $\alpha_B$ in these models. 
However, $90\%$ of all GRB have $\alpha>-1.5$ \citep{Preece(2000), Ghirlanda(2002), Kaneko(2006), Nava(2011)}. 
In about $40\%$ of the GRBs $\alpha>-2/3$ \citep{Nava(2011)}, which is impossible for synchrotron, even in the case 
of slow cooling. This problem is known as the synchrotron ``line of death'' \citep{Preece(1998)}.
However, a refined time-dependent spectral analysis in several bright bursts \citep{Guiriec(2012)} has
suggested that GRB spectra may be better fitted with a multi-component model instead of the classical ``Band function". 
In this model, the Band function component dominates the total energy but is  accompanied by weaker power-law and 
black body components. The lower energy spectral slope of the Band function is softer than in Band only fits, and is 
consistent with slow cooling synchrotron (i.e. the synchrotron ``slow cooling line of death" problem is removed).
Reconnection models may naturally allow for continuous heating of 
the electrons while they are emitting, in this way producing low-energy spectral slopes consistent with slow cooling while 
maintaining large radiative efficiencies \citep{Ghisellini(1999),Kumar(2008),Fan(2010),Beniamini(2014)}.

An anisotropic emission model has been suggested to operate in the afterglow phase of GRBs \citep{Beloborodov(2011)}.
These authors argued that synchrotron radiation is almost certainly anisotropic, and suggested that this anisotropy could help 
explain highly variable signals observed in some X-ray afterglows (such as the rapid decay phase or X-ray flares).
In this work, we explore an analytic model in which the emitting electrons are anisotropic in the jet's bulk frame 
representing the bulk motion of the jet, or more specifically, are isotropic in their local center of momentum frame, 
which moves relativistically relative to the jet's bulk frame. This could be applicable to reconnection from a striped wind where 
each pulse in the light-curve of the prompt emission is due to some macroscopic reconnection event \citep{Kumar(2015)}. 
Moreover, it  can also be applied in a broader context of anisotropic emission, such as had been suggested in the ``jet in jet" model
\citep{Levinson(1993),Pedersen(1998),Frail(2000)} and the relativistic turbulence models 
\citep{Lyutikov(2003),Kumar(2009),Lazar(2009)}.

The paper is organized as follows. In \S~\ref{motiv} we present two different physical scenarios that can allow for anisotropic emission from reconnection zones.
We discuss their different characteristics and present a simple argument, which shows that in contrast to 
``non-boosted" reconnection models (i.e. the case where the particles' velocities are not strongly an-isotropic in the jet frame), anisotropic models can be consistent with the observed variability in GRB light-curves. 
Section \ref{model} is devoted to a detailed description of the model that is considered in this work. Next, in \S~\ref{results} 
analytic expressions are derived for the resulting light-curves, and their dependence on the model parameters is discussed. 
Section \S~\ref{comparison} features an extensive comparison of our results to GRB observations. Finally, our main results
are summarized and discussed in \S~\ref{sec:diss}.

\section{Motivation}
\label{motiv}

\subsection{The Basic Setup for Relativistic Magnetic Reconnection}
\label{sec:relrec}

In GRBs and other relativistic outflow sources, magnetic reconnection is considered to be potentially most relevant 
if the outflow is initially Poynting flux dominated. In this case magnetic reconnection is expected to be most efficient 
and prominent if a high $\sigma$ is maintained in the outflow out to the reconnection site, so that the plasma that is 
inflowing into the reconnection region (or layer) is highly magnetized, with $\sigma\gg 1$. Such a situation is referred 
to as relativistic reconnection, as the total energy largely exceeds the rest energy of the particles and the outflow from 
the reconnection region is expected to be relativistic. For such relativistic magnetic reconnection, two main different 
types of reconnection models have been discussed in the literature.

The first type involves quasi-steady-state reconnection configurations with a continuous inflow of magnetized plasma 
into the reconnection layer, and a continuous outflow out of it. Analytic models of relativistic magnetic reconnection 
\citep[in which the inflowing plasma is highly magnetized with $\sigma\gg 1$, e.g.,][]{Lyubarsky(2005)} assume a 
single velocity for the plasma out-flowing from the reconnection region, which they find to be relativistic (corresponding to
a Lorentz factor of $\Gamma'\sim\sigma^{1/2}\gg 1$), and do not distinguish between individual particles of different energy.
Particle In Cell (PIC) simulations \citep[e.g.,][]{Cerutti(2012),Cerutti(2013),Sironi(2014),Guo(2014),Melzani(2014),Kagan(2015)}, 
on the other hand, find that electrons that are accelerated to higher energies also spend a longer time being accelerated in the 
electric field within the reconnection layer, near the x-points, and thus their velocities become more tightly collimated.
This leads to a positive correlation between the random Lorentz factor of the electrons $\gamma_e'$ and their bulk Lorentz factor relative 
to the jet frame, $\Gamma'$. Moreover, once the electrons are deflected out of the x-point they quickly enter magnetic islands where their
velocities are isotropized over their Larmour gyration time. Therefore, only electrons that emit a good fraction of their radiation within
less than their Larmour gyration time will produce highly beamed radiation in the jet's bulk frame. This corresponds to emitted photon 
energies above the synchrotron burn-off limit ($\gtrsim 100\;$MeV in the jet's bulk frame), which is also known as the ``maximum 
synchrotron energy" \citep{de Jager(1996)} and corresponds to an observed energy of  
$h\nu_{\rm syn,max} = E_{\rm syn,max}\approx 7 (1+z)^{-1}(\Gamma/100)\;$GeV.
Such high-energy electrons near the burn-off limit are highly "fast cooling".
We stress that this set-up is therefore likely less relevant for explaining properties of the $\lesssim\;$MeV prompt emission. 
Within this model, the low energy part of the spectrum is likely to be produced by non-boosted synchrotron emission in the jet's frame.
Since the same electrons are producing both anisotropic and non-boosted emissions, the spectrum is expected to transition smoothly from 
the former to the latter, as the observed frequency decreases.
Various implications of this scenario (at higher frequencies) are considered in detail in \S~\ref{sec:corrGammap}.
 
In this picture, in the vicinity of the x-points where $E^2>B^2$ the electric field cannot vanish in any rest frame,
and the particles are directly accelerated in the electric field.
In the magnetic islands $E^2<B^2$ and the particle velocities have been isotropized such that the center of momentum (CM)
particle velocity $\vec{\beta}'_{\rm CM}$ is the same for particles of all energies, so that in this plasma's rest frame, the 
charged particles can short-out the electric field, causing it to practically vanish (and thus approaching the ideal-MHD 
limit within the magnetic islands).
As the velocity of the islands relative to the jet's bulk rest frame is small ($\beta'_{\rm isl}\ll 1$) the electric field in the jet's
bulk frame is also very small there ($E'/B'\sim \beta'_{\rm isl}\ll 1$). This may be assumed to still hold to zeroth order also in the
intermediate region between the x-points and the magnetic islands, where most of the beamed radiation is expected to be emitted.
In this region one already has $E^2<B^2$, but the particle velocities are still highly anisotropic and the particle CM velocity is energy 
dependent, $\vec{\beta}'_{\rm CM} = \vec{\beta}'_{\rm CM}(\gamma'_e)$, so that anyway the electric field could at most vanish
in the CM frame of particles of one particular energy but not in that of particles of all other energies.

The second type of relativistic magnetic reconnection models invoke relativistic turbulence \citep{Lyutikov(2003),Kumar(2009),Lazar(2009),Inoue(2011),Zrake(2014),Zrake(2015),Lazarian(2015),East(2015),Zrake(2016)}. 
In the context of GRBs within this type of model, different regions of the jet undergo sporadic 
reconnection events that give rise to relativistic bulk motions of plasma -- ``blobs" or ``plasmoids" -- relative to the jet's frame. 
In this picture each blob is macroscopic and signifies a relativistic bulk motion relative to the jet's frame, where the blob's
bulk velocity is equal to $\vec{\beta}'_{\rm CM}(\gamma'_e)$ for electrons of any $\gamma'_e$.

Notice that although many plasmoids are found to be generated also in recent PIC simulations
\citep{Sironi(2014),Sironi(2015)}, this structure is continuously evolving in time, as the magnetic islands merge with each other and grow larger and larger. This situation could then lead to an evolution of the spectrum within the prompt phase that is not observed. However, 
if the emission from all of the magnetic islands composing each reconnection layer produces a single spike in the light-curve then it 
would predict spectral evolution within a single spike, which is observed. Moreover, since the increase in islands' sizes during this
evolution is roughly linear, this would also imply that the acceleration time-scale of the highest energy electrons is dominated by the
growth time of the largest island, which might explain the late onset of the high-energy emission (compared to the $\lesssim\;$MeV
emission) in most bright Fermi/LAT GRBs \citep[e.g.,][]{GRB080916C,GRB080916C,GRB080825C,GRB090510}.
However, in this picture during a significant fraction of the pulse's dynamical time, it is no longer possible to maintain 
$t_{\rm acc} \ll t_{\rm dyn}$ as required in order to have continuous heating that balances the radiative cooling in order to 
avoid excess optical and X-ray emissions. Furthermore, recent studies \citep{Melzani(2014)} find that the plasmoids generated 
in this setup are not necessarily traveling relativistically with respect to the jet's bulk frame. A blob-like emission described by 
the $k=0$ mode is, however, observed in numerical studies involving relativistic turbulence. In this case, the turbulence can 
enhance the reconnection rate \citep[e.g.,][]{Lazarian(2015),Zrake(2015)}.

Ideal MHD holds within each blob and the electric field within its volume vanishes in its own frame. The electrons maintain a roughly
isotropic velocity distribution within the blob's rest frame and their emission is approximately isotropic in this frame. Hence, the blob's
emission is beamed in the jet's rest frame because of the blobs relativistic motion in this frame. Previous studies have focused on 
turbulence that is isotropic in the jet's frame. However, given an initially striped wind configuration of the magnetic field, it is likely 
that this turbulence would be largely constrained to a thin layer parallel to the shell front. 
Such an anisotropic relativistic turbulence would result in a correspondingly anisotropic emission in the jet's frame.

In this scenario the electrons may be slow cooling. Moreover, second order Fermi acceleration within the turbulent region can 
continuously heat the electrons until they reach a balance between heating and cooling \citep[e.g.,][]{Murase(2012),Asano(2015)}.
This naturally allows for the electrons to remain hot while radiating, which is necessary in order to explain the observed prompt spectra 
from GRBs in a magnetically dominated emitting region \citep{Beniamini(2014)}. In addition, boosted emission is necessary in order to
explain the observed prompt variability as will be shown below. For these reasons, in this work we focus mainly on this second scenario,
while the first scenario is addressed mainly in \S~\ref{sec:corrGammap}.

Finally, we remark that although the $k=0$ and $k=1$ models involve different modes of reconnection, they both rely on a similar initial physical set-up: a magnetically dominated jet, where each pulse in the light-curve is produced by some macroscopic reconnection event in a layer perpendicular to the jet's propagation direction. While the $k=0$ is more likely for explaining the observed properties of the sub-MeV
pulse, it is possible that in at least some GRBs, the $k=1$ mode is responsible for producing the GeV emission. In this case, the two modes may be operating in the same outflow. This may occur, for instance, if the initially quasi-steady state eventually drives turbulent reconnection until the later becomes the dominant mode of reconnection (see \citealt{Lazarian(2015)}). The ratio between the GeV emission and the sub-MeV pulse would depend in this case on the fraction of the flow involved in each reconnection mode and on their respective durations. Whether or not this can reproduce the observed ratio of these spectral components could be a potential test for this scenario.

\subsection{Lightcurve Variability}
\label{sec:variability}
In a simple, non-boosted emission picture, the reconnection model cannot easily reproduce the observed variability. 
This is because in order to obtain large-scale ordered reconnection, the speed of the incoming plasma that approaches the
reconnection region from both sides is expected to be $v'_{\rm in}=\beta'_{\rm in}c\lesssim 0.1-0.25 c$ \citep{Lyubarsky(2005),Guo(2015),Liu(2015)}. 
Therefore, if the radial width of the region feeding the reconnection layer is $L$ in the lab frame and $L' = \Gamma L$
in the jet's bulk frame, then the reconnection lasts $\Delta t' = L'/v'_{\rm in}$ in the jet's bulk frame, and 
$\Delta t = \Gamma\Delta t'$ in the lab frame. The time over which the radiation that emitted from this reconnection 
event along the line of sight reaches the observer, or the radial time of the resulting pulse, 
is $\Delta T_r = (1-\beta)\Delta t \approx L/2v'_{\rm in}$. 
The increase in radius during the reconnection event is $\Delta R \approx c\Delta t$,
and since this is a lower limit to the emission radius, $R\geq \Delta R$, the angular spreading time 
in the arrival of photons to the observer, which is a lower-limit to the pulse width, is at least 
\begin{equation}
\Delta T_{\theta} = \frac{R}{2c\Gamma^2} \geq \frac{\Delta R}{2c\Gamma^2} 
\approx \frac{L}{2v'_{\rm in}} \approx \Delta T_r\ .
\end{equation}
The total pulse width accounts for both radial and angular spreading in the photon arrival times,
and is therefore 
\begin{equation}
\Delta T \sim \Delta T_r+\Delta T_\theta \sim \frac{L}{v'_{\rm in}}\ .
\end{equation}
This is a factor of $\sim 10(\beta'_{\rm in}/0.1)^{-1}$ larger than the difference in the ejection time of
neighbouring reconnecting layers or shells, $\Delta t_{\rm ej}\approx L/c$, which would also be 
the difference in the onset times of neighbouring pulses for emission from the same radius.
While fluctuations are possible around this typical difference in pulse onset times, the mean separation
does not vary unless the emission radius systematically changes throughout the GRB.
There is, however, observational evidence against this, since the gross properties of the GRB emission
do not appear to systematically change throughout the prompt emission.

The time between pulses can increase if only a small portion of the jet material would have contributed to the observed radiation
(i.e. if the distance between the reconnection layers is larger than $L$, so that $\Delta t_{\rm ej}$ would be correspondingly larger, 
where the added outflow layer does not contribute to the reconnection). However, assuming a roughly constant jet power this would
imply that only a correspondingly small fraction is dissipated and contributes to the gamma-ray emission. This is because significant 
pressure gradients tend to be washed out, so that a smooth pressure profile may be expected in the propagation direction. 
For a fixed local bulk Lorentz factor the energy flux is proportional to the proper enthalpy density, $w+B^2/4\pi = w(1+\sigma)$, 
which in the highly magnetized regions ($\sigma\gg 1$) is dominated by the magnetic term, half of which arises from the magnetic
pressure $B^2/8\pi$. The thermal pressure, on the other hand, can account for at most one quarter of the particle proper enthalpy density,
$w = \rho c^2+e_{\rm int}+p$, since even for a relativistically hot plasma $p = e_{\rm int}/3 \approx w/4$. Therefore, a comparable 
energy per unit length in the propagation direction is expected also for low magnetization regions ($\sigma< 1$) or highly magnetized
regions that do not border an oppositely oriented magnetic field region with which they can easily reconnect (e.g. if the field orientation
changes randomly due to fluctuations in an accretion disc around a black hole, rather than very orderly as in a strong MHD wind 
launched by a newly born millisecond magnetar).
This could be very challenging for the total energy budget of the GRB, as the total jet energy would be even larger than current 
in estimates, which are already considered highly demanding for viable GRB progenitors \citep{Panaitescu(2002),Granot(2006),Fan(2006),Beniamini(2015)}.

All this implies that the ``simple" reconnection model predicts pulses that are significantly broader than the intervals between them,
contrary to what is seen in observations (see also \citealt{Lazar(2009)}), namely that these times are roughly equal \citep{Nakar(2002)}.
Therefore, it is interesting to explore a slightly more complicated yet physically motivated reconnection model, 
which may be able to both account for the continuous heating, as well as reproduce the observed variability.
In this model the plasma in the thin reconnection region is ejected away from the X-point along the reconnection layer 
at a relativistic speed in the jet's bulk frame, corresponding to a Lorentz factor $\Gamma'$. The emission is assumed to be 
isotropic in a frame moving at $\Gamma'$ relative to the jet's bulk frame, and is therefore anisotropic in the jet's bulk frame for 
$\Gamma'\gg 1$.

Anisotropic emission in the jet's bulk frame only reduces $\Delta T_\theta$ by a factor of $\sim\Gamma'$, while it does not change
$\Delta T_r$.  Therefore, given that  $\Delta T \sim \Delta T_r+\Delta T_\theta$, for a smooth continuous emission as a function
of radius (and $\Gamma'\gtrsim\;$a few)  $\Delta T\sim \Delta T_r \sim R/2c\Gamma^2\approx 
\Delta t_{\rm ej}/2\beta'_{\rm in} = 5(\beta'_{\rm in}/0.1)^{-1}\Delta t_{\rm ej}$. 
However, since $\Delta T_\theta\sim \Delta T_r/\Gamma'$ is significantly smaller than $\Delta T\sim \Delta T_r$, 
and the emission at any given radius originates from a small part (a fraction $\sim 1/(\Gamma')^2$) of the visible region 
(corresponding to angles $\theta\lesssim1/\Gamma$ from the line of sight), variations in the emission intensity
with radius would be reflected in the light curve down to the shorter angular time-scale $\Delta T_r/\Gamma'$.
Moreover, such variations are arguably quite likely in efficient (high $\sigma$) magnetic reconnection (i.e. reconnection leading to dissipation of a significant fraction of the total energy),
which is rarely very smooth
and continuous. Therefore, it is likely to typically observe variability on the shorter angular time-scale $\Delta T_r/\Gamma'$.

One implication of this model is that it naturally results in larger emitting radii than those usually implied from the variability time-scale.
This is because for non-boosted emission in the jet's bulk frame the observed variability time-scale is given by:
\begin{equation}
 \Delta T_{\rm iso}=\frac{R_{\rm iso}}{2c\Gamma^2}\ ,
\end{equation}
whereas in the current model it is shorter by a factor of $\sim\Gamma'$, $\Delta T\sim\Delta T_{\rm iso}/\Gamma'$. 
This would imply a larger emission radius in order to reproduce the same observed variability time-scales. 
For $\Gamma\approx 300, \Gamma'\approx 10$ and $\Delta T=1$sec we obtain $R \approx 5 \times 10^{16}$cm.
This radius is ten times larger than would be inferred for a non-boosted emission in the jet's bulk frame, and is still a few times 
smaller than typical estimates for the deceleration radius (and therefore consistent with the interpretation that the prompt signal
originates at $R$).

\section{The model}

\label{model}
\subsection{The physical model}
We associate each pulse in the light-curve with emission from a different thin ``shell". In the context of a striped wind
each ``shell" is associated with the thin reconnection layer between regions (wider spherical shells) of oppositely oriented 
tangential (primarily toroidal) magnetic field lines. In this section we discuss the emission from a single shell, and in \S~\ref{multi}
we show that adding up the emission from many shells produce light curve variability similar to that seen in observations.
\begin{figure*}[h]
\centering
\includegraphics[scale=0.3]{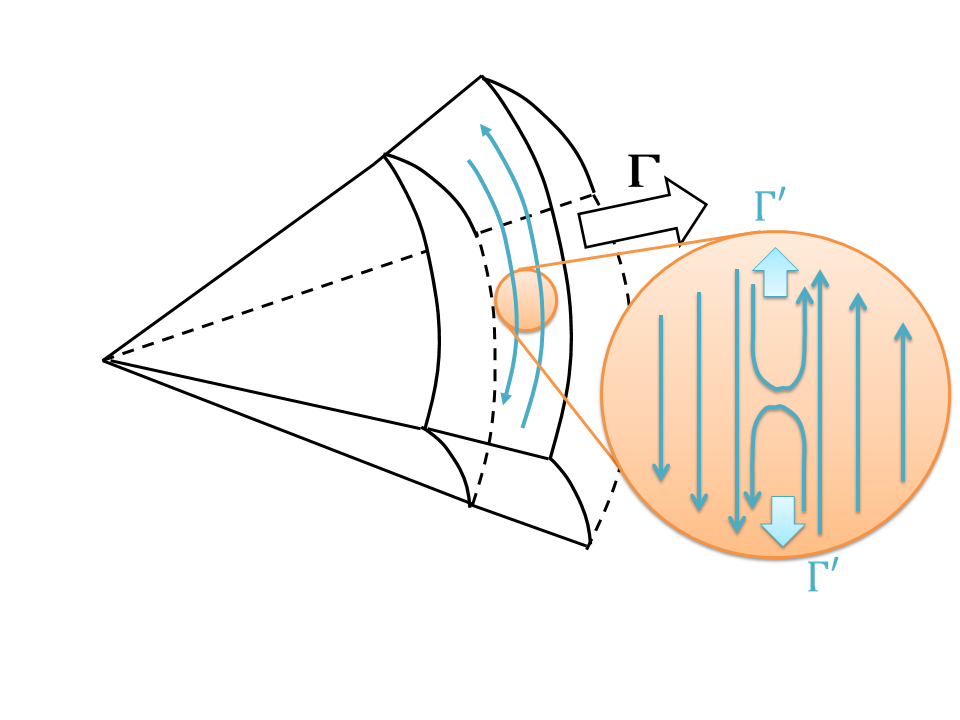}
\includegraphics[scale=0.3]{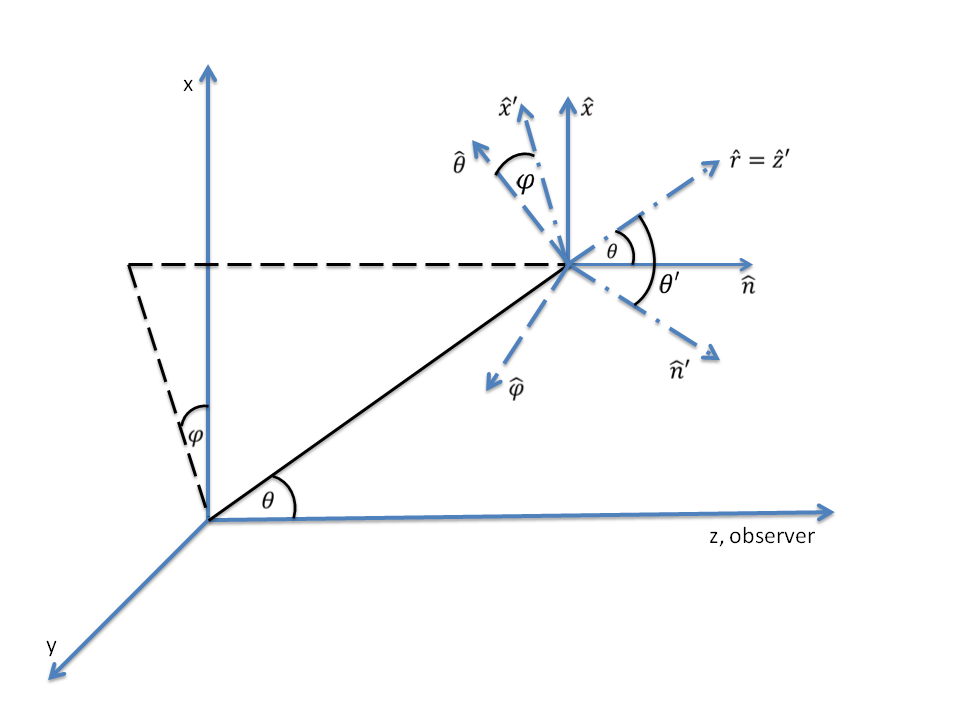}
\caption{ {\bf Left: }Schematic geometry of the model at hand. The emitting matter is moving relativistically 
out of the reconnection x-points at a Lorentz factor $\Gamma'$ in the jet's bulk frame (i.e. relative to the bulk of the outflow).
{\bf Right: } The two reference frames used in this work: the lab frame with the origin at the central source (in either Cartesian or spherical coordinates), and the jet's bulk frame (denoted by primes), travelling at a Lorentz factor 
$\Gamma$ in the $\hat{r}=\hat{z'}$ direction and centred at the emission point.}
\label{fig:angles}
\end{figure*}

Consider an ultra-relativistic (moving with $\Gamma \gg 1$ in the rest frame of the central source) thin (of radial width $\ll R/(\Gamma^2\Gamma')$)  ``spherical" (occupying an angle 
$\theta$ around the line of sight such that $\theta\Gamma\gtrsim\;$a few) shell which starts emitting at some initial radius $R_0$ 
(possibly, but not necessarily due to reconnection) and turns off at a final radius $R_f \geq R_0$.
Let the emitters be moving with a Lorentz factor $\Gamma'$ in the local bulk frame of the flow, in two opposite directions that are perpendicular to the radial direction.
There are three relevant frames in this paper: (i) the central source frame, (ii) the jet's local bulk frame, which moves with a Lorentz factor $\Gamma$ in the radial direction to the central source,
and (iii) the emitters' frame which moves with a Lorentz factor $\Gamma'$ compared to the jet's bulk frame, in which the radiation is assumed to be emitted isotropically.
Quantities in frame (i) are un-primed, those in frame (ii) are primed, while quantities in frame (iii) are double primed. 
With these definitions, the emitters' speed is given by: $\vec{\beta'}=\pm\beta' \hat{x'}$ ($\hat{z'}=\hat{r}$ is the radial direction
in the jet's bulk frame). $\hat{x'}$ is  the projection of $\hat{x}$ (the direction perpendicular to the direction to the observer 
$\hat{n}=\hat{z}$) on the plane perpendicular to the radial direction.
Notice that: $\hat{n'}=\hat{r}\cos\theta'-\hat{\theta}\sin\theta'$. In addition:
\begin{equation}
\begin{split}
 &\hat{x'}=\frac{(\hat{\theta}\cdot\hat{x})\hat{\theta}+(\hat{\phi}\cdot\hat{x})\hat{\phi}}
 {\sqrt{(\hat{\theta}\cdot\hat{x})^2+(\hat{\phi}\cdot\hat{x})^2}}=
 \frac{\hat{\theta}\cos\theta\cos\phi-\hat{\phi}\sin\phi}{\sqrt{\sin^2\phi+\cos^2\theta\cos^2\phi}}\\
& =\frac{\hat{\theta}\cos\theta\cos\phi-\hat{\phi}\sin\phi}{\sqrt{1-\sin^2\theta\cos^2\phi}}
 \approx \hat{\theta}\cos\phi-\hat{\phi}\sin\phi \ ,
\end{split}
\end{equation}
where $\hat{\beta'}=\pm \hat{x'}$ and we have approximated $\cos^2\theta=1-\sin^2\theta\approx 1$. Moreover,
\begin{equation}
\cos \phi'=\hat{x'}\cdot\hat{\theta}=\frac{\cos\theta\cos\phi}{\sqrt{1-\sin^2\theta \cos^2 \phi}}\approx \cos \phi\ .
\end{equation}
Combining this with the expression for $\hat{n'}$ we get that: $\hat{n'}\cdot \hat{\beta'}=\pm\sin\theta'\cos\phi'\ $.
For a schematic figure of the geometry see Fig.~\ref{fig:angles}.
We aim to find the resulting light-curve in the observer's frame.
Our modelling follows that of \cite{Genet(2009)} (which relies on the formalism developed in \citet{Granot(2005)} 
and \citet{Granot(2008)}), and converges to the results in this paper in the limit of $\Gamma' \to 1$  (and $\beta'\to 0$), corresponding to the case of non-boosted emission. Notice that once the motion of the local 
center of momentum velocity of the emitting particles becomes non-relativistic, their radiation is no longer beamed and even if 
they emit in a very ordered magnetic field, their emission pattern will be broadly similar to isotropic emission in the jet's bulk 
frame \citep{Granot(2002),GK(2003),Beloborodov(2011)}, and the situation reduces back to the simple ``one zone" case.

The jet may be accelerating, coasting or decelerating as it is emitting (see \S~\ref{sec:diss} for more details). To account for this, the Lorentz factor of
the bulk is assumed to evolve as a power-law with radius: $\Gamma^2=\Gamma_0^2 (R/R_0)^{-m}$ where $\Gamma_0=\Gamma(R_0)$.
The case of a coasting jet (with no acceleration or deceleration) can be easily obtained by setting $m=0$ in the following equations.
The luminosity is assumed to be isotropic in the emitter's frame (double primed frame) and depends only on the radius:
\begin{equation}
 L''_{\nu''}(R) = L''_{\nu_0''} S[\nu''/\nu_p''] f(R/R_0)\ .
\end{equation}
In the last expression, $f(x)$ is a dimensionless function of the radius that is normalized such that $\int _0^{\infty}f(x)dx=1$, 
$\nu_p''$ is the frequency where $\nu''  L''_{\nu''}$ peaks, and
$S(x)$ is the spectrum (which is assumed to be a Band function),
\begin{equation}S(x) = e^{2+\alpha_B} \left\{ \begin{array}{ll} x^{1+\alpha_B} e^{-(2+\alpha_B)x} & x
\leqslant x_b\ ,\\
x^{1+\beta_B} x_b^{\alpha_B-\beta_B} e^{-(\alpha_B-\beta_B)} & x \geqslant x_b\ ,
\end{array} \right.
\end{equation}
where $x_b = (\alpha_B-\beta_B)/(2+\alpha_B)$, while $\alpha_B$ and $\beta_B$ are the high and low energy slopes of the spectrum. 
For $\alpha_B >-2 > \beta_B$ the Band function has a peak in the $\nu F_\nu$ spectrum, at $x = 1$, and therefore since $S(x)$ is 
normalized such that $S(x) = xS(x) = 1$ at $x = 1$, it will not affect normalization of $\nu F_{\nu}$ at its peak. 
The two functional forms used in the Band function are matched at $\nu''_b = x_b \nu''_p$.

The flux is then given by:
\begin{equation}
\label{flux0}
 F_{\nu}(T) =\frac{1}{4\pi D^2} \int dL_{\nu}=\frac{1}{4\pi D^2} \int {\mathcal D}^3 dL'_{\nu'}
 =\frac{1}{4\pi D^2} \int {\mathcal D}^3 \frac{{\mathcal D}'^{\,3-k}}{\Gamma'^k} dL''_{\nu''}\ ,
\end{equation}
where for convenience we denote the effective distance to the source in terms of
the luminosity distance and red-shift by $D = d_L/\sqrt{1+z}$. In addition, ${\mathcal D}$ is the Doppler factor from the jet's bulk frame (the rest frame of the bulk) to the observer frame 
(the rest frame of the central source), ${\mathcal D}'$ is the Doppler factor from the emitter's frame to the jet's bulk frame and
$dL''_{\nu''}=L''_{\nu''}(R)d\mu d\phi / (4 \pi)$.
In the last expression, $\mu=\cos\theta$ where $\theta$ is the angle between the emitting location and line of sight and $\phi$ is the azimuthal angle (these are the angles in a spherical coordinate system
whose origin is at the central source while its $z$-axis points toward the observer).
This implies that the observed frequency ($\nu$) is related to the 
frequencies in the jet's bulk frame ($\nu'$) and in the emitter's frame ($\nu''$) by $(1+z)\nu = {\mathcal D}\,\nu' = 
{\mathcal D}\,{\mathcal D}'\nu''$. The parameter $k$ can obtain only two possible values: $k=1$ in case each 
emitter produces continuous emission (a steady state in the jet's bulk frame, e.g. as in \S \ref{sec:corrGammap}), or $k=0$ in case the emitters produce a single ``blob" (as is likely for the turbulent scenario discussed in \S \ref{motiv}). 
The observed time $T$ is set such that $T=0$ corresponds to the arrival time of a photon emitted at the central source (i.e. in the origin) 
at the ejection time of the shell. The first photons to reach the observer from the emission at radii $R_0\leq R\leq R_f$ are from
along the line of sight ($\theta=0$) from $R=R_0$, and arrive (accounting for cosmological red-shift) at
\begin{equation}\label{eq:T0}
T_0 = \frac{(1+z)R_0}{2(m+1)c\Gamma_0^2}\ .
\end{equation}
We define the equal arrival time surface (EATS) as the locus of points from which photons emitted at different radii $R$ and angles $\theta$ relative to the line of sight, reach the observer simultaneously at an observed time $T$.
The maximal radius along the EATS, $R_L(T)$ (which is obtained along the 
line of sight), and the Lorentz factor at that radius, $\Gamma_L(T)$, can be conveniently expressed in terms of this time,
\begin{equation}\label{eq:scalings}
\frac{R_L}{R_0} = \fracb{T}{T_0}^\frac{1}{m+1}\ ,\quad\quad
\frac{\Gamma_L}{\Gamma_0} = \fracb{R_L}{R_0}^{-\frac{m}{2}} = \fracb{T}{T_0}^{-\frac{m}{2(m+1)}}\ .
\end{equation}

\subsection{The observed flux}

In order to calculate the flux density $F_{\nu}$ that reaches the observer at time $T$ we integrate the luminosity $L''_{\nu''}$ 
over the EATS. This introduces a relation between 
$R$ and $\theta$, $T/(1+z) = t - R\mu/c$, for photons arriving at a given time. In addition, since the emission in our 
model is anisotropic, photons emitted at different $\phi$ will have different Doppler factors from the emitter's frame to the 
bulk frame and we cannot use symmetry arguments to integrate over $\phi$. As we assume $\Gamma\gg 1$, because of 
relativistic beaming only angles of $\theta\Gamma\lesssim\;$a few contribute significantly to the observed flux, so one can
safely assume $\theta\ll 1$. It is convenient to use the dimensionless normalized radius $y = R/R_L$ and change variables of 
integration for calculating the flux from $\mu$ to $y$ or $d\mu\to |d\mu/dy|dy$, using the relation
\begin{equation}
\left|\frac{d\mu}{dy}\right| = \frac{y^{-2}+my^{m-1}}{2(m+1)\Gamma_L^2} 
= \fracb{T}{T_0}^\frac{m}{m+1}\frac{1}{\Gamma_0^2}\frac{(m+y^{-m-1})y^{m-1}}{2(m+1)} \ .
\end{equation}
The Doppler factors and  can be expressed as
\begin{eqnarray}
\label{trigonometric}
{\mathcal D} &=& \frac{1}{\Gamma(1-\beta\mu)} = \Gamma_L\frac{2(m+1)y^{-m/2}}{m+y^{-m-1}}
= \fracb{T}{T_0}^{-\frac{m}{2(m+1)}}\Gamma_0\frac{2(m+1)y^{-m/2}}{m+y^{-m-1}}\ ,
\\ \nonumber
\\
{\mathcal D}'  &=& \frac{1}{\Gamma'(1-\beta'\sin\theta'\cos\phi')}\ ,\quad\quad
\sin\theta' = 2\frac{\sqrt{(m+1)(y^{-m-1}-1)}}{m+y^{-m-1}}\ .
\end{eqnarray}
One can also define $\nu_0\equiv 2\Gamma_0\Gamma'\nu''_0/(1+z)$ and use it in order to express
\begin{equation}\label{eq:x}
x = \frac{\nu''}{\nu''_p} = \frac{\nu}{\nu_0} \left(\frac{m\!+\!y^{-\!m\!-\!1}}{m\!+\!1}\right)
\fracb{y}{y_{\rm min}}^{\frac{m}{2}}\Gamma'^2(1\!-\!\beta'\sin\theta'\cos\phi)\ ,
\end{equation}
where the limits of integration over $y$ are from $y_{\rm min}=\min(1, R_0/R_L)$ to $y_{\rm max}=\min(1, R_f/R_L)$.

A general expression for the flux is then given by:
\begin{equation}
\label{generalFnu}
\begin{split}
F_{\nu}(T) =\,\frac{2\Gamma_0 \Gamma' L''_{\nu_0''}}{4\pi D^2 }\! \fracb{T}{T_0}^{-\frac{m}{2(m+1)}} 
&\int_{y_{\rm min}}^{y_{\rm max}}\! \! dy \bigg(\frac{m\!+\!1}{m\!+\!y^{-m-1}}\bigg)^{2} y^{-1-\frac{m}{2}}
\;f\left[ y \fracb{T}{T_0}^\frac{1}{m+1}\right] \\
&\times \frac{1}{2\pi\Gamma'^4}\int_0^{2\pi}d\phi\,(1\!-\!\beta'\sin\theta'\cos\phi)^{k-3}\,S[x(\phi,y)]\ .
\end{split}
\end{equation}

\section{Results for single-pulse emission}
\label{results}
We turn to a detailed discussion of the results of our model. In \S \ref{SplRpl}, \ref{SplRdelta} we present simple analytic expressions for the
general flux provided in Eq. \ref{generalFnu} for specific spectral and radial emission profiles. The light-curves of single pulses under various assumptions are presented in \S \ref{pulseshapes}.
Their dependence on the radial width of the jet contributing to the emission and on the emitters' Lorentz factor, $\Gamma'$, as well as estimates for their degree of asymmetry are taken up in \S \ref{paramspace}.
In \S \ref{sec:corrGammap} we deviate from the main model considered in the paper, and, motivated by PIC simulations, turn to a discussion of the predicted emission from a continuous injection of particles to
the reconnection layer resulting in a correlation between the random Lorentz factor of the electrons and the bulk Lorentz factor of the emitters.
The results derived under these assumptions are likely to be less relevant for explaining the properties of the sub-MeV pulse, but can still have interesting implications for the emission at very high frequencies.
Finally in \S \ref{multi}, we present the results for the entire light-curve using different distributions of the intrinsic parameters of the model.

\subsection{analytic estimates for power law spectra and power law function in radius}
\label{SplRpl}
For a power law spectrum, $S_{\rm PL}(x) = x^{-\alpha}=(\nu''/\nu''_0)^{-\alpha}$, one can 
obtain a somewhat simpler analytic formula for the observed flux. In this case since $f(R/R_0)=0$ for $R<R_0$ (see below), one can effectively plug $y_{\rm min}=R/R_L$ in Eq.~(\ref{eq:x}) to obtain:
\begin{equation}\label{eq:xPL}
x = \frac{\nu''}{\nu''_0} = \frac{\nu}{\nu_0}\fracb{T}{T_0}^\frac{m}{2(m+1)} 
\left(\frac{m\!+\!y^{-\!m\!-\!1}}{m\!+\!1}\right)y^\frac{m}{2}\Gamma'^2(1\!-\!\beta'\sin\theta'\cos\phi)\ .
\end{equation}
Substituting $S(x)\to S_{\rm PL}(x)=x^{-\alpha}$ and: 
\begin{equation}
\label{PLradius}
f(R/R_0) =\left\{ \begin{array}{ll} \frac{(a+1) }{(R_f/R_0)^{a+1}-1} (R/R_0)^a & R_0<R<R_f\ ,\\
0 & {\rm otherwise}\ ,
\end{array} \right.
\end{equation}
in Eq.~(\ref{generalFnu}) yields
\begin{equation}
\label{flux2}
\begin{split}
F_{\nu}(T) =\,\frac{2\Gamma_0 \Gamma' L''_{\nu''_0}}{4\pi D^2 }\!\fracb{\nu}{\nu_0}^{-\alpha}
\fracb{T}{T_0}^{2a-m(1+\alpha) \over 2(m+1)} & \frac{(a+1)}{\fracb{R_f}{R_0}^{a+1}-1} 
\int_{y_{\rm min}}^{y_{\rm max}}\! \! dy 
\left(\frac{m\!+\!1}{m\!+\!y^{-\!m\!-\!1}}\right)^{2+\alpha} y^{a-1-\frac{m}{2}(1+\alpha)}\\
&\times \frac{1}{2\pi\Gamma'^{\,4+2\alpha}}\! \int_0^{2\pi}\!\!d\phi\,(1\!-\!\beta'\sin\theta'\cos\phi)^{k-3-\alpha}\ .
\end{split}
\end{equation}

It is easy to see that Eq.~(\ref{flux2}) reduces to Eq.~13 of \citet{Granot(2008)} for $\Gamma'=1$ 
(and without the attenuation term in that paper, due to $\gamma$-$\gamma$ opacity).
At the other extreme, $\Gamma'\gg1$, Eq.~(\ref{flux2}) can be approximately estimated 
(see Appendix~\ref{approx} for a derivation) as:
\begin{equation}
\label{approxFnu2}
 F_{\nu}(T) \approx \,\frac{\Gamma_0 \Gamma'^{\,1-2k} L''_{\nu''_0}}{2^{k}\pi D^2 }\!\fracb{\nu}{\nu_0}^{-\alpha}
\fracb{T}{T_0}^{2a-m(1+\alpha) \over 2(m+1)}
\frac{(a+1)(m+2)^\frac{m(\alpha-1)-2(a+1)}{2(m+1)}}{\left(1+\frac{\Delta R}{R_0}\right)^{a+1}-1}
\frac{\Gamma(2+\alpha-k)}{\Gamma(3+\alpha-k)}\ . 
\end{equation}
We see that for a blob like emission ($k=0$) we obtain $F_{\nu}(T) \propto \Gamma'$ as expected from beaming. 
For continuous emission ($k=1$), the flux appears to be suppressed by a factor $\Gamma'^{-2}$.
This is because in continuous emission, the emission at each location in the jet's bulk frame remains constant in time
as the emitting plasma flows through it, while for a single blob of plasma the location of the emission (i.e. of the blob) 
moves in the jet's bulk frame and chases after the photons it emits, and for the viewing angle from which it appears most
luminous it trails them just by a small amount, so that the same emitted photons (over
the same time interval in the jet's bulk frame) arrive to a stationary observer (in the jet's bulk frame, and also to us)
over a shorter time interval (by a factor of $dt'_{\rm obs}/dt' = (\Gamma'{\mathcal D}')^{-1} = 1-\beta'\sin\theta'\cos\phi'$)
compared to a steady, continuous flow, resulting in a flux larger by the inverse of this factor, $\Gamma'{\mathcal D}'$. 
Alternatively, a continuous flow can be approximated as composed of a large number of blobs, in which case only a 
fraction $(\Gamma'{\mathcal D}')^{-1}$,of the $N$ blobs that are instantaneously within a given segment of the flow 
contribute to the observed emission (or photon front) from that segment, resulting in a flux smaller by this factor compared
to a single blob with the combined luminosity of the $N$ blobs. For the observed flux in our case, the viewing angle relative
to the direction of the emitting fluid in the jet's bulk frame changes with time and between the different parts of the source,
so that altogether the total flux reflect a weighted mean of $(\Gamma'{\mathcal D}')^{-1}$, which is $\sim\Gamma'^{-2}$,
for the suppression factor of a continuous flow compared to a single blob.
Stated differently, the total energy emitted in the two cases, scales as $E \propto L''_{\nu''} \Gamma'^{-2k}$, since in order to 
emit over the same observed duration, a blob needs to emit in its own rest frame over a time longer by a factor of $\sim\Gamma'^{2}$
with the same luminosity. This implies that in order to keep the total emitted energy constant in the
two scenarios, we must change the luminosity in the emitters' frame according to: $L''_{\nu''} \propto \Gamma'^{2k}$. 
Plugging this back to Eq.~(\ref{approxFnu2}) we obtain the expected scaling with $\Gamma'$, independent of $k$.

\subsection{analytic estimates for power law spectra and single emission radius}
\label{SplRdelta}

Another simple test-case is that of a power-law spectra (as in \S \ref{SplRpl}) but with a delta-function emission in radius,
\begin{equation}
f\fracb{R}{R_0}=R_0 \delta(R-R_0) = \delta\left(\frac{R}{R_0}-1\right)
= \delta\left(\frac{y}{y_{\rm min}}-1\right) = y_{\rm min}\delta(y-y_{\rm min})\ , 
\end{equation}
where $y_{\rm min}=(T/T_0)^{-1/(m+1)}$. In this case, one can analytically perform the integral over $y$ and obtain:
\begin{equation}
\label{appdeltaF}
 \begin{split}
 &F_{\nu}(T)=\,\frac{2\Gamma_0 \Gamma' L''_{\nu''_0}}{4\pi D^2 }\!\fracb{\nu}{\nu_0}^{-\alpha}
\bigg(\frac{m\!+\!1}{m\!+\!\frac{T}{T_0}}\bigg)^{2+\alpha}
\frac{\Gamma'^{-4-2\alpha}}{2\pi}\! 
\int_0^{2\pi}\!\!d\phi\!\left(1\!-\!2\beta' \frac{\sqrt{(m\!+\!1)\left(\frac{T}{T_0}-1\right)}}{m\!+\frac{T}{T_0}}\cos\phi\right)^{k-3-\alpha}\ .
\end{split}
\end{equation}

It is convenient to use here the initial radial time $T_0=T_r(R_0)$ given by Eq.~(\ref{eq:T0}), which is the time when the first photons reach the observer (for our definition of $T=0$ as the arrival time of photons from the source when the shell is ejected), 
and the corresponding angular time, $T_{\theta}=T_{\theta}(R_0)=R_0/(2c\Gamma_0^2)=(m+1)T_r(R_0)=(m+1)T_0$ that determines the width of the pulse. We also define $T_s=T_0-T_{\theta}=-mT_0$. We may now write
the expression for the flux as:
\begin{equation}
\label{approxTnu} 
\begin{split}
 F_{\nu}(T)\!=\,\frac{2\Gamma_0 \Gamma' L''_{\nu''_0}}{4\pi D^2 }\!\fracb{\nu}{\nu_0}^{-\alpha}
 \fracb{T\!\!-\!\!T_s}{T_{\theta}}^{-2\!-\!\alpha}
 \frac{\Gamma'^{-4-2\alpha}}{2\pi}\! 
\int_0^{2\pi}\!\!d\phi\!\left(1\!-\!2\beta' \frac{\sqrt{(T\!-\!T_0)T_\theta}}{T-T_s}\cos\phi\right)^{k-3-\alpha}\ .
\end{split}
\end{equation}
For $\beta'=0$ the integrand becomes 1 and we obtain a power law decay with a temporal index $2+\alpha$ as is well known 
for high latitude emission. At the other extreme, for $\beta'\rightarrow1$ the integrand becomes maximal when 
$\sin\theta' = 2\sqrt{(m+1)\bar{T}}/(m+1+\bar{T})\approx 1$ where $\bar{T}\equiv(T-T_0)/T_0$, which occurs
for $\bar{T}\approx m+1$, and $\cos\phi\approx 1$, which occurs for $\phi\approx 0$. Therefore, the peak in the light curve 
occurs at small values of $\bar{t}\equiv [\bar{T}/(m+1)]-1$, and for estimating the fluence (i.e. evaluating the light-curve 
around its peak) we may assume that $\bar{t}\ll 1$, where in this limit $\sin\theta'\approx 1-\bar{t}^2/8$, and 
$(T-T_s)/T_\theta = (m+T/T_0)/(m+1)\approx 2$.  In terms of $\phi$, the width of the region where the dominant contribution 
comes from is roughly $\sim 1/\Gamma'\ll 1$ around $\phi = 0$ or $|\phi|\Gamma'\lesssim\;$a few, so that we can safely assume 
$|\phi|\ll 1$, $\cos\phi\approx 1-\phi^2/2$, and take the limits of integration to be
from $-\infty$ to $\infty$, where $1-\beta'\sin\theta'\cos\phi \approx [1+(\Gamma'\bar{t}/2)^2+(\Gamma'\phi)^2]/2\Gamma'^2 
= (1+\tilde{t}^2+\tilde{\phi}^2)/2\Gamma'^2$ (where $\tilde{t}\equiv\Gamma'\bar{t}/2$ and $\tilde{\phi}\equiv\Gamma'\phi$),
\begin{equation}\label{eq:R0flux}
\begin{split}
 F_{\nu}\left(\frac{T}{T_0}\approx m+2\right)\!&
 \approx\,\frac{2\Gamma_0 \Gamma' L''_{\nu''_0}}{4\pi D^2 }\!\fracb{\nu}{\nu_0}^{-\alpha}
 \fracb{T\!\!-\!\!T_s}{T_{\theta}}^{-2\!-\!\alpha}\frac{\Gamma'^{-4-2\alpha}}{2\pi}\! 
\int_{-\infty}^{\infty}\!\!d\phi\!\fracb{1\;+(\Gamma'\bar{t}/2)^2\!+\!(\Gamma'\phi)^2}{2\Gamma'^2}^{k-3-\alpha}\\
& =  \,\frac{\Gamma_0 \Gamma'^{2-2k} L''_{\nu''_0}}{2^{k-1-\alpha}\pi^2 D^2 }\!\fracb{\nu}{\nu_0}^{-\alpha}
 \int_{-\infty}^{\infty}\!\!d\tilde{\phi}\,(1\!+\tilde{t}^2\!+\!\tilde{\phi}^2)^{k-3-\alpha}\\
& = \,\frac{\Gamma_0 \Gamma'^{2-2k} L''_{\nu''_0}}{2^{k-1-\alpha}\pi^{3/2} D^2 }\!\fracb{\nu}{\nu_0}^{-\alpha}
\!\frac{\Gamma(2.5+\alpha-k)}{\Gamma(3+\alpha-k)}(1\!+\tilde{t}^2)^{k-\frac{5}{2}-\alpha}\ .
\end{split}
\end{equation}
We have $dT =(2T_\theta/\Gamma')d\tilde{t}$, and near the peak $F_\nu(\tilde{t})\approx K(1+\tilde{t}^2)^{k-2.5-\alpha}$ 
so that the fluence is given by
\begin{equation}\label{eq:R0fluence}
\begin{split}
f_\nu = \int dT\,F_\nu(T) \approx 
\frac{2KT_\theta}{\Gamma'}\int_{-\infty}^{\infty}\!\!d\tilde{t}\, (1\!+\tilde{t}^2)^{k-\frac{5}{2}-\alpha}
= \frac{\Gamma_0 \Gamma'^{1-2k} L''_{\nu''_0}T_\theta}{2^{k-2-\alpha}\pi D^2 }\!\fracb{\nu}{\nu_0}^{-\alpha}
\!\frac{\Gamma(2+\alpha-k)}{\Gamma(3+\alpha-k)}\ .
\end{split}
\end{equation}
The dependence on $\Gamma_0$ and $\Gamma'$ in Eq.~(\ref{eq:R0fluence}) is the same as in Eq.~(\ref{approxFnu2}),
while in Eq.~(\ref{eq:R0flux}) there is an extra power of $\Gamma'$. This arises since the width of the peak in the
light-curve for the emission from a single radius is $\sim T_\theta/\Gamma'$, i.e. $\sim 1/\Gamma'\ll 1$ times the 
angular time from that radius, due to the beaming of the emission in the jet's bulk frame. The time dependence of the 
emission near the peak, in Eq.~(\ref{eq:R0flux}) is dominated by the beaming in the jet's bulk frame, while at late times
$T\gg T_\theta$ it becomes dominated by the usual high-latitude emission, as the beaming in the jet's bulk frame approaches
a constant (i.e. the $\hat{n}'\approx -\hat{r}$ and the integral in Eq.~(\ref{approxTnu}) approaches a constant, leaving only 
the explicit time dependence in front of the integral). While for $\Delta R = R_f-R_0 \sim R_0$ the pulse width is $\sim T_\theta$,
the normalization of the emission from a single radius is such that the fluence in both cases is similar.
\begin{figure*}[h]
\centering
\includegraphics[scale=0.28]{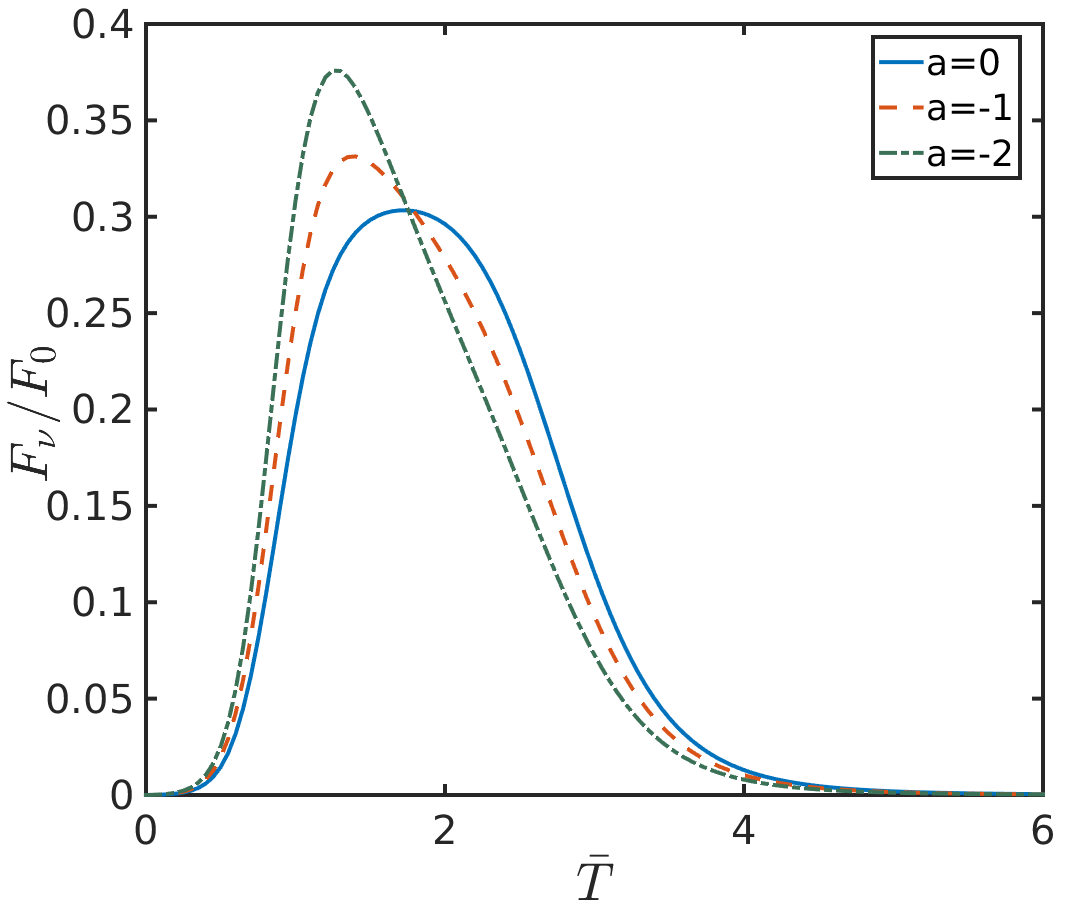}
\includegraphics[scale=0.28]{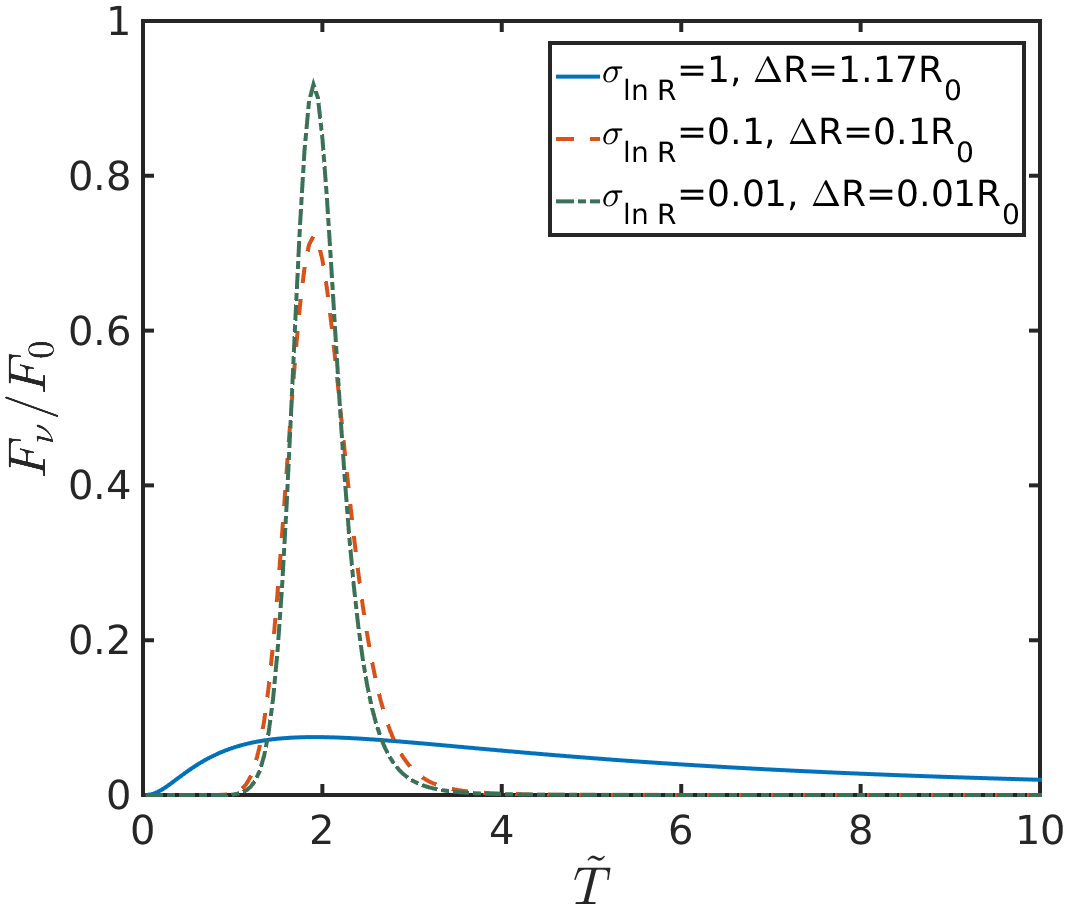}
\caption{Lightcurves of a single pulse with a power-law between $R_0$ and $R_f=2R_0$ ({\it left panel}) and 
log-normal ({\it right panel}) dependence of the luminosity on radius. All cases are plotted for a typical Band function spectral 
emissivity and $\Gamma'=3, m=0, k=0$. The flux is given in units of the ``typical" peak flux $F_0$ (giving in Eq.~[\ref{eq:F0}]), 
and shown as a function of the normalized time $\bar{T}$ or $\tilde{T}$ (given in Eq.~[\ref{eq:T}]). $\tilde{T}=0$ corresponds 
to the ejection time, and $\bar{T}=0$ corresponds to the time when the first photons reach the observer. We show results for power-law and log-normal luminosities as a function of radii for three different values of the power-law index $a$ and the standard deviation $\sigR$, respectively.}
\label{fig:fR}
\end{figure*}

\subsection{Band spectrum emission for different dependencies of the luminosity on radius}
\label{pulseshapes}
Using Eq.~(\ref{generalFnu}) we calculate the resulting light-curve for a Band spectrum emission function with a low-energy photon index $\alpha_B=-1$ and a high-energy photon index $\beta_B=-2.3$, typical values obtained 
from fitting of GRB data. For the radial dependence of the emissivity we use some specific functions for $f(R/R_0)$.
For clarity, in all cases we take $\Gamma'=3, m=0, k=0$. We plot two typical cases in Fig.~\ref{fig:fR}: a power law in 
radius (Eq. \ref{PLradius}) with $R_f=2R_0$ and a Gaussian in $\ln(R)$:
\begin{equation}
\label{Gaussianradius}
f(R/R_0) =\frac{R_0/R}{\sqrt{2 \pi}\,\sigR}\,\exp\!\left(-\frac{[\ln(R/R_{0,{\rm eff}})]^2}{2\sigR^2}\right)\ ,
\end{equation}
where $R_{0,{\rm eff}}\equiv \exp(\sigR^2)R_0$ is a slightly different effective radius chosen such that $f(R/R_0)$ would peak at $R=R_0$. Note that $\sigR$ can be related to the typical width in $R$ through: $\Delta R\approx\sinh(\sigR)R_0$
(which for $\sigR \ll 1$ yields $\Delta R\approx \sigR R_0$). 
We plot the results at the frequency $\nu_p$. The flux is given in units of the ``typical" peak flux, 
\begin{equation}\label{eq:F0}
F_0\equiv \frac{2\Gamma_0 \Gamma'^{1-2k} L''_{\nu''_0}}{4\pi D^2 }\ ,
\end{equation}
and the time in units of the typical time $T_0$, either starting at the observed ejection time $T_{\rm ej}$ (denoted as $\tilde{T}$) 
or starting at $T_{\rm ej}+T_0$, the time 
when the first photons reach the observer (this is defined as $\bar{T}$),
\begin{equation}\label{eq:T}
\tilde{T}\equiv \frac{T-T_{\rm ej}}{T_0}\ ,\quad\quad \bar{T}\equiv\frac{T-T_0-T_{\rm ej}}{T_0}=\tilde{T}-1\ .
\end{equation}
For a power-law emissivity, $f(R/R_0)\propto R^a$, the pulse shape becomes less symmetric and peaks at earlier times 
as $a$ decreases. For a Gaussian $f(R/R_0)$, on the other hand, the peak occurs later for smaller typical widths ($\Delta R$) 
but becomes more symmetric. The power-law and Gaussian cases approach the case of a delta function in radius and reproduces 
the result from  \S \ref{SplRdelta} in the limits $a\to-\infty$ and $\Delta R \rightarrow 0$, respectively. We note also that 
in the Gaussian case the typical width of the pulse is $ \sim\max(T_0/\Gamma', T_0\Delta R/R_0)$ in accordance with 
Eqs.~(\ref{appdeltaF}), (\ref{approxFnu2}).

\begin{figure*}[h]
\centering
\includegraphics[scale=0.28]{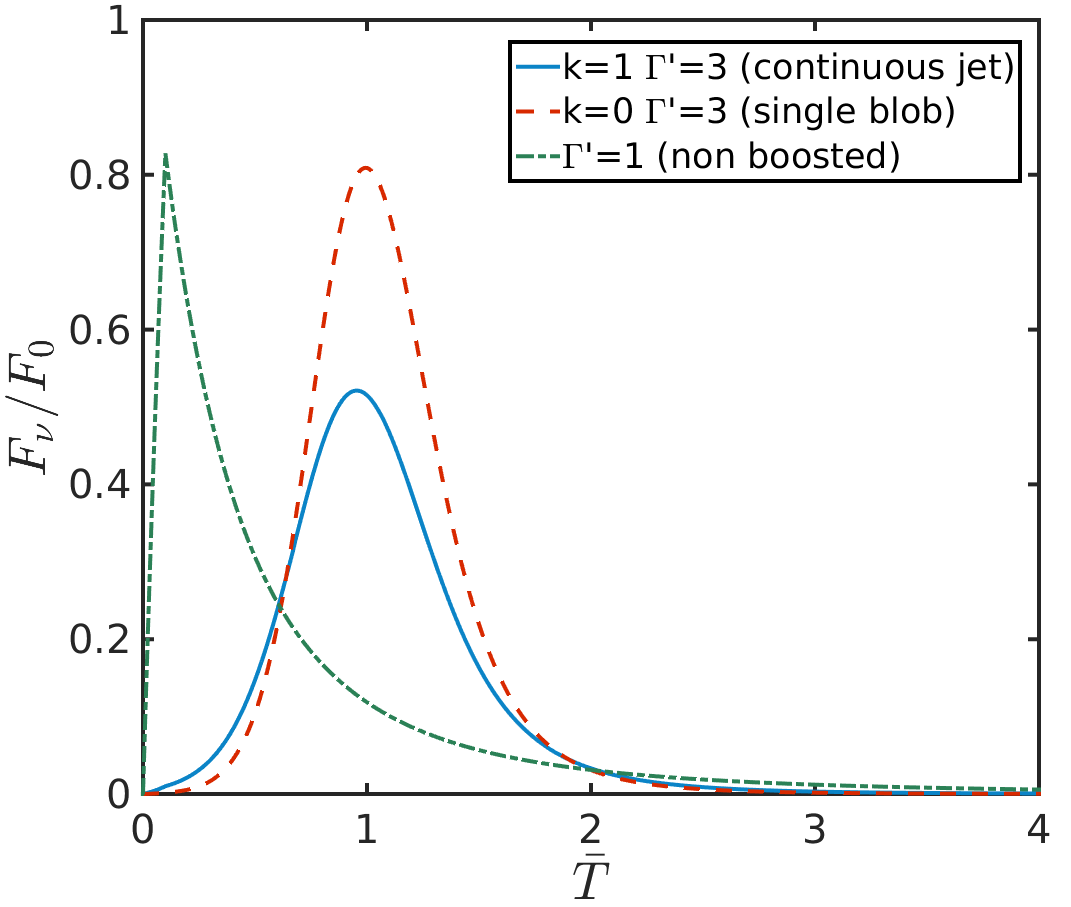}\includegraphics[scale=0.28]{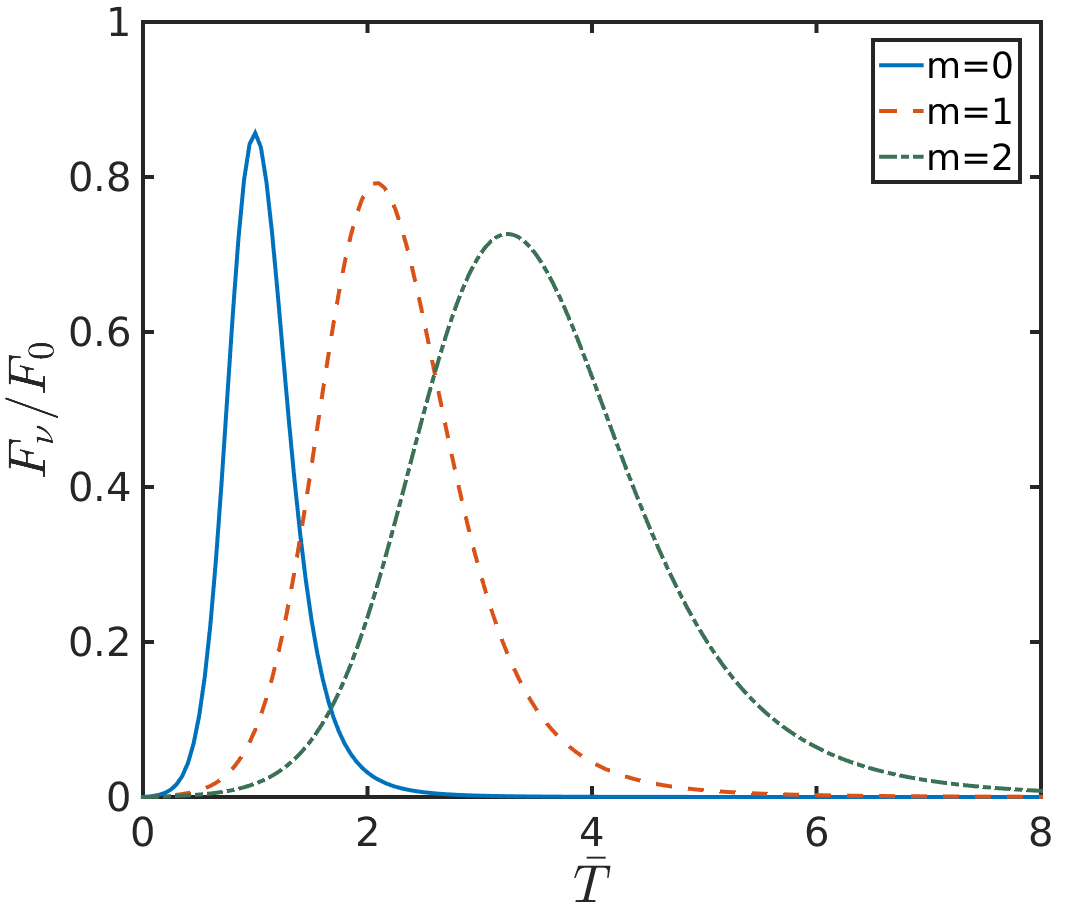}
\caption{{\bf Left}: single pulse light-curves for different assumptions on the emission (continuous and 
steady in the jet's bulk frame vs. single blob vs. non-boosted emission). All cases are plotted with a typical 
Band function spectral emissivity, a constant dependence of luminosity on the radius with $R_f=1.1R_0$ 
as well as $m=0$. The flux is given in units of the ``typical" peak flux $F_0$ as a function of the 
normalized time $\bar{T}$. {\bf Right}: light-curves with $k=0$, $\Gamma'=3$, a constant 
dependence of luminosity on the radius with $R_f=1.1R_0$ and different values of $m$.}
\label{fig:contvsblob}
\end{figure*}

\subsection{Exploring the single-pulse parameter space}
\label{paramspace}
\begin{figure*}[h]
\centering
\includegraphics[scale=0.35]{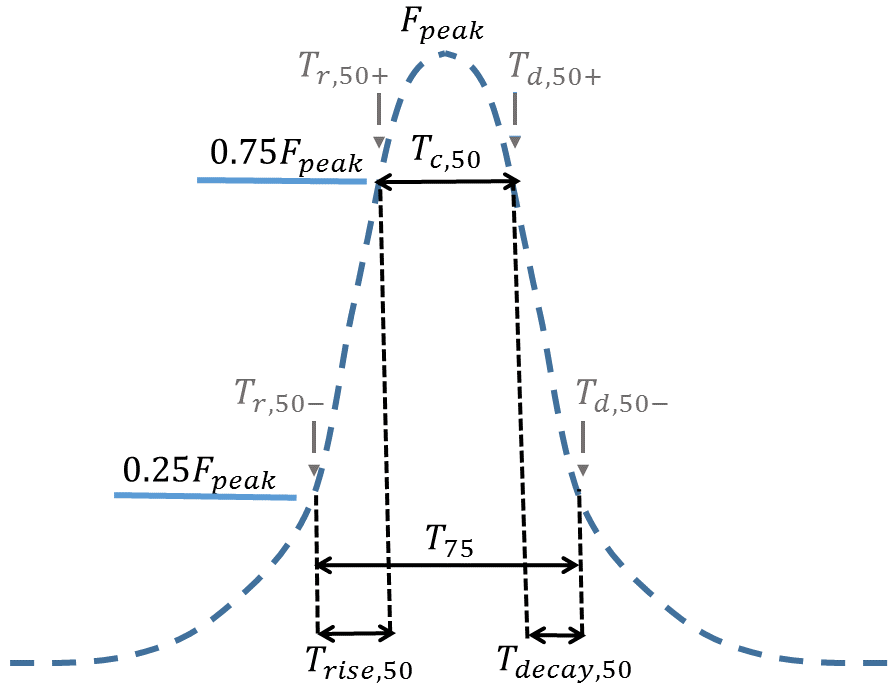}
\caption{Schematic figure of the rise, decay and ``central" times defined in the text for $X=50$.
$T_{{\rm rise},X}$ ($T_{{\rm decay},X}$) is the time it takes the flux to rise (decay) from $X_{-}\%$ ($X_{+}\%$) to $X_{+}\%$ ($X_{-}\%$) of the peak value (where $X_\pm = (100\pm X)/2$).
$T_{c,X}$ is the time the flux stays above $X_{+}\%$ of its peak value.}
\label{fig:times}
\end{figure*}

Here we explore the dependence of the single pulse light-curve on the various model parameters.
We define $T_{r,Y}$ ($T_{d,Y}$) as the first time when the flux first (last) reaches $Y\%$ of its peak value. The rise (decay) time, is then defined by:
$T_{{\rm rise},X}\equiv T_{r,X_+}-T_{r,X_-}$ ($T_{{\rm decay},X}=T_{d,X_-}-T_{d,X_+}$) where $X_\pm = (100\pm X)/2$.
In addition, we define the ``central" time, $T_{c,X}\equiv T_{d,X_+}-T_{r,X_+}$ (see Fig.\ref{fig:times}).
Notice that if there is a significant ``plateau" stage then it is possible that $T_{\rm rise}, T_{\rm decay} \ll T_c$.
If the rise and decay of a pulse are monotonic,
then $T_{100-X_-}\equiv T_{{\rm rise},X}+T_{c,X}+T_{{\rm decay},X} = T_{d,X_-}-T_{r,X_-}$ is the time spent at a flux above 
$X_-\%$ of the peak flux (analogous to the definition of GRB durations but with flux instead of fluence).

First, let us consider the more familiar case of non-boosted emission in the jet's bulk frame.
If $\Delta R \lesssim R_0$ then the flux peaks at approximately:
$T = T_{\rm ej}+T_f$, where $T_f = (R_f/R_0)^{m+1}T_0$, or $\bar{T}_f=(R_f/R_0)^{m+1}-1=[1+(\Delta R/R_0)]^{m+1}-1$.
The rise time is then given by:
\begin{equation}\label{eq:T_rise}
\frac{T_{\rm rise}}{T_0}=\bar{T}_f  =\fracb{R_f}{R_0}^{m+1}-1 = \left[1+\frac{\Delta R}{R_0}\right]^{m+1}-1
\ \ \xrightarrow[{\Delta R\ll R_0}]\ \  (m+1)\frac{\Delta R}{R_0}\ ,
\end{equation}
and the decay time $T_{\rm decay}\approx T_{\theta}(R_f)$ is given by:
\begin{equation}\label{eq:T_decay}
\frac{T_{\rm decay}}{T_0} = (m+1)\fracb{R_f}{R_0}^{m+1} = (m+1)\left[1+\frac{\Delta R}{R_0}\right]^{m+1}
\ \ \xrightarrow[{\Delta R\ll R_0}]\ \  (m+1)\left[1+(m+1)\frac{\Delta R}{R_0}\right]\ ,
\end{equation}
leading to a total pulse width $\Delta T = T_{\rm rise}+T_{\rm decay}$ of:
\begin{equation}\label{eq:pulse_width}
\frac{\Delta T}{T_0} = (m+2)\fracb{R_f}{R_0}^{m+1}-1 = (m+2)\left[1+\frac{\Delta R}{R_0}\right]^{m+1}-1
\ \ \xrightarrow[{\Delta R\ll R_0}]\ \  (m+1) \left[1+(m+2)\frac{\Delta R}{R_0}\right]\ ,
\end{equation}
which for $m=0$ leads to $\Delta T/T_0 \sim R_f/R_0$.
For $\Delta R \gg R_0$ both the rise time and the decay time may be altered due to the emissivity function (see Table \ref{tbl:cases} for details),
which (for the case of power law luminosity as a function of radius and frequency) dominates the emission so long as $0.5 \lesssim \bar{T} <\bar{T}_f$.
The relevant parameter in this case is the power law index $q$ in the expression for $F_{\nu}(T)\propto T^q$. Using Eq. \ref{approxFnu2} 
we define:
\begin{equation}
 q = \frac{2a-m(1+\alpha)}{2(m+1)}\ .
\end{equation}
Specifically for $q=0$ the flux at this interval reaches an approximate plateau.
This is an illustration of a situation in which $T_c \gg T_{\rm rise}$.
We define a quantitative measure of pulse asymmetry as $\Lambda \equiv T_{\rm rise}/T_{\rm decay}$. Pulses are asymmetric if $\Lambda \gg 1$ or $\Lambda \ll 1$.
In fact, even for $\Lambda=1$, the shape of the light-curve could still
be different between the rise and decay parts. For the non-boosted emission we find that unless the emissivity is increasing with radius, then
$\Lambda \ll 1$ (estimates for individual cases are listed in Table \ref{tbl:cases}) implying shorter rise than decay time and a large degree of asymmetry regardless of the relation between $\Delta R$ and $R_0$.
In fact it is hard to produce very symmetric pulses ($\Lambda \approx 1$) for the case of such non-boosted emission, since the mechanisms controlling the rise and decay times are different.
We now compare these results to the anisotropic case.

\begin{figure*}[h]
\begin{center}
\includegraphics[scale=0.4]{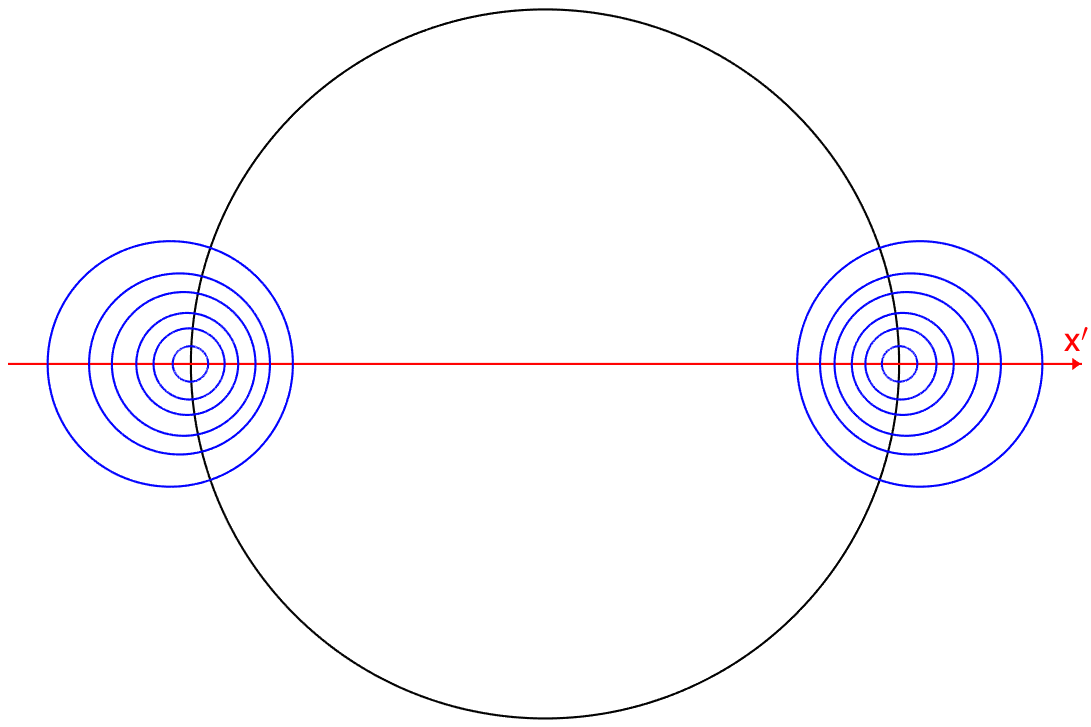}
\caption{The emission region -- the blue contours show the area within the emitting spherical shell that is 
within the beaming cone of the emitting material in the jet's bulk frame, i.e. where $\hat{n}'$ is within an angle 
of $1/\Gamma'$ from the direction of its velocity $\hat{\beta}' = \pm\hat{x}'$, for $\Gamma' = 3,4,5,7,10,20$.
The big black circle indicates an angle of $\theta=1/\Gamma$ around the line of sight to the central source,
while the red arrow shows the $x'$-axis.}
\label{fig:emission_region}
\end{center}
\end{figure*}

In Fig.~\ref{fig:contvsblob} we plot the dependence of the pulse shape on the nature of the emission: continuous, single blob 
and non-boosted emission, for a constant co-moving luminosity with $\Delta R/R_0 = 0.1$.
For $\Delta R <R_0$ 
the flux peaks earlier for non-boosted emission than for the cases with anisotropy.
The light-curve is sharper and less symmetric for the non-boosted emission so long as $\Delta R<R_0/\Gamma'$. This arises since 
for $\Delta R<R_0/\Gamma'$ in the non-boosted case the pulse rise time is determined by $\Delta R$ 
($\bar{T}_{\rm rise} \sim \Delta R/R_0$) while its decay time is $T_{\rm decay}\sim T_0$, whereas in 
the anisotropic case the pulse is rather symmetric ($\Lambda\approx 1$), since $\bar{T}_{\rm rise}\sim \bar{T}_{\rm decay}\sim 1/\Gamma'$ (and
its width reflects that for the emission from a single radius).
To easily understand the latter expression, consider that the emission from
each radius is dominated by two regions of angular size $\sim 1/(\Gamma'\Gamma)$ around the two points of the intersection 
of the $x'$-axis (where $\vec{\beta'}=\pm\beta' \hat{x'}$) and the circle $\theta = 1/\Gamma$, or $\xi = (\Gamma\theta)^2=1$
(see Fig.~\ref{fig:emission_region} and \S~\ref{sec:angtimescale}). This dominant contribution arrives over a time
$\sim T_0/\Gamma'$, which is thus the ``smearing time-scale" compared to the very large $\Gamma'$ limit.
In the right panel of Fig. \ref{fig:contvsblob} we plot also the dependence of the light-curve on $m$ for the anisotropic case.
As shown in \S \ref{SplRdelta} (and can be seen for the non-boosted case from Eq. \ref{eq:T_rise}), in terms of the normalized time 
$\bar{T}$ the peak occurs at $\bar{T}\approx m+1$, which is when the beaming cone points toward the observer,
after accounting for the aberration of light for the jet's frame to the observer's frame. 
This behaviour, of the peak occurring at a later time for a larger $m$, can indeed be seen in the right panel of 
Fig.~\ref{fig:contvsblob}.

\begin{figure*}[h]
\includegraphics[height=0.3\textwidth,width = 0.3\textwidth]{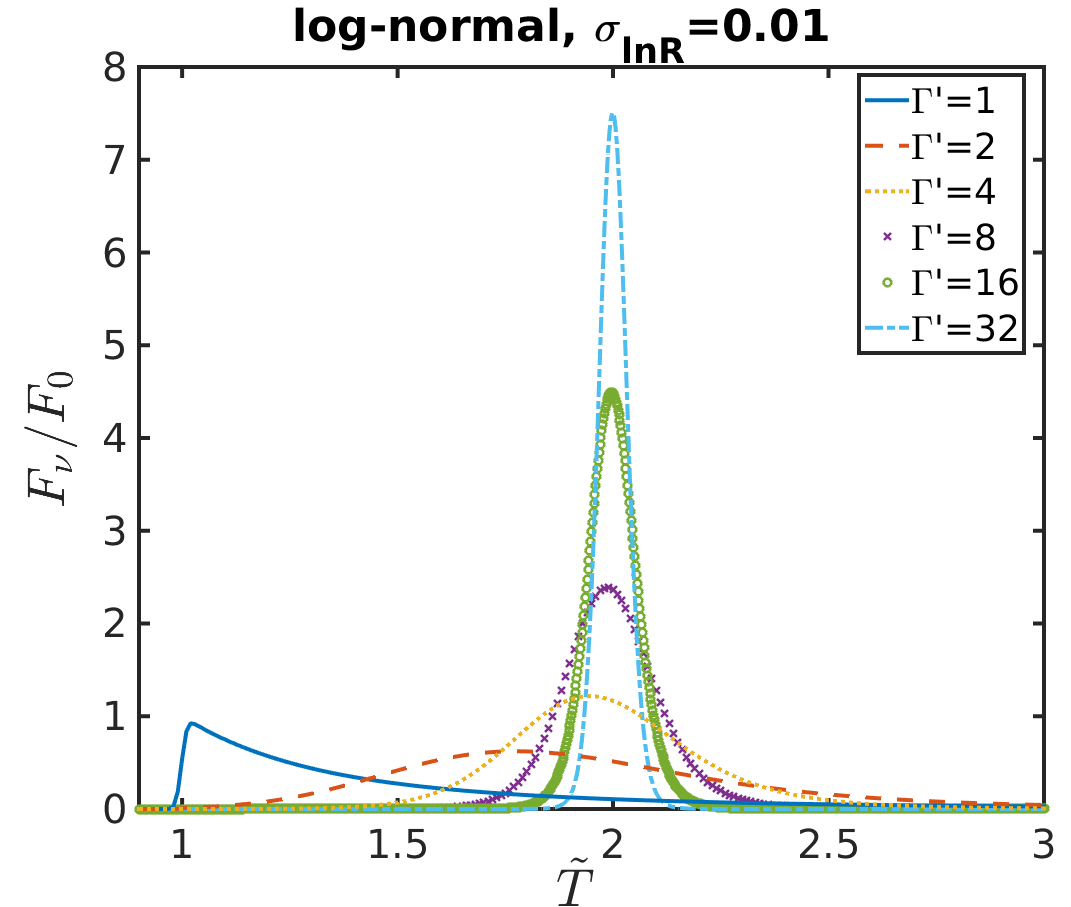}
\includegraphics[height=0.3\textwidth,width = 0.3\textwidth]{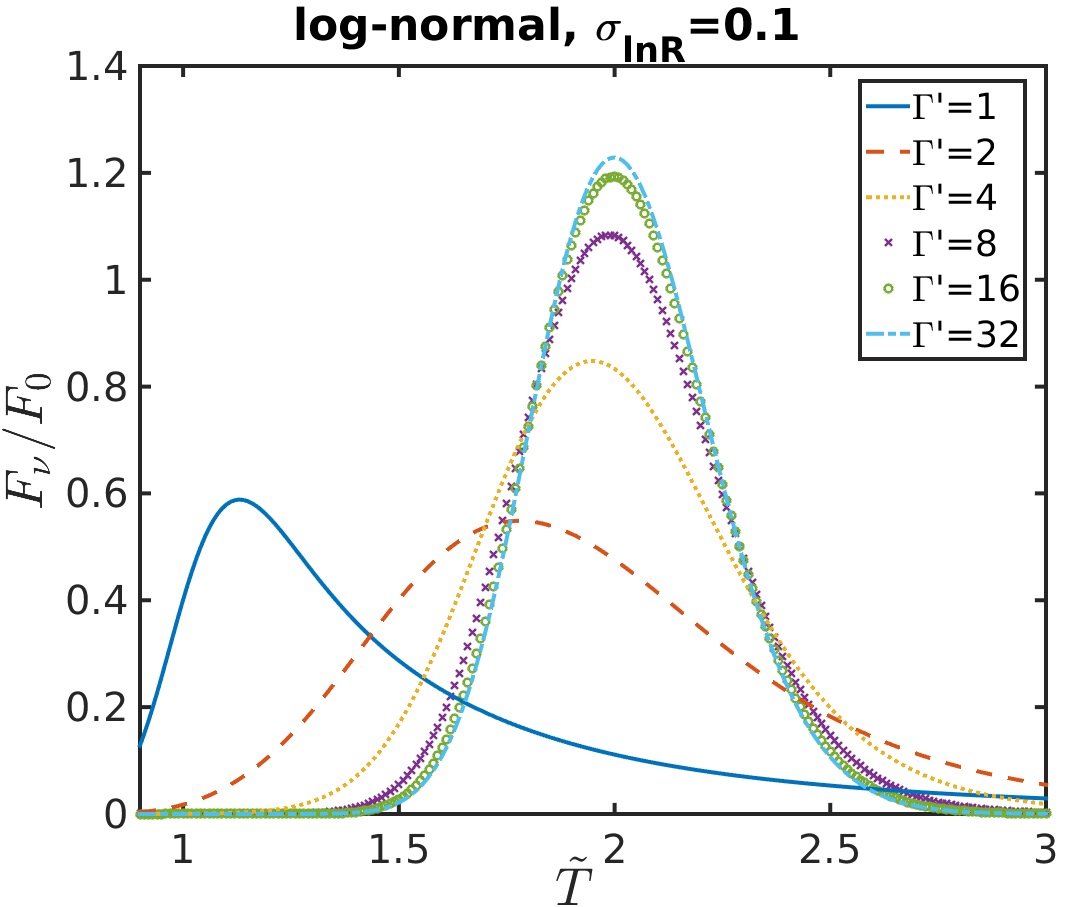}
\includegraphics[height=0.3\textwidth,width = 0.3\textwidth]{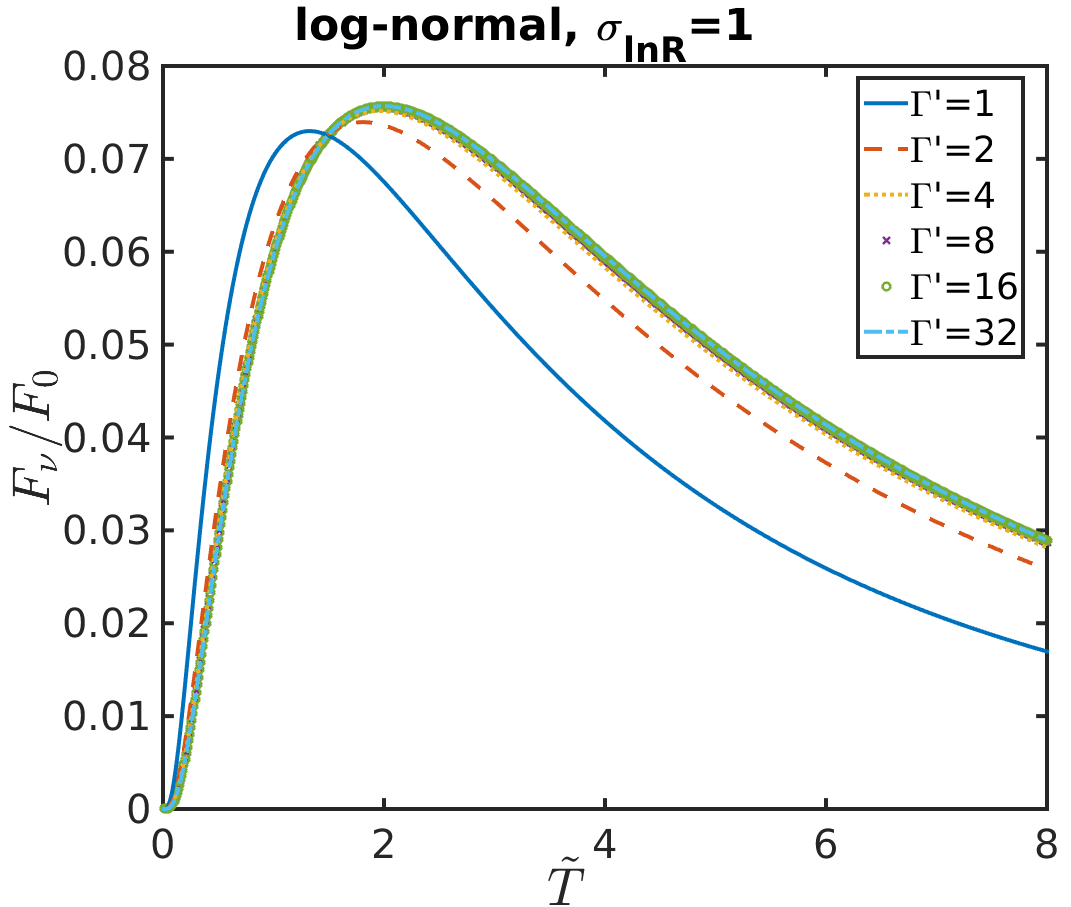}\\ \vspace{0.0cm}\\
\includegraphics[height=0.3\textwidth,width = 0.3\textwidth]{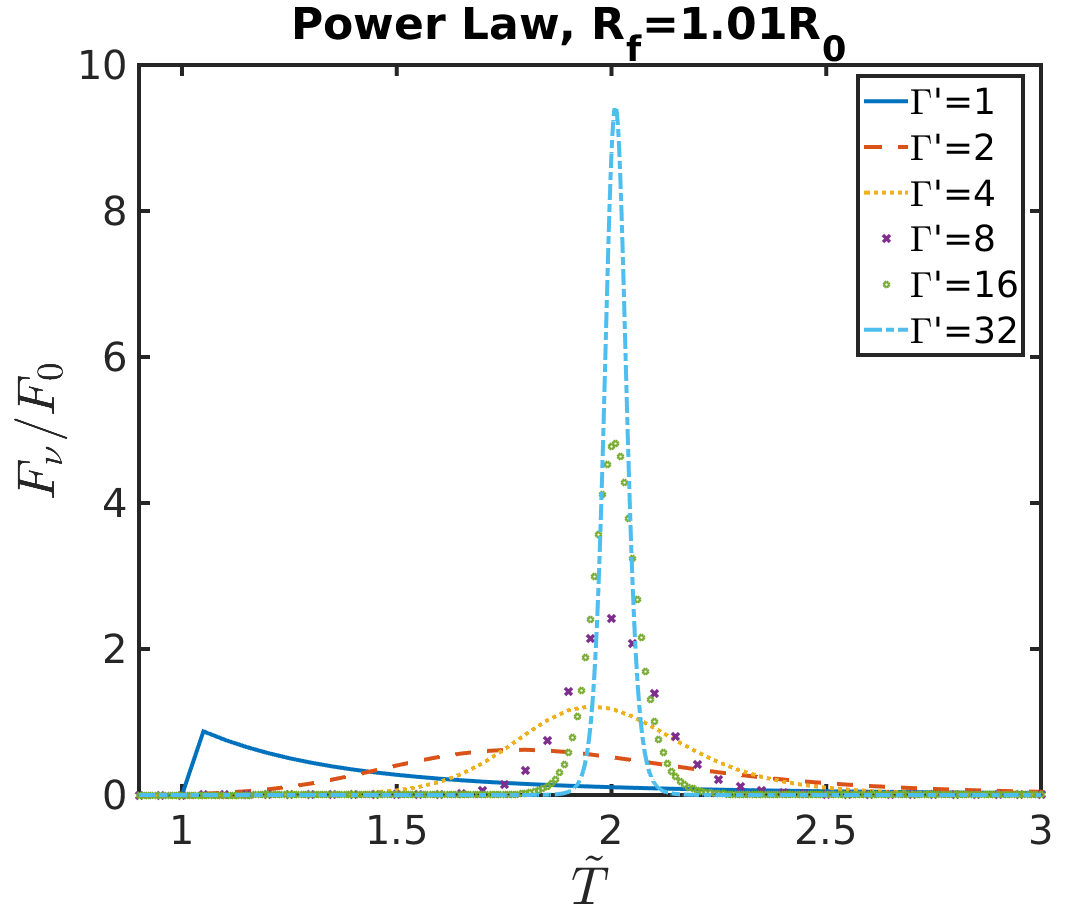}
\includegraphics[height=0.3\textwidth,width = 0.3\textwidth]{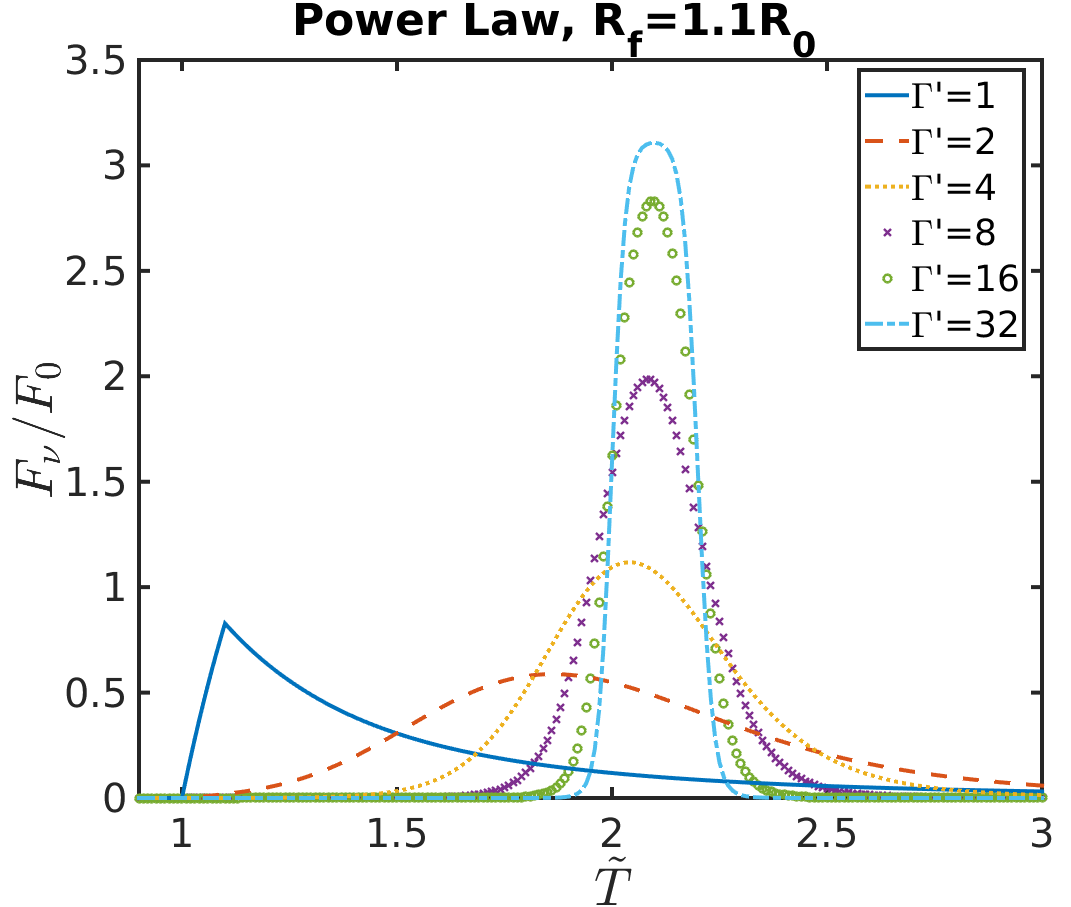}
\includegraphics[height=0.3\textwidth,width = 0.3\textwidth]{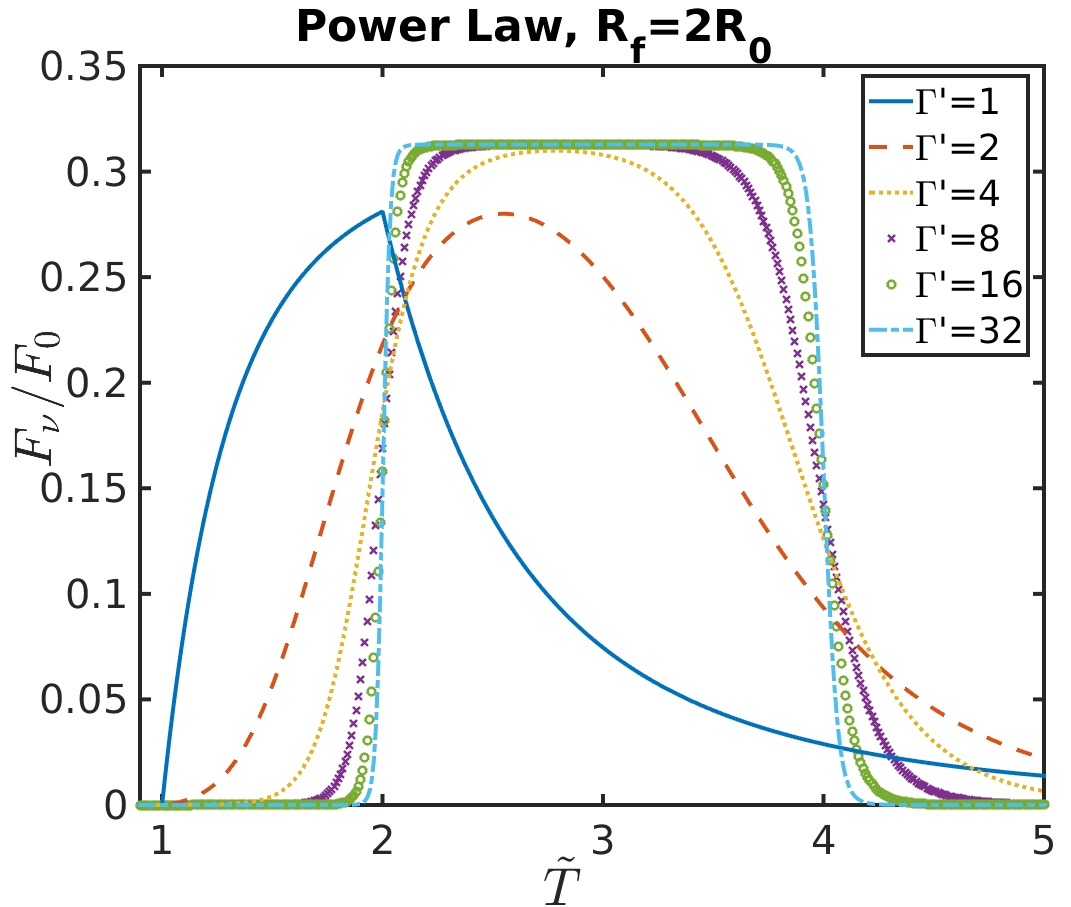}\\
\caption{Light-curves of a pulse assuming different values of the emitter's Lorentz factor $\Gamma'$ 
($\log_{2}(\Gamma')=0,\,1,\,2,\,3,\,4,\,5$) with $m=0$ and $k=0$. 
The flux is given in units of $F_0$ as a function of the normalized time $\tilde{T}$. 
{\bf Top panels}: results for a log-normal dependence of luminosity on radius
with different typical widths ($\sigma_{lnR}=0.01,\,0.1,\,1$ from left to right). 
{\bf Bottom panels}: results for a constant dependence of luminosity on radius with different typical widths
($\Delta R/R_0=(R_f-R_0)/R_0=0.01,\,0.1,\,1$ from left to right).}
\label{fig:Gammap}
\end{figure*}

\begin{figure*}[h]
\begin{center}
\includegraphics[scale=0.24]{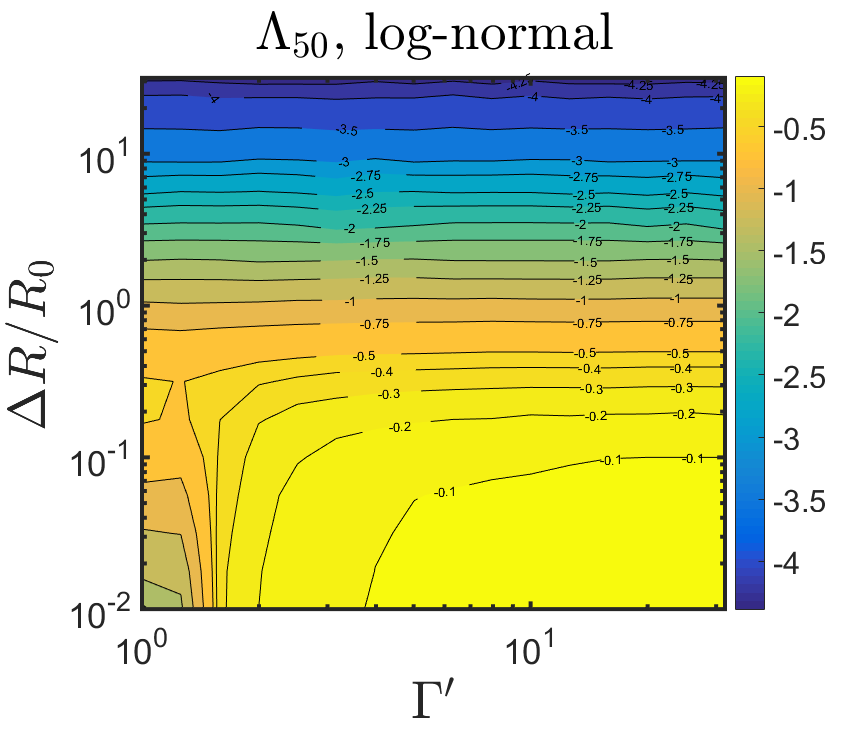}
\includegraphics[scale=0.24]{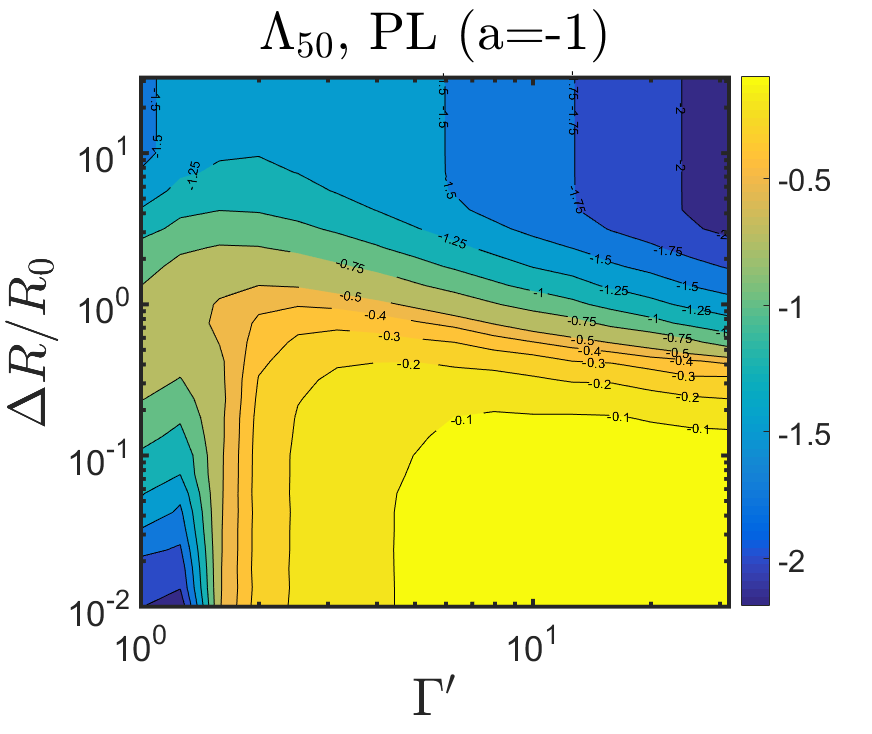}\\
\includegraphics[scale=0.24]{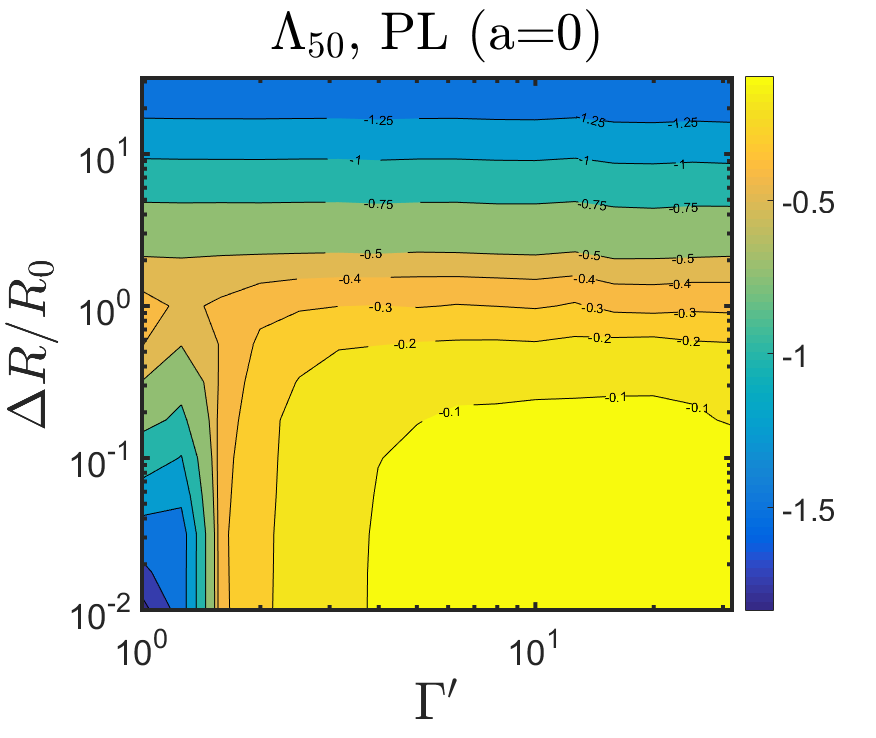}
\includegraphics[scale=0.24]{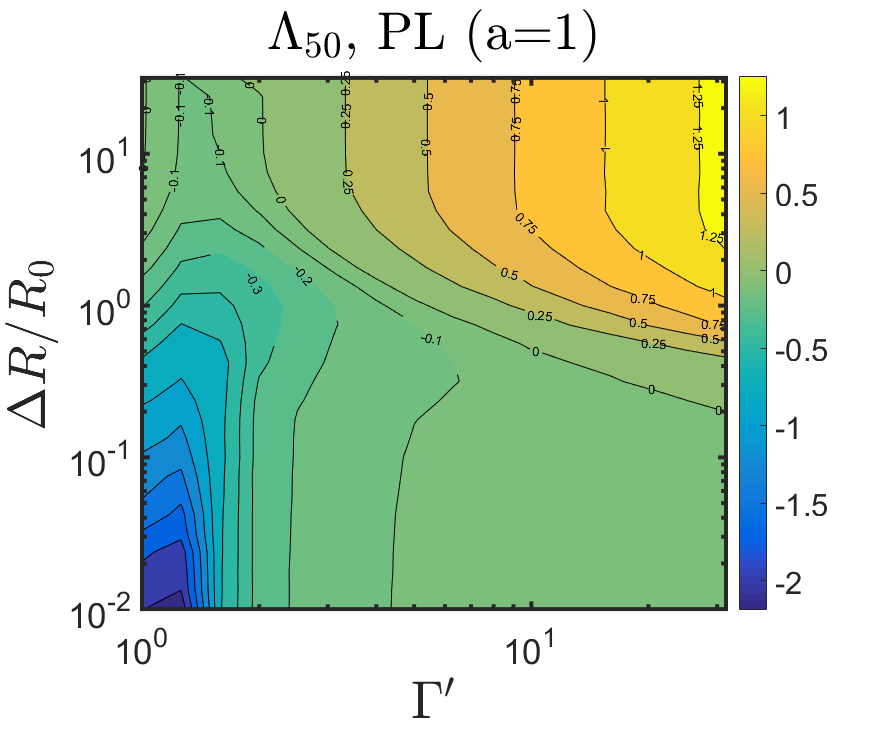}\\
\caption{The asymmetry ratio: $\Lambda_{50} \equiv T_{\rm rise,50}/T_{\rm decay,50}$ as a function of $\Gamma'$ and $\Delta R/R_0$.
All plots are for $k=0,m=0$ and a power law spectrum with $\alpha=1$. 
{\bf Top Left:} results for log-normal emissivity as a function of radius.
{\bf Top Right:} results for PL emissivity as a function of radius with $a=-1$ (corresponding to $q=-1$).
{\bf Bottom Left:} results for PL emissivity as a function of radius with $a=0$ (corresponding to $q=0$).
{\bf Bottom Right:} results for PL emissivity as a function of radius with $a=1$ (corresponding to $q=1$).}
\label{fig:Lambda2}
\end{center}
\end{figure*}

In Fig.~\ref{fig:Gammap} we plot the dependence of the pulse shape on the values of $\Gamma'$. 
Once $\Gamma' \gg 1$ and $1/\Gamma' \ll \Delta R/R_0$, the pulse's shape does not depend strongly on $\Gamma'$, and is set instead by $f(R/R_0)$.
This is because in this case, the typical times related to the emissivity function are larger than 
the pulse width from a single radius, $\Delta T\sim T_0/\Gamma'$. 
The top-right panel of Fig.~\ref{fig:Gammap} shows light-curves for $\Delta R/R_0=1$ and a log-normal 
radial dependence of the co-moving emissivity. It demonstrates that for $\Delta R \sim R_0$ with a smooth
turn-on and turn-off of the emissivity, as might typically be expected, the pulse-shape for anisotropic emission
is not very different from that of non-boosted emission, showing a similarly noticeable degree of asymmetry
($\Lambda_{50}\approx 0.107$, $\Lambda_{90}\approx 0.096$ for the non-boosted case as opposed to 
$\Lambda_{50} \approx 0.118$, $\Lambda_{90} \approx 0.173$ for the anisotropic case).
The asymmetry becomes larger for $\Delta R \gg R_0$ (and remains independent of $\Gamma'$). This is because the rise time is dominated by the radial time-scale at the point where the emissivity function is within $1\sigma$ of the peak, 
i.e. $R_i \approx R_0^2/\Delta R \Longrightarrow T_{\rm rise}\sim (R_0/\Delta{R})^{\!m\!+\!1}$, 
whereas the decay time is dominated by the radial time-scale where the emissivity function is $1\sigma$ above the peak: 
$R_f \approx \Delta R \Longrightarrow T_{\rm decay}\sim (\Delta{R}/R_0)^{\!m\!+\!1}$. 
This leads to $\Lambda\sim (R_0/\Delta{R})^{2(m\!+\!1)}$ (in fact, there is a numerical pre-factor here of order 0.1 arising from the smooth shape of the log-normal function's decline, see table \ref{tbl:cases}).
Slightly asymmetric pulses are in broad agreement with the typical
observed pulse shape. On the other hand, symmetric pulses are also sometimes observed, and are hard to produce
with non-boosted emission, but rather easy to produce for anisotropic emission.
A summary of the typical times for the various cases is given in table \ref{tbl:cases}.
The expected asymmetry measures in our model can be seen explicitly in Fig. \ref{fig:Lambda2} where we plot the value of $\Lambda$ for different values of $\Gamma', \Delta R/R_0$ and for different
assumptions on the emissivity as a function of radius. 
\begin{landscape}
\begin{table}[h]\tiny
\renewcommand{\arraystretch}{2.2}
\begin{tabular}{ |c|c |c |c | c | c | c | c |c |}
\hline
case & width &  $\Gamma'$ &  $f(R)$ & $\bar{T}_{\rm rise,X}$ &  $\bar{T}_{\rm decay,X}$ & $\Lambda$ & $\bar{T}_{c,X}$  & description \\
\hline
1 & $\Delta R \!\ll \!R_0$ & $\Gamma'=1$ & any & $\bar{T}_f\!\approx \!(m\!+\!1)\frac{\Delta R}{R_0}$ & $\tilde{T}_{\theta}(R_f)$ & $\frac{\Delta R}{R_0}\ll 1$ &$\bar{T}_{\rm rise} \ll \bar{T}_c \ll \bar{T}_{\rm decay}$ & { asymmetric, no plateau}\\
2 & $\Delta R\! \ll \!R_0$ & $\frac{\Delta R}{R_0}\!<\!\frac{1}{\Gamma'}<1$ & any & $\frac{1}{\Gamma'}$ & $\frac{1}{\Gamma'}$ & 1 &$\bar{T}_c\lesssim \bar{T}_{\rm rise}+\bar{T}_{\rm decay}$ & {symmetric, no plateau}\\ 
3 & $\Delta R \!\ll\! R_0$ & $\frac{1}{\Gamma'}\!<\!\frac{\Delta R}{R_0}<1$ & PL; any $q$ & $\frac{1}{\Gamma'}$ & $\frac{1}{\Gamma'}$ & 1 & $\bar{T}_c \!\approx\! \frac{\Delta R}{R_0}\!\gg\! \bar{T}_{\rm rise}\!\sim\! \bar{T}_{\rm decay}$ & {symmetric, short plateau}\\ 
4 & $\Delta R \!\ll \!R_0$ & $\frac{1}{\Gamma'}\!<\!\frac{\Delta R}{R_0}<1$ & log-normal & $\bar{T}_f\!\approx \!(m\!+\!1)\frac{\Delta R}{R_0}$ & $\bar{T}_{f}\approx (m\!+\!1)\frac{\Delta R}{R_0}$ & 1 & $\bar{T}_c\lesssim \bar{T}_{\rm rise}+\bar{T}_{\rm decay}$ & {symmetric, no plateau}\\ 
5 & $\Delta R\!\gg \!R_0$ & $\Gamma'=1$ & PL; $q\!<\!0$~$^\dagger$ & $\tilde{T}_{\theta}(R_0)=1$ & $\bar{T}_f$ & $\fracb{R_0}{R_f}^{m\!+\!1}$ & $\bar{T}_{\rm rise}\!\ll\!\bar{T}_c\!\approx\!\bar{T}_{\rm decay}\!\approx\!\fracb{\Delta R}{R_0}^{m\!+\!1}$ & { asymmetric, no plateau}\\
6 & $\Delta R \!\gg \!R_0$ & $\Gamma'=1$ & PL; $q\!=\!0$ & $\tilde{T}_{\theta}(R_0)=1$ & $\bar{T}_{\theta}(R_f)$ & $\frac{(R_0/R_f)^{m\!+\!1}}{m+1}\!\ll\!1$ & $\bar{T}_{\rm rise}\!\ll\!\bar{T}_c\!\approx\!\bar{T}_{\rm decay}\!\approx\!\fracb{\Delta R}{R_0}^{m\!+\!1}$ & { asymmetric, with plateau}\\ 
7 & $\Delta R\!\gg \!R_0$ & $\Gamma'=1$ & PL; $q\!>\!0$~$^\ddagger$ & $\bar{T}_f$ & $\bar{T}_{\theta}(R_f)$ & $\frac{1}{m+1}$ & $\bar{T}_c\lesssim\bar{T}_{\rm rise}+\bar{T}_{\rm decay}$ & { asymmetric, no plateau}\\
8 & $\Delta R \!\gg \!R_0$ & $\Gamma' > 1$ & PL; $q\!<\!0$~$^\dagger$ & $\frac{1}{\Gamma'}$ & $\tilde{T}_f$ & $\frac{1}{\Gamma'}\!\fracb{R_0}{R_f}^{m\!+\!1}\!\ll\! 1$ & $\bar{T}_c\ll\bar{T}_{\rm rise}+\bar{T}_{\rm decay}$ & {asymmetric, no plateau}\\ 
9 & $\Delta R\! \gg \!R_0$ & $\Gamma' > 1$ & PL; $q\!=\!0$ & $\frac{1}{\Gamma'}$ & $\frac{\bar{T}_{\theta}(R_f)}{\Gamma'}$ & $\frac{(R_0/R_f)^{m\!+\!1}}{m\!+\!1}\!\ll\!1$ & $\bar{T}_c\!\approx\!\fracb{\Delta R}{R_0}^{m\!+\!1}\!\gg \!\bar{T}_{\rm decay} \!\gg\! \bar{T}_{\rm decay}$ & {asymmetric, with plateau}\\
10 & $\Delta R \!\gg \!R_0$ & $\Gamma' > 1$ & PL; $q\!>\!0$~$^\ddagger$ & $\bar{T}_f$ & $\frac{\tilde{T}_{\theta}(R_f)}{\Gamma'}$ & $\frac{\Gamma'}{m\!+\!1}$ & $\bar{T}_c\!\approx\!\bar{T}_{\rm rise}\!\gg\!\bar{T}_{\rm decay}$ & {asymmetric, no plateau}\\
11 & $\Delta R\! \gg \!R_0$ & any $\Gamma'$ & log-normal & $\bar{T}_i\!\approx\!\bigg(\frac{R_0}{\Delta R}\bigg)^{\!m\!+\!1}$ & $^\ast 10\bar{T}_f\!\approx\!10\bigg(\frac{\Delta R}{R_0}\bigg)^{\!m\!+\!1}$ & $0.1\bigg(\frac{R_0}{\Delta R}\bigg)^{2(m\!+\!1)}\!\ll\! 1$& $\bar{T}_{\rm rise}\! \ll\!\bar{T}_c\!\ll\!\bar{T}_{\rm decay}$ & {asymmetric, no plateau}\\
\hline
\end{tabular}
\\
$^\dagger$assuming that $(100-X)/2<f(\bar{T}_f)/f(\bar{T}_0)<(100+X)/2$. In case $f(\bar{T}_f)/f(\bar{T}_0)>(100+X)/2$ the situation reduces back to that obtained at $q=0$, while if $f(\bar{T}_f)/f(\bar{T}_0)<(100-X)/2$ then the decline time
stops increasing with $\Delta R$ and instead remains fixed on the value $\sim\bar{T}_{f*}$ obtained for 
$f(\bar{T}_f)/f(\bar{T}_0)=(100-X)/2$ with $\Lambda\sim (\Gamma'\bar{T}_{f*})^{-1} \propto 1/\Gamma'$.\\
$^\ddagger$assuming that $f(\bar{T}_f)/f(\bar{T}_0)>\frac{100+X}{100-X}$. In case the increase with radius is shallower, the situation reduces back to that obtained at $q=0$.\\
$^\ast$as it turns out, due to the smooth shape of the log-normal function, the decay time is approximately 10 times larger than $\bar{T}_f$.
 \caption{\small Properties of observed light-curves for single pulses with different assumptions on the Lorentz factor of the emitters ($\Gamma'$) and $q$ the power law index of $T$ in the expression for $F_{\nu}(T)$.}
\label{tbl:cases}
\end{table}
\end{landscape}


\subsection{Correlation between $\Gamma'$ and $\gamma_e'$}
\label{sec:corrGammap}
Here we explore the implications of a possible 
correlation between $\Gamma'$ and the random Lorentz factor of the electrons in the bulk frame $\gamma_e'$
such that $\Gamma'=K\gamma_e'^{\,\eta}$.
Correlations of this kind may result in case the reconnection layer is continuously fed by incoming particles (i.e. $k=1$) and a quasi steady state is achieved between incoming and outgoing particles.
This behaviour has previously been found in PIC simulations 
\citep[e.g.,][]{Cerutti(2012),Cerutti(2013)}. There is a physical motivation for this since the electrons that are
accelerated to higher energies are preferentially those that spend a longer time being accelerated in the 
reconnection layer, and thus their velocities tend to be more tightly collimated, down to a scatter of 
$\Delta\theta\sim 1/\gamma_e'$.
This can be understood in the following way. The electric field in the reconnection layers is perpendicular to both the 
original magnetic field's direction and to the incoming particles' velocities.
Electrons are only accelerated as long as they have not deflected significantly from  the $X$-point and move along the electric field $E$.
If this acceleration is over a distance $l_\parallel$, then they acquire an energy of 
$\gamma_e'm_ec^2\approx eEl_\parallel\approx\beta_{\rm in}eB_0l_\parallel$ where $B_0$ is the magnetic field strength in the bulk, 
before the reconnection layer. On the other hand, they exit the acceleration site after traversing a distance $l_\perp$ in the perpendicular direction, at which point their angle relative to the electric field direction is 
\begin{equation}
\theta \sim \frac{l_\perp}{l_\parallel} \sim \frac{\beta_{\rm in}l_\perp eB_0}{\gamma_e'm_ec^2}
\approx\frac{\beta_{\rm in}l_\perp}{R_L(B_0)}\propto \frac{1}{\gamma_e'}\ .
\end{equation}
It follows that $\Gamma'$ may be up to linear in $\gamma_e'$ (see Eq. \ref{theta_0} in \S \ref{sec:zeromomentum}) and that one might 
naively expect $0\leq\eta\leq 1$.

In the presence of such a 
correlation, electrons of different typical energies may dominate the emission at different observed frequency bands.
This leads to a qualitatively different physical scenario than that presented elsewhere in the text.
The main difference from the case with no $\Gamma'-\gamma_e'$ correlation has to do with determining the frame in which the electric field vanishes. In case a correlation between $\Gamma'$ and $\gamma_e'$ exists,
electrons of different energies are experiencing different fields and, contrary to the case with no correlation,
there is no single primed frame where the electric field vanishes and where electrons can gyrate freely around the magnetic field 
without their velocities getting isotropized in the jet's frame. Here the role of the vanishing electric field frame is played by the jet's 
frame.\footnote{In the acceleration region (or X point), where $E^2>B^2$, there is no frame where the electric field vanishes. 
However, the radiation occurs mostly outside of this region, in the jet's bulk frame.} As a result, in order for the electrons' emission to remain beamed,
they must radiate a significant fraction of their energy before their velocities get isotropized in the jet (rather than the plasmoids') frame (i.e. within less than a single Larmour gyration). 
Therefore, such a correlation can be expected to manifest itself only at very high frequencies - at a band near the classical ``maximal synchrotron energy".
We stress that the results obtained under these assumptions are therefore likely not to be relevant for understanding the observed properties of sub-MeV pulses. 
However, they may still have interesting observable implications for the high energy emission.

Notice that the emitters with different $\gamma_e'$ have different $\Gamma'$ corresponding to different center 
of momentum frames.
Moreover, the average electrons' Lorentz factor in the emitters' frame, $\gamma_e''$, is related to $\gamma_e'$ through $\Gamma'$: 
$\gamma_e'\approx \Gamma' \gamma_e''$ (see \S \ref{sec:zeromomentum} for a derivation). It follows that:
\begin{equation}
\label{Kdef}
\gamma_e'=\bigg(\frac{\Gamma'}{K} \bigg)^{1/\eta}=\Gamma' \gamma_e'' 
\quad\Longrightarrow\quad \Gamma'=K^{1 \over1- \eta} {\gamma_e'' }^{\,\frac{\eta}{1-\eta}}\ .
\end{equation}
In addition, if we assume that the emission is dominated by synchrotron radiation (as expected in a magnetically dominated 
emission region), then $ \nu_{syn}'=\frac{eB'}{2\pi m_e c}\gamma_e'^{\,2}$, where $B'$ is the magnetic field strength in the 
jet's bulk frame (which is expected to be roughly constant for 
particles of different energy, as they are all assumed to radiate in a similar location).
It can be related to the magnetic field in the emitter's frame through\footnote{This is different from the usual relation that is derived under the
assumption that there is a single primed frame where the electric field vanishes.}
 $B''\approx B' \Gamma'$.
It is useful to relate the observed frequency to $\Gamma'$:
\begin{equation}\label{eq:nuSyn}
\nu = \nu_{\rm syn}(\gamma_e')\approx \Gamma \nu_{\rm syn}'(\gamma_e')
=\frac{\Gamma eB'}{2\pi m_e c} \bigg( \frac{\Gamma'}{K} \bigg)^{2 \over \eta}.
\end{equation}
Since the peak spectral emissivity is determined in the jet's bulk frame. The spectrum in this frame is given by the regular synchrotron formulas.
As detailed above, a correlation between $\gamma_e'$ and $\Gamma'$ implies that the electric field vanishes only in the jet's (rather than the plasmoids') frame and as a result we should consider electrons in the ``fast cooling" regime
(i.e. that cool significantly in less than a dynamical time), and so the particles' spectrum is:
\begin{equation}
 \frac{dN_e}{d\gamma_e'}=A\gamma_e'^{\,-p-1}\quad\Longleftrightarrow\quad \frac{dN_e}{d\ln\gamma'_e} = A\gamma_e'^{\,-p}\ .
\end{equation}
The resulting spectrum is therefore: $P_{\nu}\propto \nu ^{-p / 2}$ as for the case of synchrotron without boosted emission in the jet's bulk frame.

Although the spectrum does not change for $\Gamma'\gg 1$, there will be an angle dependent cut-off to the spectrum in this case.
Notice that electrons with different Lorentz factors $\gamma_e'$ are emitted by emitters with different $\Gamma'$. 
Emission from an emitter moving at $\Gamma'$ can be seen up to an angle $\theta'_{\rm obs}\approx1/\Gamma'(\gamma_e')$
where $\cos\theta'_{\rm obs} = \hat{n}'\cdot\hat{\beta}'$, 
this implies a cut-off of the observed spectrum that is determined by the observation angle:
\begin{equation}
\label{cutoff}
 \nu_{\rm max}(\theta'_{\rm obs})=\nu_{\rm obs}(\Gamma'=1/\theta'_{\rm obs})
 =\frac{\Gamma e B'}{2\pi m_e c} (K \theta'_{\rm obs})^{-2/\eta}\ .
\end{equation}
As mentioned above, in order to maintain the anisotropy, electrons are doing less than one gyration around the magnetic field. Therefore, this effect will take place at frequencies above the ``regular"
maximal synchrotron frequency $\nu_{\rm syn,max}$.

Another effect introduced by a correlation of the type  $\Gamma' \propto \gamma_e'^{\,\eta}$ relates to the typical width of the pulses. So long as the latter are determined by $\Gamma'$
(see \S \ref{paramspace} for details), the typical width will be different at different frequencies (which are dominated by emission from electrons at different energies). As seen by Eq. \ref{eq:nuSyn},
this will lead to $\Delta T \propto \nu^{-\eta/2}$. We return to this point in \S \ref{spectralevolution}.

As is the usual case for synchrotron, we can safely assume that each frequency is dominated by electrons of a single 
energy.\footnote{This is true so long as the spectrum obtained in this way is softer than the synchrotron spectrum for 
single electrons: $F_{\nu}\propto \nu^{1/3}$.}
As mentioned above, the only frame with a purely magnetic field in case of a $\Gamma'-\gamma_e'$ correlation, is the jet's frame. 
We therefore start with the peak spectral emissivity of individual electrons in the jet's frame 
($P'_{\nu'\rm ,e,max}\approx\sigma_T m_e c^2 B'/(3e)\propto B'$) 
and use the regular Lorentz transformation (as in Eq.~(\ref{flux0}), but with the direction of the transformation reversed) 
to obtain the (isotropic) luminosity in the emitters' frame:
\begin{equation}
\label{dLdgamma}
\begin{split}
L''_{\nu''}&\!=\!\int \!d\Omega'\int \! d\gamma_e'\delta(\gamma_e'\!-\!\gamma'_{\rm e,syn})
L'_{\nu'}(\gamma'_{\rm e,syn})\mathcal{D}'(\gamma_e')^{k-3}\Gamma'(\gamma_e')^k   
\\ &\approx \! \int \! d\Omega'\frac{dN_e}{d\ln\gamma_e'}(\gamma'_{\rm e,syn}) \, L'_{\nu'\rm ,e,max}\, 
\mathcal{D}'(\gamma'_{\rm e,syn})^{k-3}\Gamma'(\gamma'_{\rm e,syn})^k  
\ \sim\  A\gamma_{\rm e,syn}^{\prime\,-p}\frac{P'_{\nu'\rm ,e,max}}{\Gamma'(\gamma'_{\rm e,syn})}\ ,
\end{split}
\end{equation}
where $\gamma'_{e,{\rm syn}}$ is the Lorentz factor (in the jet's frame) of electrons for which the synchrotron frequency 
is the observed frequency $\nu$:
\begin{equation}
 \gamma_{\rm e,syn}'=\bigg( \frac{\nu 2 \pi m_e c K}{{\mathcal D}(y) {\mathcal D}'(y,\phi) eB'} \bigg)^{1\over 2-\eta}\ .
\end{equation}
Note that in Eq.~(\ref{dLdgamma}) we have expressed the spectral luminosity in the jet's frame ($L'_{\nu'}$) in terms of  the electrons'
energy distribution ($dN_e/d\gamma_e'$) and the peak luminosity of individual electrons ($L'_{\nu'\rm ,e,max}$), where
$L'_{\nu'}(\gamma'_{\rm e,syn})/L'_{\nu'\rm ,e,max}\approx\frac{dN_e}{d\ln\gamma_e'}(\gamma'_{\rm e,syn}) \approx N_e(\gamma'_{\rm e,syn})$ is the number of electrons whose synchrotron frequency is close to the observed frequency,
so that each of these electrons radiates at that frequency near its peak spectral power $P'_{\nu'\rm ,e,max}$ or its peak 
isotropic equivalent spectral luminosity $L'_{\nu'\rm ,e,max}$. The latter are related through 
$L'_{\nu'\rm ,e,max}\sim \Gamma^{\prime\,2}P'_{\nu'\rm ,e,max}$, since
$P'_{\nu'} = dE'/d\nu'dt'$ and $L'_{\nu'} = 4\pi(dE'/d\nu'dt'd\Omega') =4\pi(dP'_{\nu'}/d\Omega')\sim (4\pi/\Delta\Omega')P'_{\nu'}$
and in our case $4\pi/\Delta\Omega'\approx 4\Gamma^{\prime\,2}\sim\Gamma^{\prime\,2}$ as because of relativistic beaming, 
in the jet's frame, the contribution of the luminosity is significant only within a cone of opening angle $\sim 1/\Gamma'$ in which 
$\mathcal{D}'\sim\Gamma'$.
The basic result of Eq.~(\ref{dLdgamma}) is that $L''_{\nu''}\sim N_e(\gamma'_{\rm e,syn})P'_{\nu'\rm ,e,max}/\Gamma'(\gamma'_{\rm e,syn})$ or that in the emitters' frame the spectral power of a single electron whose synchrotron frequency is close to the observed one,
satisfies $P''_{\nu'\rm ,e,max}\approx L''_{\nu''}/N_e(\gamma'_{\rm e,syn}) \sim P'_{\nu'\rm ,e,max}/\Gamma'(\gamma'_{\rm e,syn})$,
which is the usual familiar Lorentz transformation for the received emission in the jet's frame, which corresponds to a single propagating 
electron, e.g. that is part of a blob and so corresponds to $k=0$. The correction for a steady state in the jet's frame, $k=1$, is the extra power
of  $1\!-\!\beta'(\gamma_{\rm e,syn})\sin\theta'\cos\phi$ (which is the ratio of the received and emitted times for a point source such as 
an individual electron) that appears in Eq.~(\ref{gGlightcurve}), which accounts for the fact that in this case there is no distinction 
between the emitted and received times, so that the radiation is not received over a shorter time than it is emitted, as is the case for a blob.

\begin{figure*}[h]
\centering
\includegraphics[scale=0.3]{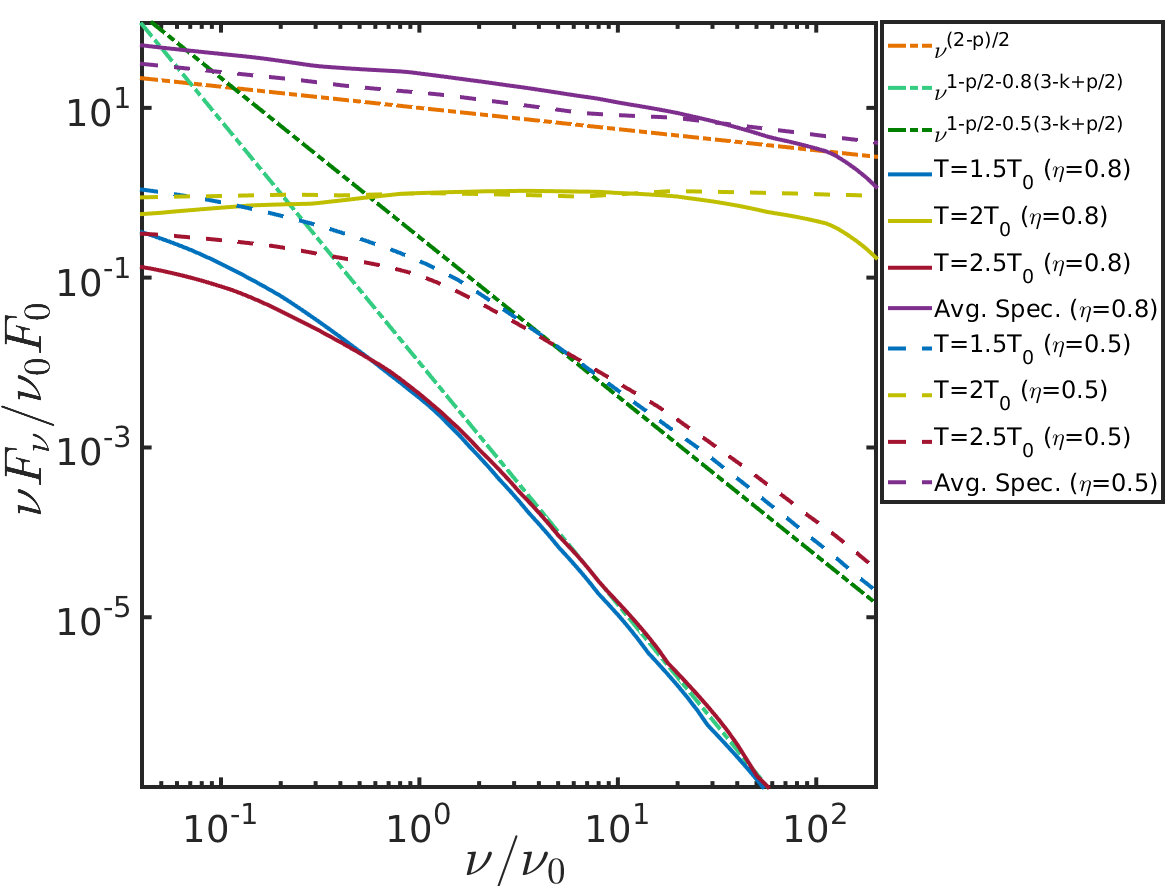}
\caption{Spectra of a single pulse at different times: $T=[1.5,2,2.5]T_0$, alongside the time-averaged spectrum,
assuming a correlation of the type $\Gamma'\propto \gamma_e'^{\eta}$.
The electrons' energy distribution is assumed to be $\frac{dN_e}{d\gamma_e'}\propto \gamma_e'^{-1-p}$, with $p=2.5$. We also assume $m=0, k=1$ and a log-normal emissivity with $\sigma_{\ln R}=0.01$. The peak of the light-curve
is at $T=2T_0$. The spectra is plotted for $\eta=0.8$ (solid lines) and $\eta=0.5$ (dashed lines). 
For comparison we also plot the expected spectrum in this case, for non-boosted emission (and for the time-averaged spectrum): $\nu F_{\nu}\propto \nu^{2-p \over 2}$,
and the asymptotic spectrum expected in this case for $\nu>\nu_{\rm max}$. The frequency is in units of $\nu_0$,
which is the typical frequency at which electrons with $\Gamma'=100$ radiate (see \ref{eq:nuSyn}), and below which the time-scale is dominated by the radial instead of the angular one.
$\nu F_{\nu}$ is given in the units of $\nu_0 F_0$, where $F_0$ is the peak of the time-resolved flux at $\nu_0$.}
\label{fig:spectrumcorrelation}
\end{figure*}

In order to obtain the observed flux, we plug  Eq.~(\ref{dLdgamma}) into Eq.~(\ref{generalFnu}) and integrate over $\gamma_e'$:
\begin{equation}
\label{gGlightcurve}
\begin{split}
F_{\nu}(T) =\,\frac{2\Gamma_0 AP'_{\nu'\rm ,e,max}}{(4\pi)^2 D^2 }\! \fracb{T}{T_0}^{-\frac{m}{2(m+1)}} 
&\int_{y_{\rm min}}^{y_{\rm max}}\! \! dy \bigg(\frac{m\!+\!1}{m\!+\!y^{-m-1}}\bigg)^{2} y^{-1-\frac{m}{2}}
\;f\left[ y \fracb{T}{T_0}^\frac{1}{m+1}\right] \\
&\times \int_0^{2\pi}d\phi\, \frac{\gamma_{\rm e,syn}^{-p}}{2 K^4\gamma_{\rm e,syn}^{4\eta}}(1\!-\!\beta'(\gamma_{\rm e,syn})\sin\theta'\cos\phi)^{k-3}\,\ .
\end{split}
\end{equation}
The resulting spectrum is shown in Fig. \ref{fig:spectrumcorrelation}. The time integrated spectrum, as well as the spectrum 
close to the time of the peak flux approach $\nu F_{\nu}\propto \nu^{2-p \over 2}$ as expected for non-boosted emission.
Results are plotted for $p=2.5$ which is consistent with expectations from theory. Note that although for acceleration in 
reconnection sites a large range of $p$-values is found, depending on the exact setup and initial magnetization \citep{Sironi(2014),Guo(2014),Guo(2015),Werner(2014),Kagan(2015)} down to values of the order of $p\approx 1.5$, 
this of course cannot hold up to very large energies as they would require a divergence of the total energy, and is also 
inconsistent with both afterglow observations and observations of the high-energy spectral slope of the Band function.
At either early or late times, and frequencies above $\nu_{\rm max}$, the flux will become
steeper, due to the angle-dependent cut-off described in Eq. \ref{cutoff}. For these, the emission is dominated by large angles from the 
line of sight, with $\theta'\Gamma'(\nu)>1$, and one can relate the observed spectrum $F_{\nu}(\theta')$ 
to the one along the line of sight, $F_{\nu}(\theta'=0)=F_\nu(0)$ \citep[e.g.,][]{Granot(2002),GR-RP(2005)},
\begin{equation}
\begin{split}
&\frac{F_{\nu}(\theta)}{F_{\nu}(0)}= \bigg(\frac{{\mathcal D}'(\theta')}{{\mathcal D}'(0)}\bigg)^{3-k+\frac{p}{2}}
\approx \left[1+(\Gamma'\theta')^2\right]^{k-3-\frac{p}{2}}
\propto\gamma_e'^{2\eta(k-3-\frac{p}{2})}\propto\nu^{-\eta(3-k+\frac{p}{2})}\ ,
\\ \\
&\Longrightarrow\quad F_{\nu}(\theta)\propto F_{\nu}(0)\nu^{-\eta(3-k+\frac{p}{2})}\propto\nu^{-p/2-\eta(3-k+p/2)}\ .
\end{split}
\end{equation}
Fig. \ref{fig:spectrumcorrelation} clearly shows that at early and late times, the flux approaches this slope. The transition from the $\nu^{-p/2}$ slope to this steeper slope, is at $\nu_{\rm max}$.

\subsection{Results for the entire light-curve}
\label{multi}
\begin{figure*}[h]
\centering
\includegraphics[scale=0.19]{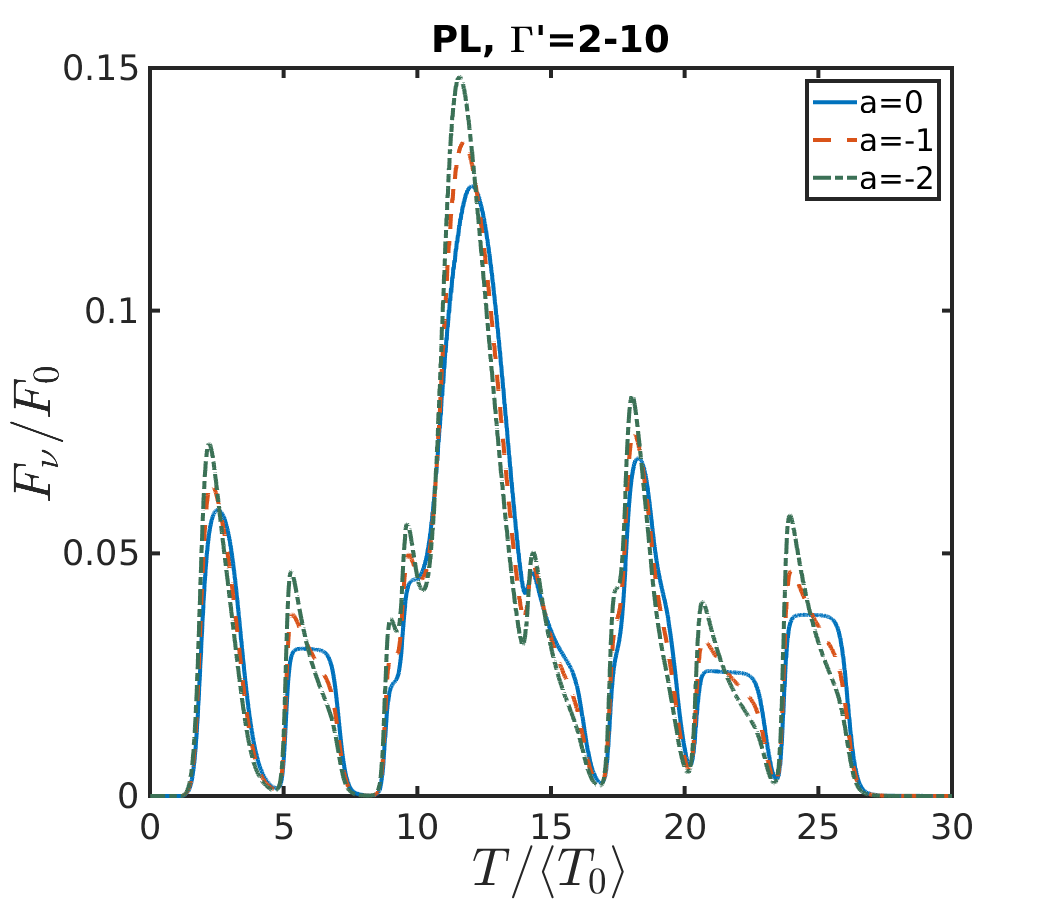}
\includegraphics[scale=0.19]{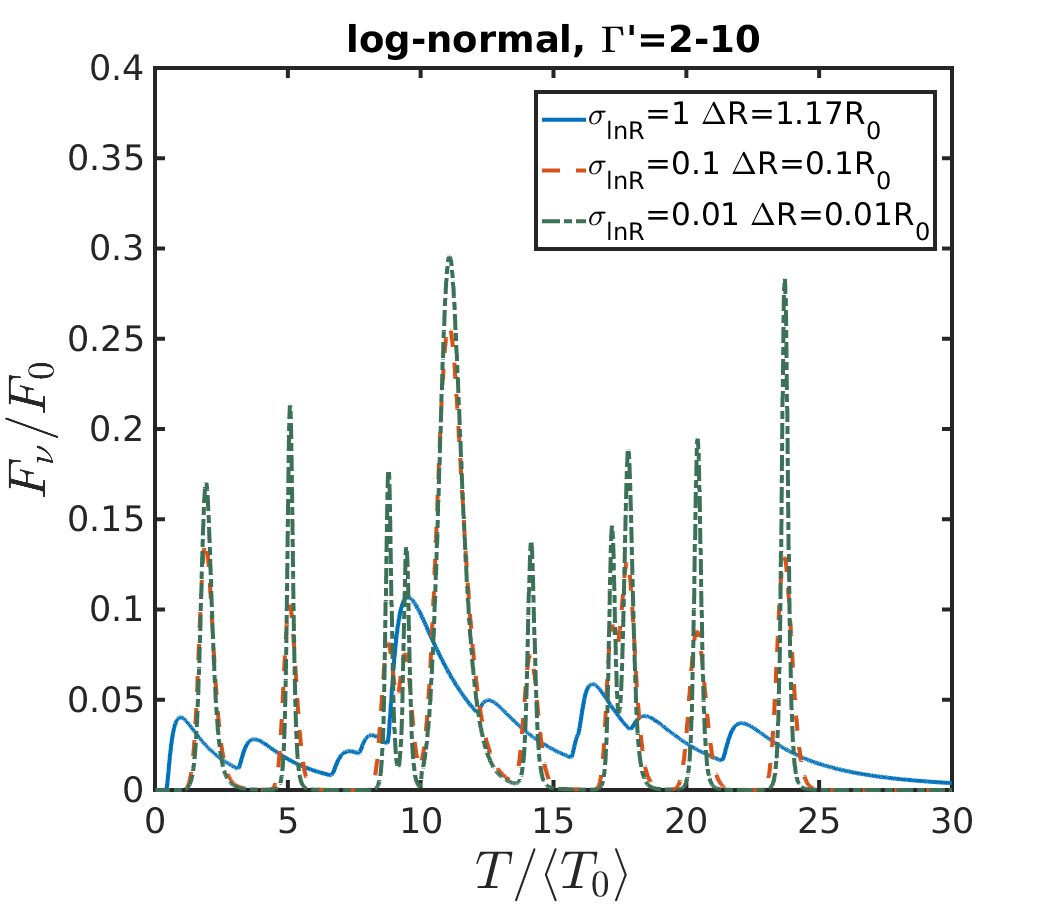}\\
\includegraphics[scale=0.19]{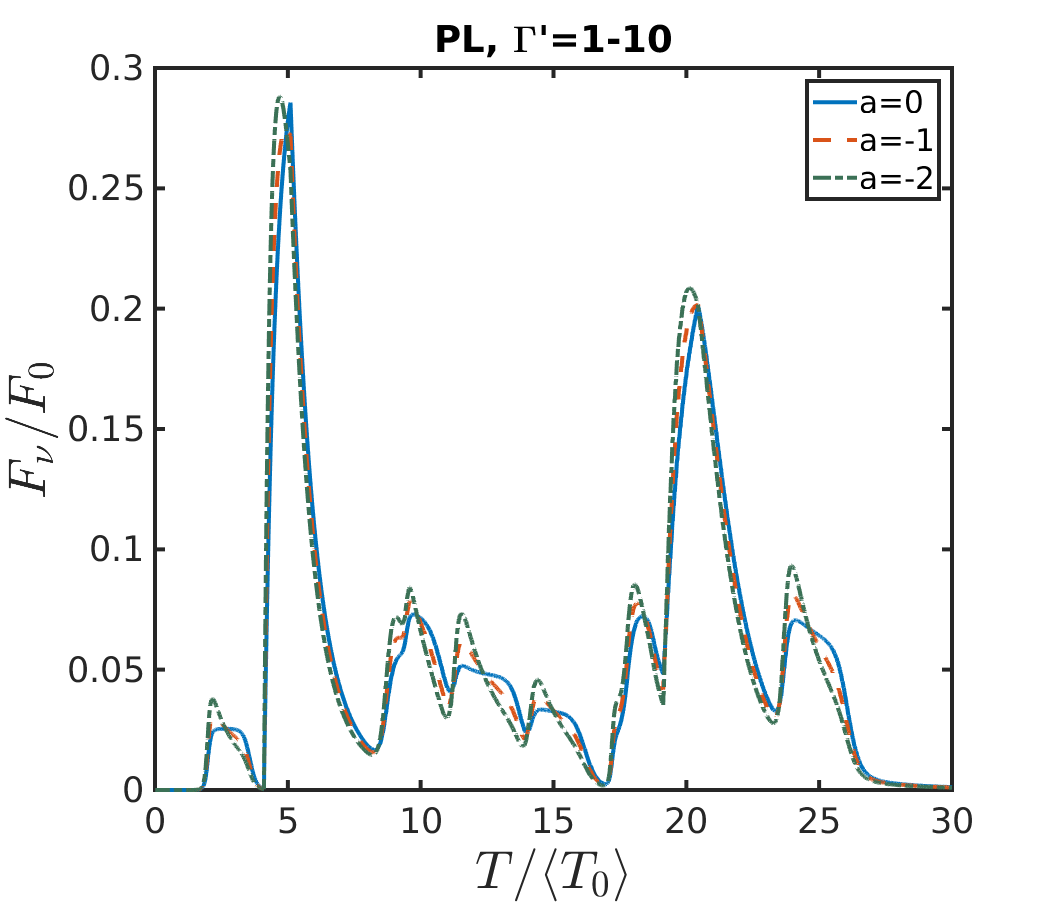}
\includegraphics[scale=0.19]{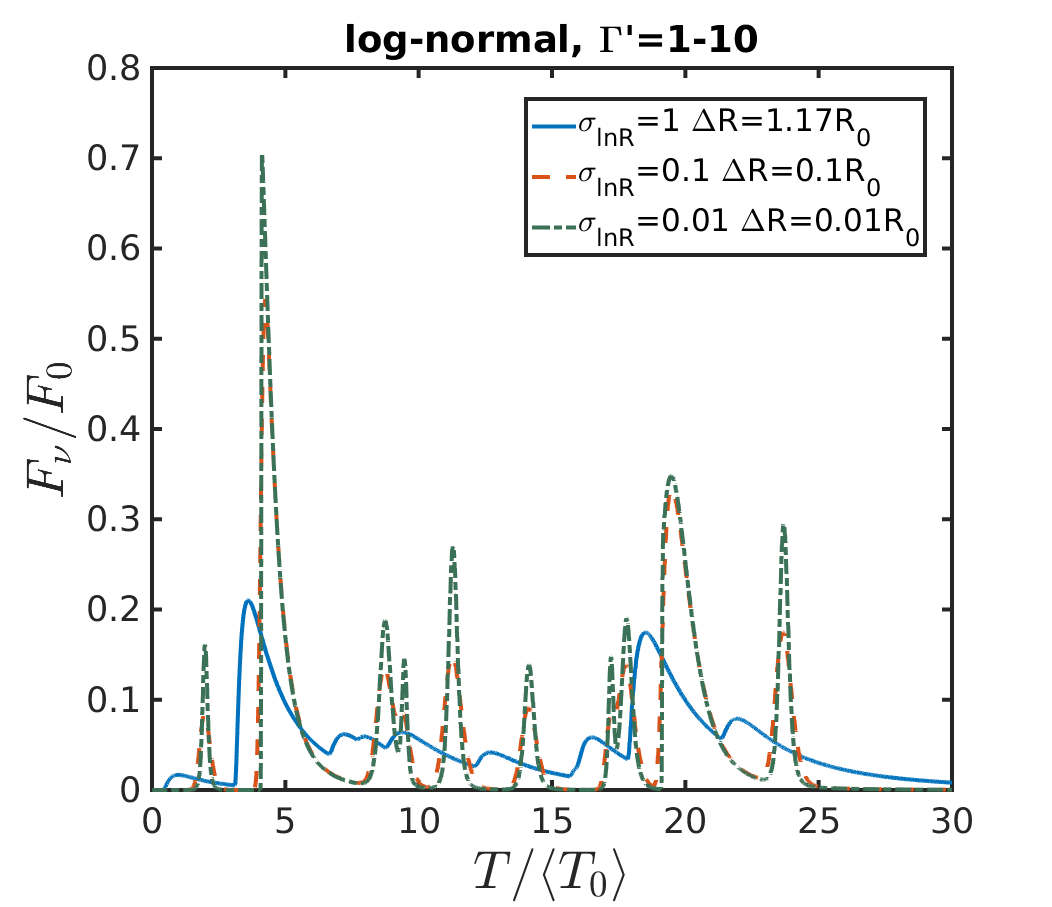}
\caption{Light-curves of ten consecutive pulses with different $L(R)$ 
(but the same for different pulses within a light-curve). 
We assume a Band function spectrum and $m=0, k=0$.
The flux is in units of the ``typical" peak flux $F_0(\Gamma'=1)$ as a function of the
average normalized time $T/\langle T_0 \rangle$. 
$T_0, L''_{\nu''_0}$ for different shells are log-normally distributed with widths of $0.3, 0.5$ dex correspondingly.
The ejection times are uniformly distributed in the range
$0\leq T_{\rm ej}\leq 20\langle T_0\rangle$. Finally, $\Gamma'$ is uniformly distributed in the range 2-10 (1-10) in the top (bottom) panels.
{\bf Left}: Power-law $L(R)$ with different power-law slopes $a$ and $R_f=2R_0$. 
{\bf Right}: log-normal $L(R)$ with different standard deviations $\sigR$.
All parameters (except for $\Gamma'$) are held constant for different light-curves within any panel. The values of $\Gamma'$ in the top (bottom) panel, from left to right, are: $3.4,9,9.6,2.6,8,7.6,9.8,5.5,9.7,6.4$
($8,1.1,4.6,7.4,9.3,5.5,1.3,4.6,6.7,8.1$).}
\label{fig:multifR}
\end{figure*}

Under the same assumptions as in \S~\ref{pulseshapes}, and for the same dependencies of the luminosity on radius 
we calculate the resulting light-curve.
In principle, the distributions of luminosities, Lorentz factors and typical radii of emission, are model dependent, and are not uniquely determined by the fact that there is anisotropic emission in the jet's bulk frame. These can reflect the inner workings of the central engine 
(see \S \ref{sec:diss} for more details). Since we do not want to invoke a specific central engine model, these distributions were chosen
phenomenologically, motivated by typical light-curve observations.
We assume that individual shells (leading to different pulses) are characterized 
by different typical duration $T_0$, different ejection times $T_{\rm ej}$, different $\Gamma'$ and different peak luminosities 
(i.e. different $F_0$). We assumed ten spikes in the light-curve, corresponding to ten emitting shells. 
In order to calculate the corresponding light-curve from ten shells, we randomly draw the parameters of each shell out of the following probability 
distributions. We draw the values of $T_0$ for different shells from a log-normal distribution with a central value denoted as 
$\langle T_0\rangle$ and a typical width of 0.3 dex.
The ejection times $T_{\rm ej}$ are randomly drawn from a uniform distribution in the range 
$0\leq T_{\rm ej}\leq 20\langle T_0\rangle$. We draw the values of $\Gamma'$ for different shells from either a uniform distribution between $2$ and $10$ (or between $1$ and $10$).
Finally, we draw the luminosities ($L''_{\nu''_0}$) from a log-normal distribution with a central value denoted as $\langle L''_{\nu''_0}\rangle$ and a typical width of 0.5 dex.
The results are plotted in Fig.~\ref{fig:multifR}. 
For larger $\Delta R/R_0$ and for $a=0$, the pulses tend to be wider (and the peak flux is weaker in order to conserve the total luminosity of the pulse). In this case, it is harder to resolve individual pulses and there is more overlapping of pulses. Generally, the resulting light-curves can be quite complex and have a wide range of typical behaviours, as seen in observations. However, it is basically a reflection of the choice of distributions above and therefore a true comparison with observations will depend on the specific model assumed.

\section{comparison with observations}
\label{comparison}
\subsection{peak Luminosity-variability correlation}
\label{LumVar}
Several studies have explored the possibility that
the variability of GRB light-curves is related to their peak luminosities \citep{Stern(1999),Fenimore(2000),Reichart(2001)}. We use the definition for the variability presented in \citep{Kobayashi(2002)},
which is a simplified version of that in \citep{Reichart(2001)}:
\begin{equation}
\label{eq:var}
 V=\frac{\sum (F_{\nu}(T)-\langle F_{\nu}(T)\rangle)^2}{\sum F_{\nu}(T)^2}
\end{equation}
where $\langle F_{\nu}(T)\rangle$ is the flux smoothed with a boxcar window with a time scale equal to the smallest fraction 
of the burst time history that contains a fraction 0.45 of the total flux. \cite{Reichart(2001)} have found that the peak luminosities
$L$ are correlated with the variability measures: $L_{\rm max} \propto V^{3.3_{-0.9}^{+1.1}}$. It should however be noted that other studies \citep{Guidorzi(2005a),Guidorzi(2005b)} have claimed
a correlation with a much shallower slope, of the order of $1.30_{-0.44}^{+0.84}$ and $0.85\pm 0.02$ respectively. The difference between the results has been attributed to their respective treatment of sample variance \citep{Guidorzi(2006)}. These authors use a Monte Carlo method to show that when the sample variance is of order the total variance the latter methods become more reliable estimators of the slope in the $L_{\rm max} - V$ correlation.

\begin{figure*}[h]
\centering
\includegraphics[scale=0.28]{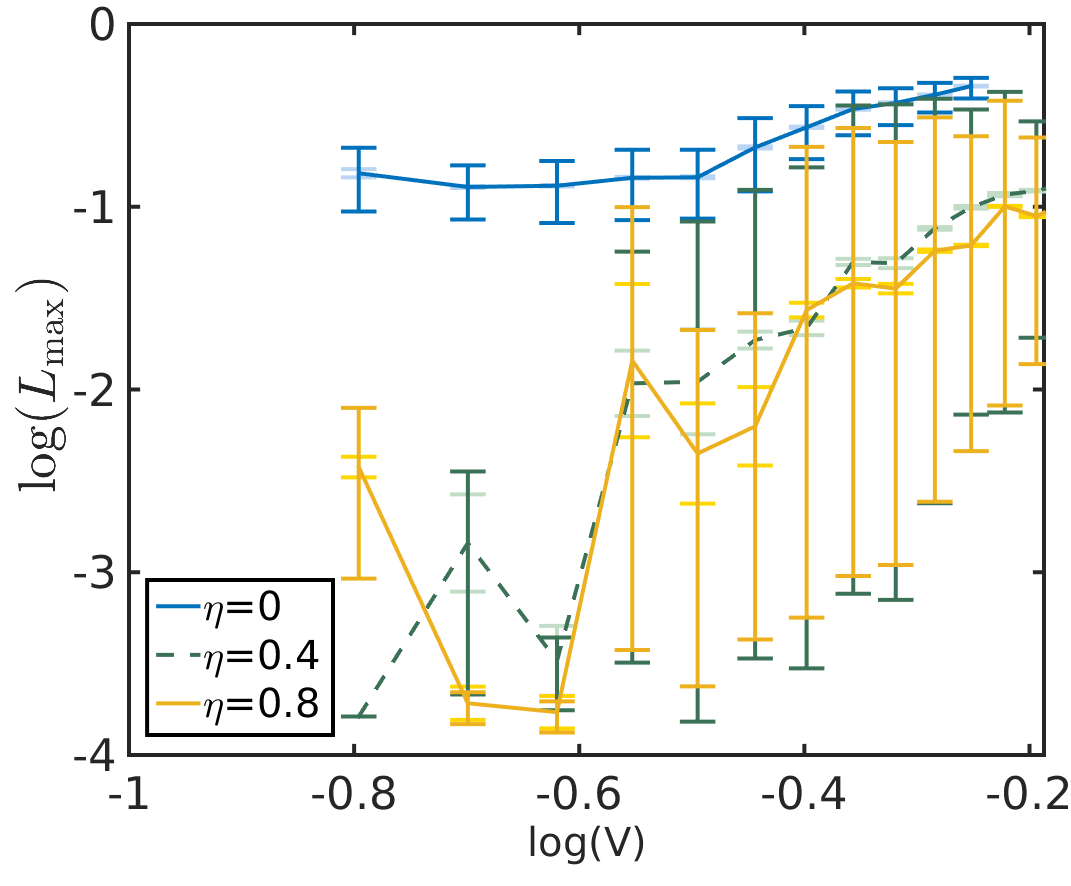}
\includegraphics[scale=0.28]{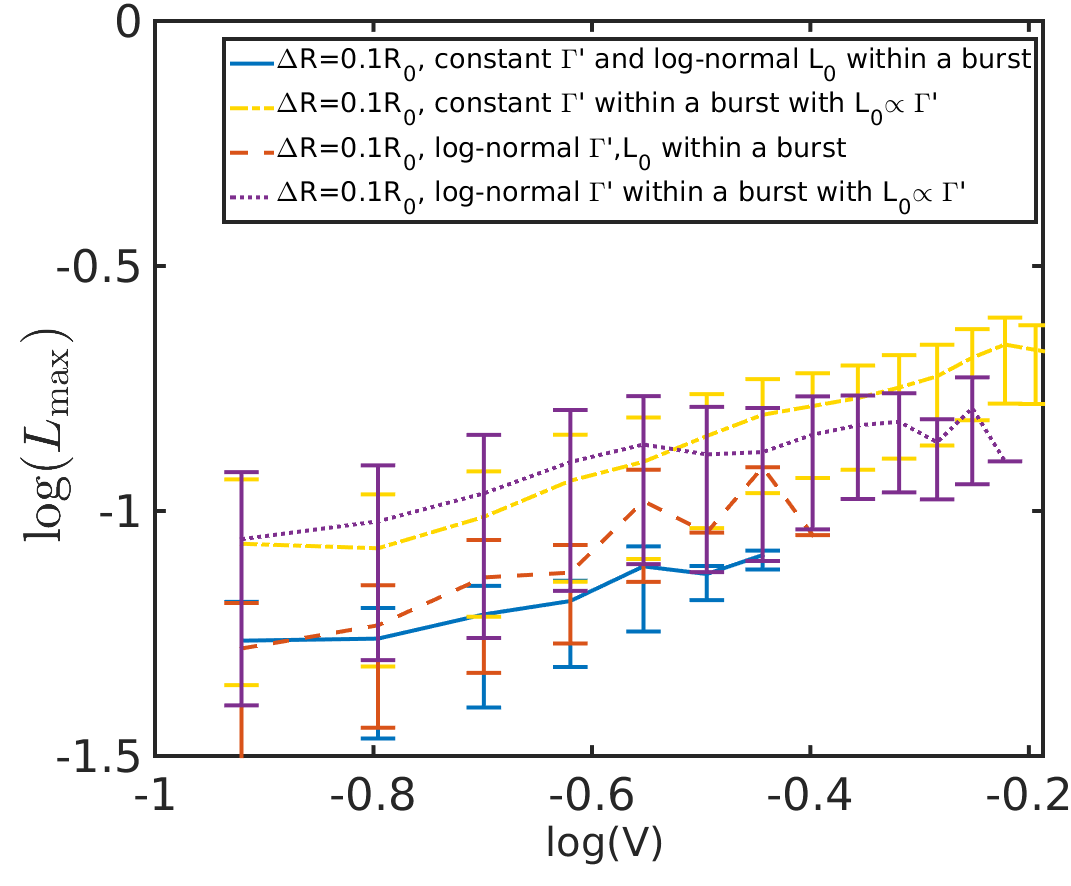}
\caption{{\bf Left:} Maximal luminosity of a simulated light-curve plotted against the variabilities associated 
with the same light-curves using a simplified version of the definition in \citep{Reichart(2001)} assuming $\Delta R=0.01R_0$
for a general emission mechanism ($\eta=0$) and for synchrotron emission with a correlation 
between $\Gamma'$ and $\gamma_e'$ as described in \S~\ref{sec:corrGammap}. The inner error bars on each plot are the statistical errors.
Notice, that these are small compared with the intrinsic errors, as described in \S \ref{LumVar}.
In the latter, we explore different possibilities for the power-law index of the correlation, $\eta$. 
{\bf Right: } Maximal luminosity as a function of variability for different assumptions on the width of the shells,
$\Gamma'$ (either varying independently between shells or constant within a burst), and the values of $F_0$ 
for different pulses (varying log-normally or correlated with $\Gamma'$).}
\label{fig:variability}
\end{figure*}

A correlation between peak luminosity and variability may also be expected in our model, since the 
peak luminosity increases with $\Gamma'$ and the width of pulses decreases with $\Gamma'$ so long as 
$\Delta R /R \lesssim 1/\Gamma' $. Notice, however, that this trend is reversed for $\Gamma'\lesssim2$ 
(see Fig.~\ref{fig:Gammap}), which may cause a negative correlation if the typical values of $\Gamma'$ 
are in the range: $1<\Gamma'\lesssim 2$. In Fig.~\ref{fig:variability} we plot the results for the 
$L_{\rm max}\,$--$\,V$ correlation for simulated light-curves in our model. In calculating the light-curve we 
assume a typical Band function spectrum emissivity ($\alpha_B=0, \beta_B=-2.3$) as well as $m=0$, $k=0$
and a Gaussian dependence of luminosity on the logarithm of the radius.

We first let $T_{\rm ej}, \Gamma'$ vary between different pulses (according to the distributions assumed in 
\S~\ref{multi}, see Fig.~\ref{fig:multifR}). Next, we consider synchrotron emission with a correlation 
between $\Gamma'$ and $\gamma_e'$ as described in \S~\ref{sec:corrGammap}. As mentioned above, such a correlation is
less likely to hold for electrons radiating at frequencies corresponding to the sub-MeV peak. However, it still remains an open question whether
a similar relation between variability and luminosity also exists at the GeV range. It is therefore important to provide theoretical predictions
for this situation which could be tested by observations at a later stage.
Again, we let $T_{\rm ej}$ vary randomly for different pulses. We calculate the light-curve at a constant observed frequency. 
As can be seen from Eq. \ref{eq:nuSyn} this means that we should consider $\gamma_e'={\rm constant}$ between different
pulses, implying $K\propto \Gamma'$. We therefore vary $K$ using the same distribution of $\Gamma'$ as for the case with no correlation.
In addition, we assume $k=1$ for this scenario, as described in \S~\ref{sec:corrGammap}.
The light-curve is then calculated using Eq.~(\ref{gGlightcurve}). In both cases, we find that for $\sigR=0.01$ there is evidence for a positive correlation between $L_{\rm max}$ and $V$
with a power law index in the range $0.8-4.8$ (see left panel of Fig.~\ref{fig:variability}), consistent with the observed values.
To quantify the degree of correlation, we calculate the Pearson correlation coefficient for the generated light-curve distributions described above, and find it to be 0.7-0.9.
The errors in Fig. \ref{fig:variability} denote the smallest range within which lie 68.3$\%$ of the values of $\log_{10}(L_{\rm max})$ obtained from different realizations of the light-curve.
Note that the large errors in Fig. \ref{fig:variability} are systematic and not due to insufficient statistics (as is shown explicitly in the left panel of this figure, where we compare
the systematic and total errors). They reflect the fact that the variability is set by a combination of many different factors,
such as the exact $\Gamma'$ distribution for the different pulses and whether or not there are overlapping pulses in the light-curve. For this reason
a given degree of variability can be obtained by light-curves with different intrinsic properties, and different values of $L_{\rm max}$. Note also that the obtained distribution of $L_{\rm max}$
does not follow a simple log-normal distribution and the errors in general may be asymmetric.

Variation in the dissipation rate and in the associated co-moving luminosity over $\Delta R\ll R$ as reflected
in $\sigR=0.01$ may be present and have some contribution to the overall light-curve. However,
it is probably more realistic to consider wider shells, or larger $\Delta R/R$. When considering $\sigR=0.1$ 
in the {\it right panel} of Fig.~\ref{fig:variability} (and also by letting $F_0$ vary log-normally, 
as in \S~\ref{multi}, and independently of $\Gamma'$), and leaving the other parameters as before, 
we find that the positive correlation is wiped out. This is because shells with the largest values of $\Gamma'$ 
are no longer necessarily the narrowest. In addition, whereas $L_{\rm max}$ depends on the maximal 
$\Gamma'$ in the burst, the variability depends on the width of all the pulses in the light-curve. 
Since we have assumed no correlation between $\Gamma'$ values in different pulses, 
it is natural that no strong correlation is seen between $L_{\rm max}\,$--$\,V$.

Nonetheless, reconnection simulations suggest that $\Gamma'$ increases with the magnetization 
parameter $\sigma$ \citep{Guo(2014),Werner(2014)},  albeit very slowly, 
and it is not unreasonable that the typical magnetization would vary between different GRBs. 
Thus motivated, we also explore a simple model in which $\Gamma'$ varies between different 
GRBs as in \S~\ref{multi} (i.e. uniformly between 2 and 10) but is constant for a given GRB.
We also explore the possibility that $F_0\propto \Gamma'$. 
The correlations for these simulated light-curves can be seen in the {\it right panel} of Fig.~\ref{fig:variability}.
Even with these modifications, a positive correlation between variability and $L_{\rm max}$, 
for shells of width $\Delta R\approx 0.1R_0$ is only obtained when $F_0\propto \Gamma'$ (resulting in a correlation with a power law index in the range $0.4-1.2$ and a Pearson coefficient of 0.3-0.6).
This power law index is smaller than the observed value of ${3.3_{-0.9}^{+1.1}}$ reported by \cite{Reichart(2001)}, however it is consistent with
the later measurements of $1.3_{-0.4}^{+0.8}$ and $0.8\pm 0.02$ found by \citep{Guidorzi(2005a,2005b)} (see discussion above).

\subsection{peak luminosity - peak energy correlation}
There have been various claims in the literature that
$\nu_p$ may be correlated with the peak luminosity, $L_p$, \citep{Yonetoku(2004),Ghirlanda(2005),Yonetoku(2010)}.
The reported correlations vary in different studies and range between: $L_p \propto \nu_p ^{1.5-2}$.
We aim to test the viability of such a correlation in the context of our model.

First, recall that:
\begin{equation}\label{eq:nu_p-L_p}
 \begin{split}
 & \nu_p\approx\Gamma \nu_p' \\
 & L_p\approx \nu_p L_{\nu_p}  \approx (\Gamma \nu_p')(\Gamma L'_{\nu_p'}) \approx\Gamma^2 L'_p\ .
 \end{split}
\end{equation}
The factor of $\Gamma^2$ for the luminosity can be understood as follows. The emitted power in the jet's bulk and observer frames
is equal, $dE/dt = dE'/dt'$, but the received power, $dE/dT$ (where $T$ is the observed time), is larger by a factor
of $dt/dT = 1/(1-\beta\mu)\sim\Gamma^2$ in the observer frame. This consideration accounts for the angular distribution 
of the emitted radiation through the estimate  $dt/dT \sim\Gamma^2$, which effectively assumes that the observed power is 
dominated by material whose radiation is beamed towards the observer.
Alternatively, the emitted power per solid angle per particle ($L_{{\rm iso},e}=dE/dtd\Omega$) scales as 
${\mathcal D}^4\sim\Gamma^4$, but the fraction of the spherical emitting shell size that is within the visible region is 
$\sim 1/\Gamma^2$ so that together we once more obtain a factor of $\sim\Gamma^2$. Combining the two relations in 
Eq.~(\ref{eq:nu_p-L_p}), we obtain:
\begin{equation}
 L_p= \bigg(\frac{\nu_p}{\nu_p'}\bigg)^2 L_p' \longrightarrow \frac{L_p}{\nu_p^2}= \frac{L_p'}{\nu_p'^2}.
\end{equation}
This implies that if the ratio between the total luminosity in and square of the peak energy in the jet's bulk frame, $L'_p/\nu_p'^2$, 
is roughly constant, then a similar relation holds also in the observed frame, $L_p \propto \nu_p^2 $, which reproduces the 
observed correlation (see also \citealt{Wijers(1999),Ghirlanda(2012)}). This can naturally occur if the properties of the 
emission are set in the jet's bulk frame.  Even more importantly, the result is independent of $\Gamma$ which is likely to 
vary significantly from burst  to burst (and even within a single burst) and is unlikely to be directly related to the properties 
of the flow in the co-moving frame.

In case of anisotropic emission in the jet's bulk frame, where the emission properties are set in the emitters' frame, as considered in this paper, the situation is more complex. It is most instructive to examine the behaviour for $\Gamma'\gg 1$. 
From Eq.~(\ref{approxFnu2}) that holds for  $\Delta R/R_0>1/\Gamma'$ (and motivated the definition of 
$F_0\propto \Gamma_0 \Gamma'^{1-2k} L''_{\nu''_0}$   in Eq.~[\ref{eq:F0}]) it can be seen that in this regime
$L_{\nu_p}/L''_{\nu''_p}\sim \Gamma_0 \Gamma'^{1-2k}\max(1,R_0/\Delta R)$ where we identify 
$\nu''_0\to\nu''_p\sim\nu_p/\Gamma\Gamma'$ and $L''_{\nu''_0}\to L''_{\nu''_p}\sim L''_p/\nu''_p$.
On the other hand,  Eq.~(\ref{eq:R0flux}) that is for a delta function emission in radius and reasonably holds for
$\Delta R/R_0<1/\Gamma'$ shows that in this regime $L_{\nu_p}/L''_{\nu''_p}\sim \Gamma_0 \Gamma'^{2-2k}$.
Since $L_p/L''_p \sim (\nu''_p/\nu_p)L_{\nu_p}/L''_{\nu''_p}\sim (L_{\nu_p}/L''_{\nu''_p})/(\Gamma\Gamma')$,
this implies that
\begin{equation}
\frac{L_p}{\nu^2_p} = \frac{L_p'}{\nu'^{\,2}} = \frac{L_p''}{\nu''^{\,2}}\Gamma'^{\,-2k}f_*\ \ \ ,\quad\quad
f_* \approx \left\{ \begin{array}{cl} 1 & \Delta R/R_0>1\ ,\\
R_0/\Delta R\ \  & 1/\Gamma'<\Delta R/R_0<1\ ,\\
\Gamma' & \Delta R/R_0<1/\Gamma'\ .
\end{array} \right.
\end{equation}
That is, $L_p''/\nu''^{\,2}_p = L_p/\nu_p^2 \Longleftrightarrow \Gamma'^{\,-2k}f_*=1$ is satisfied (i.e. this scaling 
extends to the emitters' frame) only for $k=0$ (blob-like emission) and $\Delta R\gtrsim R_0$, but not for $k=1$
(continuous emission) or $\Delta R<R_0$. In the latter cases a residual dependence remains on $\Gamma'$ (for $\Delta R<R_0$),
and sometimes also on $\Delta R/R_0$ (for $1/\Gamma'<\Delta R/R_0<1$). Nevertheless, it is still reasonable that some of 
the properties of the emission in the emitters' frame would be determined by processes that occur in the jet's bulk frame,
in which the emitters are energized and launched at $\Gamma'$, so that they could in principle correlate with $\Gamma'$
and still produce the observed correlations.

We stress that whether or not the ratio $L'_p/\nu_p'^2$ is indeed constant, depends on the specifics of the emission mechanism, 
which are not uniquely determined by the model at hand, namely, anisotropic relativistic velocity distribution of particles in the 
jet's bulk frame. It is nonetheless intriguing that if such a correlation indeed holds, then it will not be wiped out 
by variability in $\Gamma$ (and possibly also in $\Gamma'$).

\begin{figure*}[h]
\centering
\includegraphics[scale=0.2]{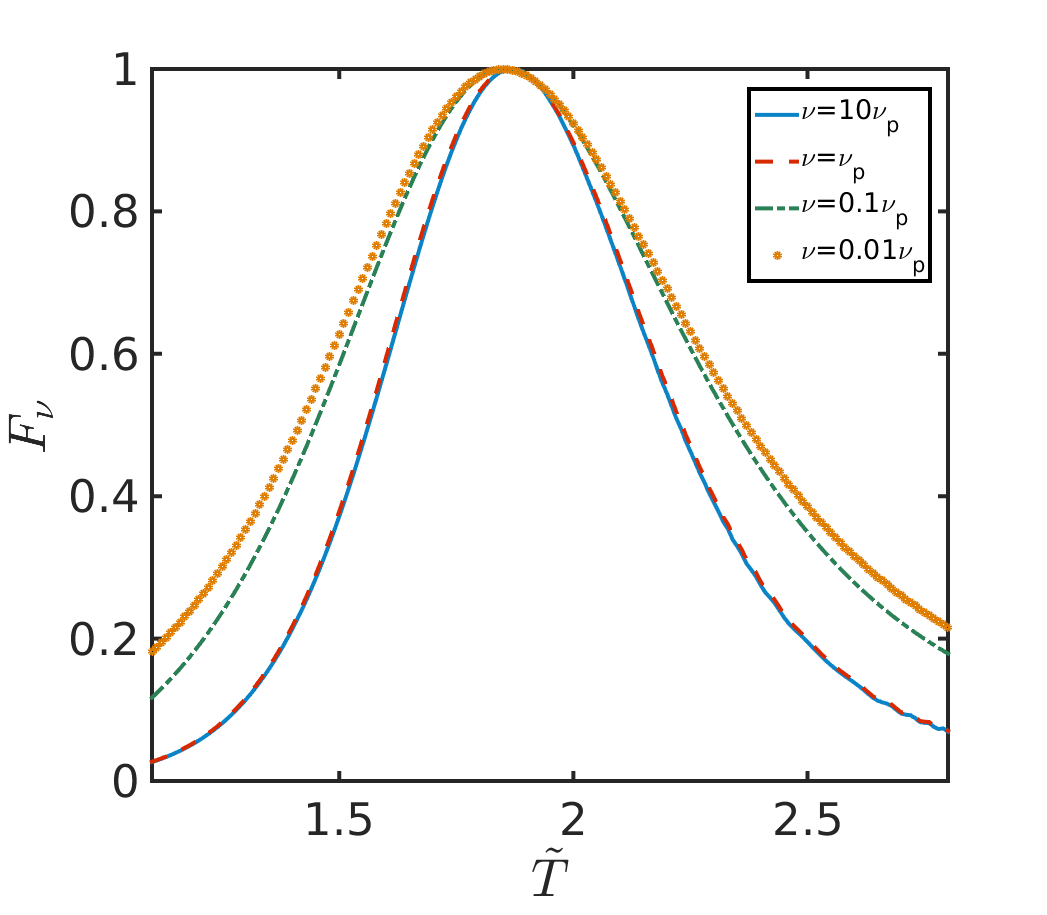}
\includegraphics[scale=0.2]{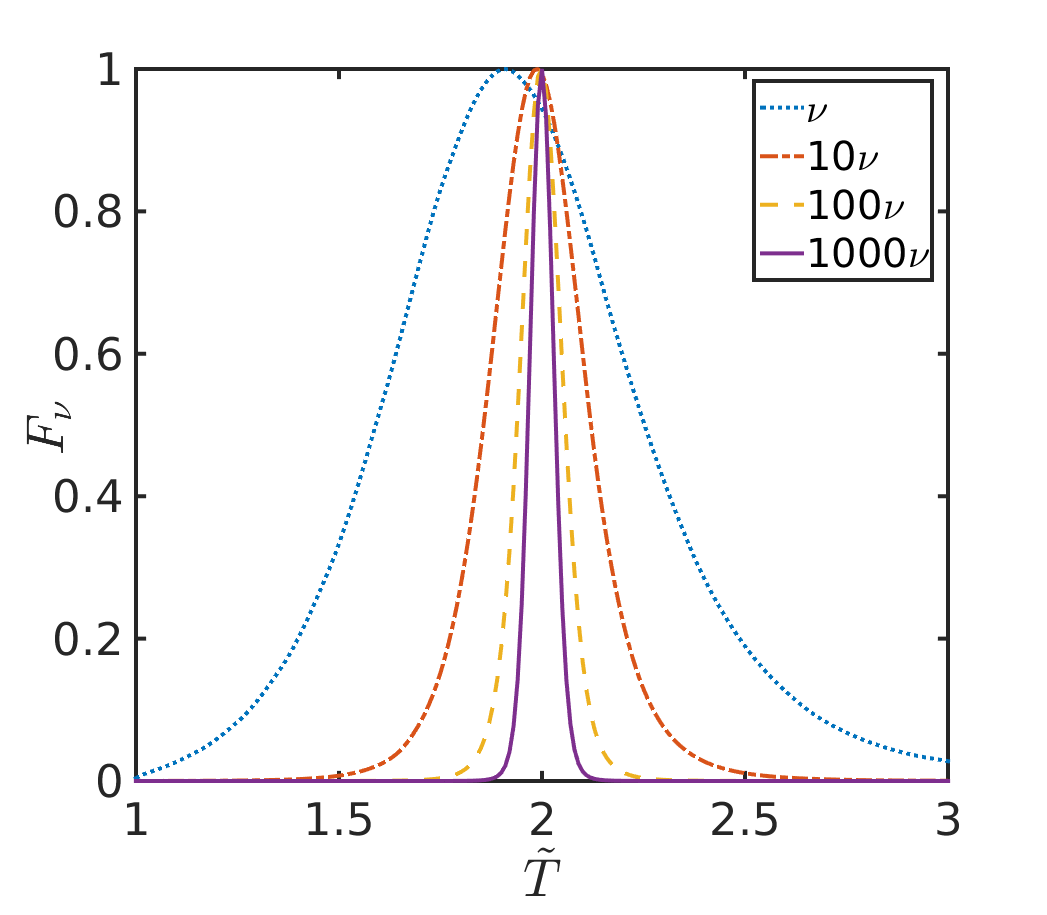}
\caption{light-curves of a single pulse at different frequencies. 
All cases are plotted with a typical Band function 
($\alpha_B=-1, \beta_B=-2.3$), $m=0$ and a log-normal dependence of luminosity 
on radius with $\Delta R /R_0=0.01$. The maximal flux is normalized to 1, and the time is in units of the 
typical time $T_0$ and starting at $T_0$ (when the first photons reach the observer).
{\bf Left}: No $\gamma_e'-\Gamma'$ correlation. In this case we assume $\Gamma'=3, k=0$ and a typical Band function.
The shape of the peak is roughly 
constant at different frequencies, but tends to becomes slightly broader below the peak.
{\bf right:} A $\gamma_e'-\Gamma'$ correlation with $\eta=0.8, k=1$ and with the ``typical" values of $\Gamma'$ corresponding to the observed frequency (see Eq. \ref{eq:nuSyn}),
being: $3.4,8.5,21.5,54$ from the lowest to highest frequency. In this case, the spectrum is $\nu^{-p/2}$ (see \S \ref{sec:corrGammap}).
Here the pulse clearly becomes narrower with lower frequencies, approximately following the relation 
$\Delta T\propto \nu^{-0.4}$ implied by Eq.\ref{eq:nuSyn}.}
\label{fig:difffreq}
\end{figure*}

\subsection{spectral evolution in a single pulse}
\label{spectralevolution}
Using the results of \S~\ref{SplRdelta} we can estimate the typical width of a spike in the light-curve at different frequencies.
As can be seen in Eq.~(\ref{approxTnu}), so long as there is no $\gamma_e'-\Gamma'$ correlation, the width of the pulse is mostly independent of frequency.
However, switching between frequencies at different parts of the spectrum (say from the peak to the lower energy spectral
slope) can somewhat change the pulse width.

\begin{figure*}[h]
\centering
\includegraphics[scale=0.22]{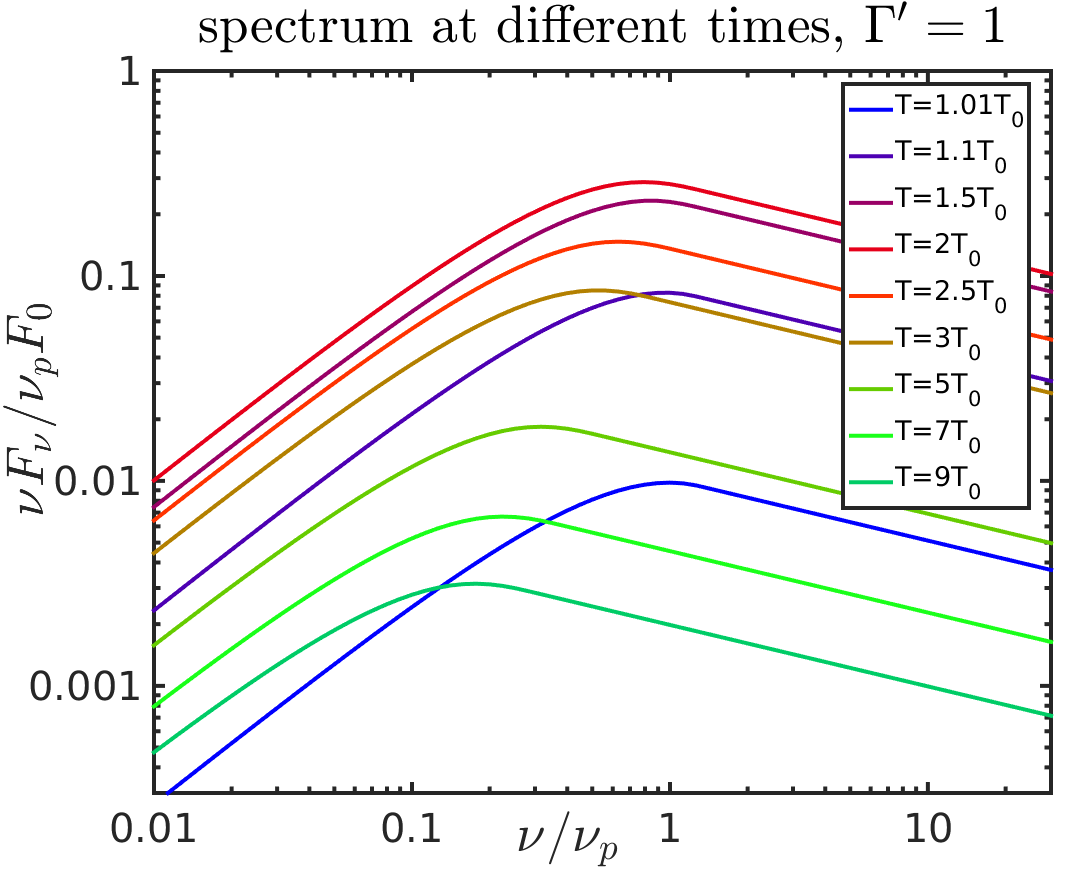}
\includegraphics[scale=0.22]{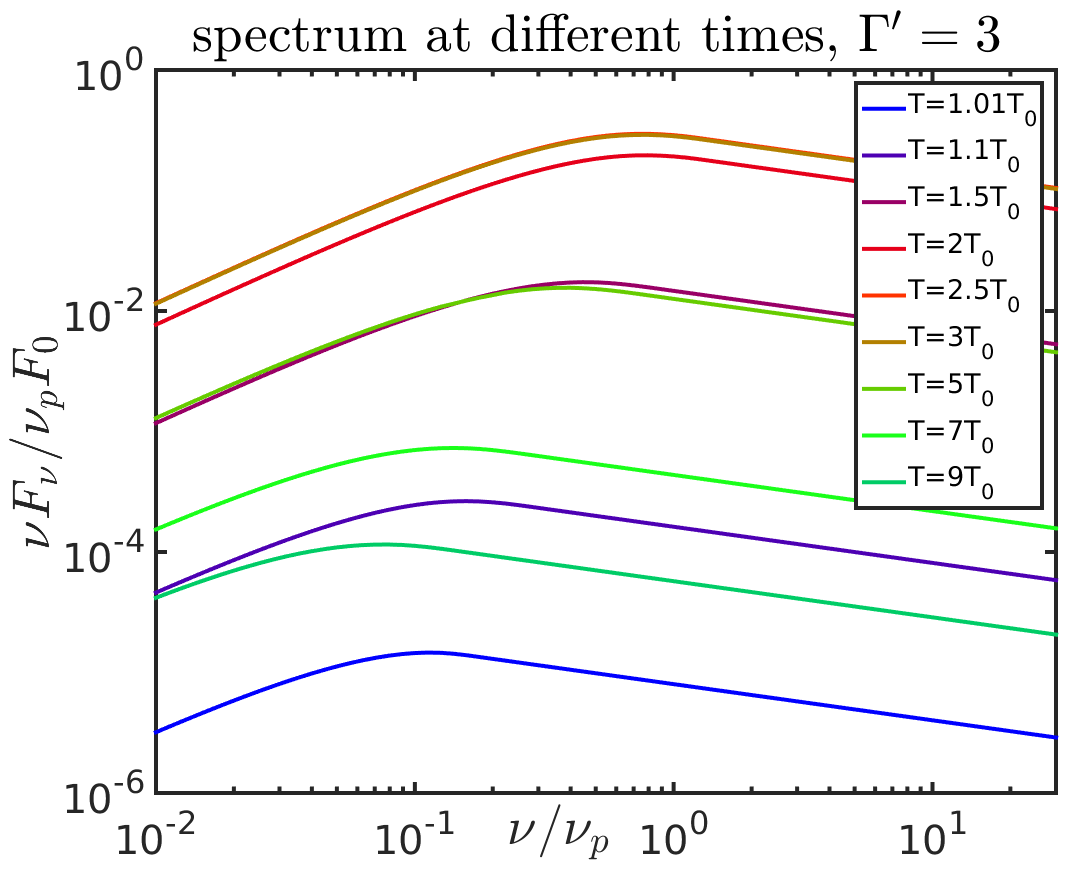}
\caption{Spectra of a single pulse at nine different times: $T=[1.01,1.1,1.5,2,2.5,3,5,7,9]T_0$ for $\Gamma'=1$ (left) and $\Gamma'=3$ (right). 
The intrinsic spectra is assumed to be a typical Band function ($\alpha_B=-1, \beta_B=-2.3$) with $m=0, k=0$ and a constant emissivity from $R_0$ to $R_f=2R_0$. The peak of the light-curve
is at $T=2T_0$. The spectra is plotted for both $\Gamma'=1$ ({\it left}) and $\Gamma'=3$ ({\it right}).
$\nu F_{\nu}$ is given in the ``typical" units of $\nu_p F_0$ and the frequency is in units of the intrinsic peak frequency $\nu_p$.
The spectrum evolves from soft and weak to hard and strong and back again.}
\label{fig:difftime}
\end{figure*}

\begin{figure*}[h]
\centering
\includegraphics[scale=0.19]{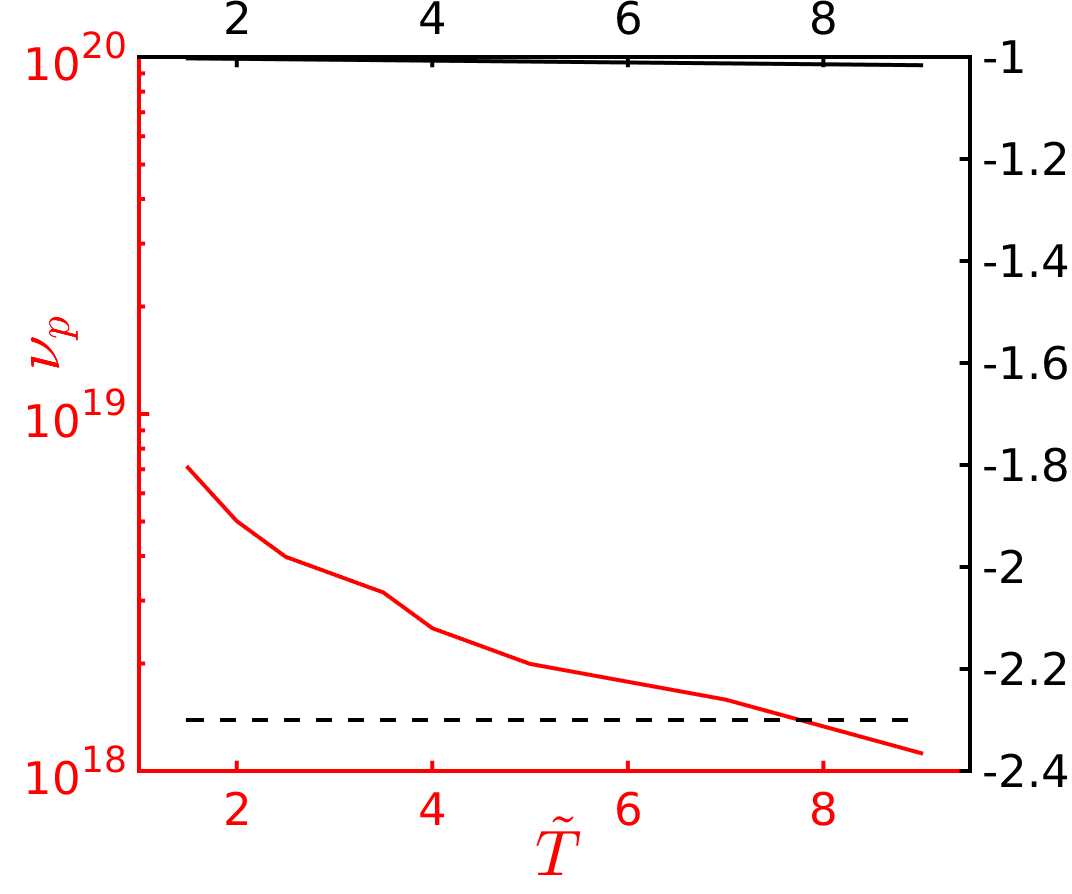}
\includegraphics[scale=0.19]{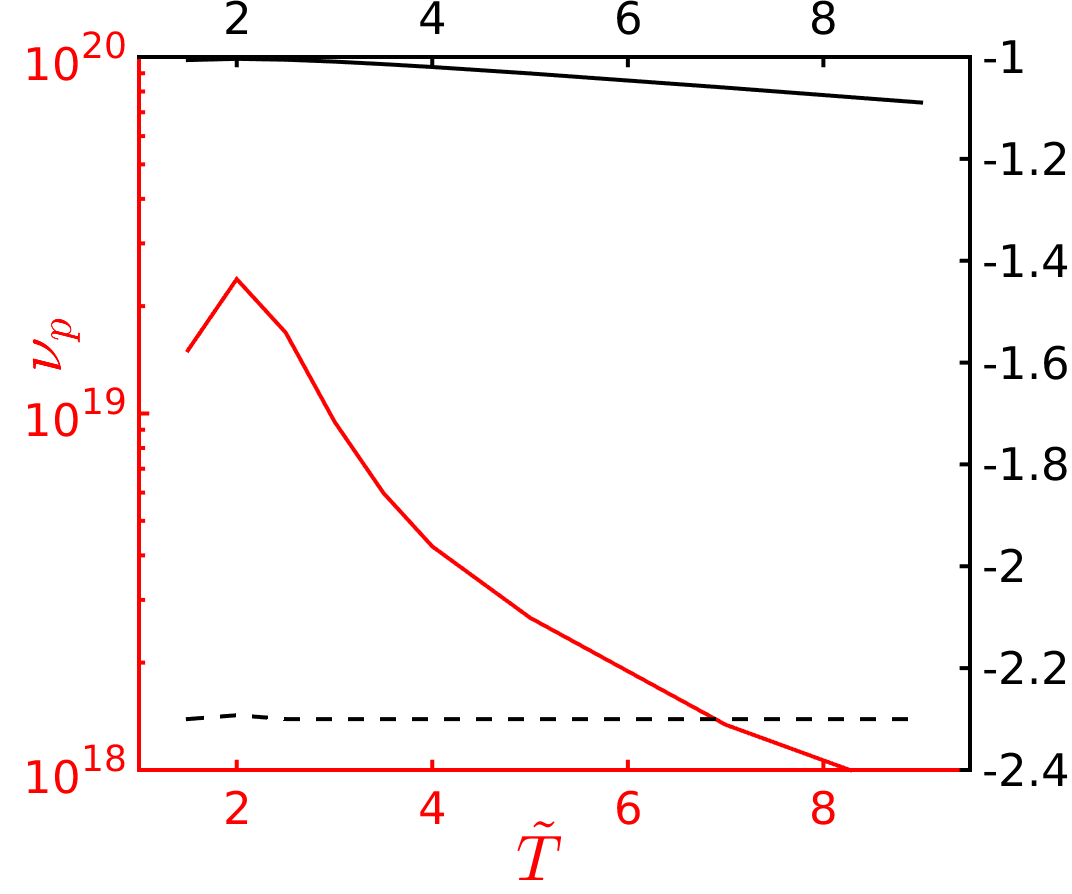}
\includegraphics[scale=0.19]{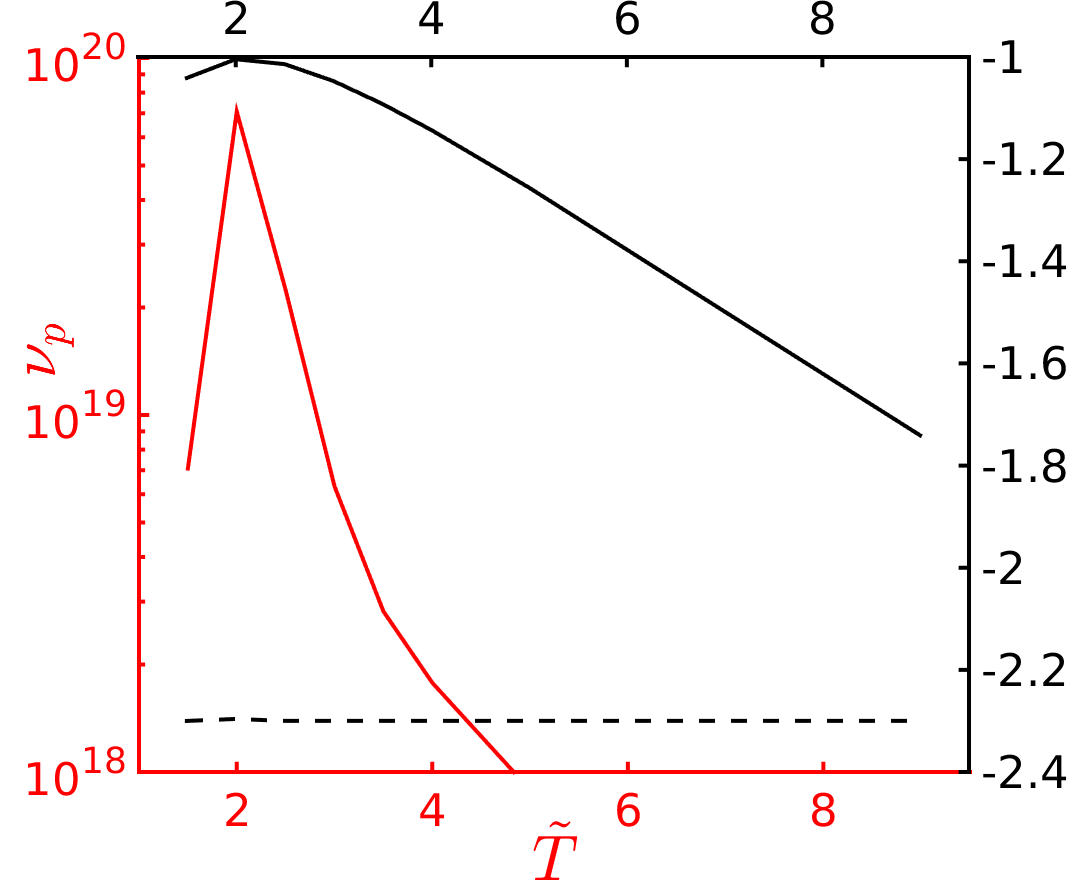}
\caption{Spectral parameters of a single pulse at different times (Left: $\Gamma'=1$, Center: $\Gamma'=3$, Right: $\Gamma'=10$).
The figures are plotted with a constant emissivity from $R_0$ to $R_f=2R_0$.
The peak frequency is plotted in red and the Band photon indices $\alpha_B,\beta_B$ are in black (solid and dashed lines accordingly).
The intrinsic spectra is assumed to be a typical Band function ($\alpha_B=-1, \beta_B=-2.3$) and $m=0, k=0$.
At first the peak energy increases for a short while, and later it decreases significantly.
This behaviour is more pronounced for larger values of $\Gamma'$.
The spectral indexes are basically equal to their intrinsic values and are almost constant in time.}
\label{fig:specparamtime}
\end{figure*}

In Fig.~\ref{fig:difffreq} we plot the resulting light-curve at different frequencies, for a Band spectrum emission.
We assume here a typical Band function ($\alpha_B=-1, \beta_B=-2.3$), since the intrinsic spectra is model dependent and not uniquely defined by the anisotropic emission model.
We note however, that the results presented in this section are not strongly dependent on this assumption, since they are mainly governed by the peak frequency and peak flux which
change more as a function of time (as will be shown in Fig. \ref{fig:specparamtime}).
The light-curves of a single pulse at four different frequencies compared with
the peak: $\nu=0.01\nu_p,0.1\nu_p,\nu_p,10\nu_p$. All cases are plotted with $\Gamma'=3,m=0$ and a Gaussian dependence of luminosity on the logarithm of the radius $\sigR=0.01$.
For the case with no $\Gamma'-\gamma_e'$ correlation, the shape of the peak is roughly constant at different frequencies, but tends to becomes smoother and somewhat broader at lower frequencies.
However, if a correlation between $\Gamma'$ and $\gamma_e'$ exists, then higher $\gamma_e'$ 
can correspond to a larger $\nu_{\rm syn}$ and a larger $\Gamma'$ and thus a narrower spike, of width (see Eq. \ref{eq:nuSyn})
$\Delta T \propto 1/\Gamma'\propto(\nu/\Gamma)^{-\eta/2}$. The observed power law index of $\sim 0.4-0.5$ \citep{Fenimore(1995),Norris(1996),Norris(2005),Bhat(2012)}
would suggest $\eta\sim 0.8-1$. However, we stress that due to isotropization of the electrons' orbit due to Larmour gyration,
this effect is expected to be mostly dominant near $\nu_{\rm syn,max}\approx few\times$GeV, whereas in observations, the pulse widening can be seen at much lower frequencies, down to X-rays.

\begin{figure*}[h]
\centering
\includegraphics[scale=0.25]{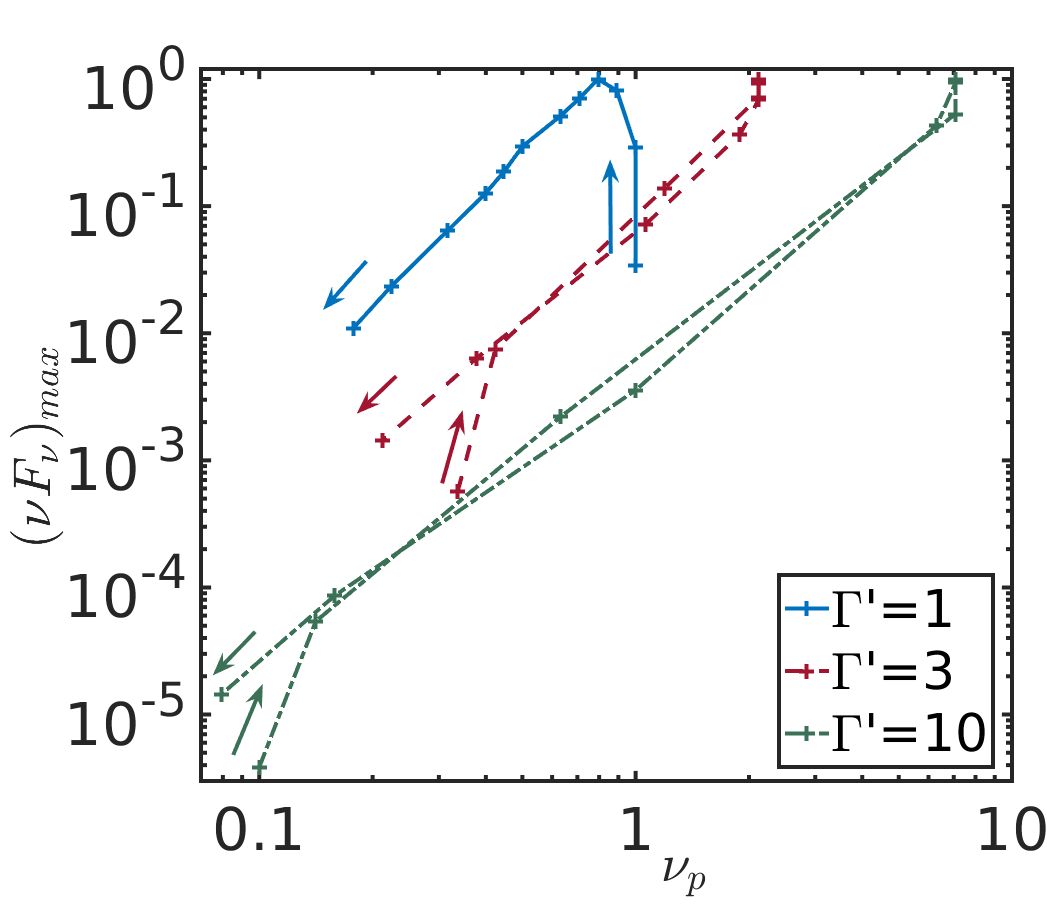}
\caption{peak of $\nu F_{\nu}$ (normalized such that its maximum value is 1) as a function of $\nu_p$ (normalized such that the initial peak of the spectrum in the bulk frame is 1) for a single pulse at different times
($\Gamma'=1$ - solid line, $\Gamma'=3$ - dashed line and $\Gamma'=10$ - dashed-dotted line).
The figures are plotted with a typical Band function ($\alpha_B=-1, \beta_B=-2.3$),
$m=0, k=0$ and a constant emissivity from $R_0$ to $R_f=2R_0$. Arrows denote the direction of progressing time along the curves. $\Gamma'=1$ shows a ``hard to soft'' behaviour, whereas $\Gamma'=3,10$ show an ``intensity tracking behaviour.}
\label{fig:FmaxEp}
\end{figure*}

Another result of this is a change in the spectral shape over time. In Fig.~\ref{fig:difftime} we plot the spectra of 
a single pulse at nine different times: $T=[1.01,1.1,1.5,2,2.5,3,5,7,9]T_0$ for $m=0, k=0$, a constant dependence of luminosity on radius from $R_0$ to $R_f=2R_0$, and for both $\Gamma'=1$ and $\Gamma'=3$. The spectrum evolves from soft 
and weak to hard and strong and back again.
Observations of GRB pulses have shown two typical behaviors: intensity tracking and hard to soft evolution \citep{Ford(1995),Preece(2000),Kaneko(2006),Lu(2012)}. 
Indeed, Figure~\ref{fig:difftime} (as well as Figs.~\ref{fig:specparamtime} and \ref{fig:FmaxEp}) demonstrates 
that our model predicts intensity tracking evolution for sufficiently large $\Gamma'$ (approximately $\Gamma'>2$), 
while a hard to soft evolution appears for low enough $\Gamma'$ (approximately $\Gamma'<2$).
Since pulses in our model are asymmetric, with the rise time much faster than the decay time, this will make 
the hard to soft trend much easier to detect than the initial soft to hard (or intensity tracking) evolution, 
and may account for the observed population of GRB pulses exhibiting ``hard to soft" evolution 
(for instance this was shown for GRB 130427A in the best analysed pulse to date \citep{Preece(2014)}).
Note that a hard to soft evolution is also obtained for $\Gamma'\lesssim 2$.
The evolution of the spectral parameters can be seen more directly in Fig.~\ref{fig:specparamtime} 
where we plot $\nu_p$ and the Band photon indices $\alpha_B, \beta_B$ as a function of time.
The latter do not change significantly in time and remain constant on their intrinsic values.
We also note that the change in spectrum becomes significantly more pronounced for as $\Gamma'$ increases. 
For the latter, the softening and weakening of the spectrum in time follows the behaviour predicted for regular 
high-latitude emission: $F_{\nu}=t^{-2-\alpha} \nu ^{-\alpha}$, where $\alpha$ is the relevant spectral slope of $F_{\nu}$ 
depending on the observed portion of the Band spectrum (either $\alpha=-(\alpha_B+1)$ or $\alpha=-(\beta_B+1)$). 
This behaviour can be seen more clearly in Fig.~\ref{fig:FmaxEp} for $\Gamma'=1$ and to a lesser extent 
also for $\Gamma'=3,\,10$ at late times (which would be harder to observe, as the flux is by then much lower
than its peak value). Initially, the decay for $\Gamma'=3$ is much steeper and might be able to account for the 
observed steep decay phase at the beginning of GRB afterglows \citep[e.g.,][]{Tagliaferri(2005)}.

\section{Discussion}
\label{sec:diss}
In this work we have considered a model for a relativistic source that emits anisotropically in its own bulk rest frame.
Our main motivation was to calculate the most prominent expected observational signatures for prompt GRB emission 
in which the dissipation that powers this emission is driven by magnetic reconnection. This motivated our choice of
the geometry of the magnetic reconnection sites (thin spherical shells -- reconnection layers) and the bulk Lorentz factor 
($\Gamma\gg\Gamma'$). This model may, however, also be applied to other anisotropic emission models in GRBs, 
as suggested for instance in the relativistic turbulence models and the ``jet in jet" models.
This model may also be relevant to other astrophysical systems, such as active galactic nuclei, where very short time-scale 
variability (minutes) has been discovered \citep[e.g.,][]{Aharonian(2007),Albert(2007)}
possibly suggesting emission from mini-jets within the main relativistic jet \citep{Giannios(2009)}. 
Another possible context for relativistic motions in the bulk frame are the GeV flares in the Crab nebula.
These flares were suggested to be Doppler boosted emission from the termination shock \citep{Komissarov(2011)}
or on the hot spot at the base of the jet feature in the nebulae \citep{Lyubarsky(2012)}.
In both cases, however, the bulk frame is not moving relativistically towards us, but is instead effectively at rest.

We have considered a thin spherical shell propagating at a Lorentz factor $\Gamma \gg 1$, which emits as it expands 
outwards and either (i) turns on and off abruptly at radii $R_0$ and $R_f=R_0+\Delta R$, respectively, with the co-moving 
luminosity scaling as a power-law (of index $a$) with radius within this range, or (ii) turns on and off more gradually where 
the co-moving luminosity has a log-normal dependence on radius (centred on $R_0$ with a standard deviation $\sigR$).
The emitting material is assumed to move at a Lorentz factor $\Gamma'$ relative to the shell's local rest frame, 
in a direction transverse to the shell's velocity (i.e. normal to the radial direction). This direction of motion is associated 
with the anti-parallel reconnecting magnetic field lines, and equal amounts of material are assumed to move along the 
corresponding two opposite directions (e.g., on either side of an x-point within the thin reconnection layer).

Strong magnetic fields may also play an important role in the dynamics of relativistic jets.
In AGN jets, the Thompson optical depth at their base is insufficient for efficient thermal acceleration by radiation pressure,
suggesting that Blazars' jets must be magnetically accelerated and therefore initially Poynting-flux dominated. 
In GRBs thermal acceleration is a viable alternative, but hydromagnetic jet launching is expected to dominate, and the 
required high magnetization near the source helps avoid excessive baryon loading, which might prevent the jet from achieving the 
required high-Lorentz factors inferred from the GRB prompt emission \citep[for a recent review on the role of magnetic fields 
in GRBs see][]{Granot(2015)}. Unconfined steady-state, ideal MHD axi-symmetric outflows that start highly magnetized
(with $\sigma_0\gg 1$) near the central source remain highly magnetized ($\sigma\sim\sigma_0^{2/3}$) far from the source, 
while converting only a small fraction of their initial electromagnetic energy to bulk kinetic energy ($\Gamma\sim\sigma_0^{1/3}$).
This may be alleviated if the outflow is gradually collimated into a narrow jet of half-opening angle $\theta_j$, 
which could asymptotically lead to $\sigma\sim\sigma_0^{2/3}\theta_j^{2/3}$ and $\Gamma\sim\sigma_0^{1/3}\theta_j^{-2/3}$,
or due to an abrupt drop in the external pressure (e.g. when a long-duration GRB jet breaks out of its progenitor star), but
in all cases the asymptotic magnetization is still fairly high, $\sigma\geq 1$.
However, the strong Inverse Compton component observed in Blazar spectra suggest a low $\sigma$ in the emission region
\citep[see e.g.][]{GhiselliniTavecchio09}, while in GRBs different lines of evidence suggest a similar conclusion
\citep[e.g.,][]{Beniamini(2014)}. This result is commonly refereed to as the $\sigma$ problem \citep{Goldreich(1969)}.
This problem may be solved by relaxing one or more of the underlying assumptions of steady-state, ideal MHD and axi-symmetry.
Strong time-dependence can lead to a much more efficient conversion of magnetic to kinetic energy, and to $\sigma<1$ 
far from the source that allows efficient dissipation in internal shocks \citep{GKS11}. 
Alternatively, magnetic reconnection can play an important role in the jet's acceleration and in the reduction of $\sigma$,
and may be greatly enhanced by various instabilities, some of which or non-axisymmetric (such as the kink instability).
The dissipation of magnetic energy generates heat, some of which can be channelled to radiation, but most of which can 
contribute to the (now thermal) acceleration of the jet. The distance from the central engine where efficient reconnection 
takes place, will determine whether the jet is still accelerating \citep{Drenkhahn(2002),Lyubarsky(2010)}, coasting, 
or decelerating \citep{Lyutikov(2003),Giannios(2006)} at that stage. In GRBs such late reconnection during the jet's deceleration 
phase might be expected for the following reason. The jet loses lateral causal contact during its acceleration, eventually reaching 
$\Gamma\theta_j\gg 1$, so that once it starts being decelerated by the external medium increasingly larger angular scales 
($\sim 1/\Gamma$) in the jet come into causal contact and allow reconnection of magnetic structures on such scales, 
which were not possible at earlier times.

Efficient magnetic reconnection may naturally be achieved in a striped wind magnetic field geometry.
In this case, the light-curve variability time may reflect the time for the flipping of the magnetic field at the source (i.e. the 
base of the outflow) over the (dimensionless) co-moving flow velocity into the reconnection layer, $\beta'_{\rm in}$.    
For a magnetar central engine, a periodic flipping of the field's direction is expected with a sign flipping time equal to 
half of the magnetar's rotational period $P$, as the local field switches sign twice in each rotation \citep[e.g.,][]{Usov(1992)}.
This may lead to relatively ``ordered" light-curves with short variability time-scales, 
$\Delta T\sim 5(P/1\,{\rm ms})(\beta'_{\rm in}/0.1)^{-1}\;$ms. These become even shorter for a relativistic inflow
velocity $\beta'_{\rm in}\sim 1$ into the reconnection layer that may arise for large $\sigma$-values
\citep[e.g.,][]{Guo(2015)}, which also leads to acceleration of the emitters to larger values of $\Gamma'$. 
Such short variability times are typically significantly shorter than the dominant variability time in most GRB light-curves.

An alternative option that can naturally lead to larger variability times as observed in most GRB light-curves is that the 
central engine is an accreting black hole, where the magnetic field that is advected inwards with the accreted material 
randomly switches sign either through some disk instability or by accreting blobs of plasma with a randomly oriented 
frozen-in magnetic field \citep[e.g.,][]{Blandford(1982),Rees(1984),Uzdensky(2006),Barkov(2008)}. 
In this case the typical time-scale for a flip in the magnetic field 
orientation is more uncertain and in addition, due to the stochastic nature of the process, the resulting scale for the flipping 
of the magnetic field will be much more variable. Therefore, it can potentially account both for the much larger observed
typical variability times compared to the naive expectation for a millisecond magnetar, as well as for the large range of 
variability time-scales that are observed in most GRB light-curves and their stochastic nature in general. 
The resulting efficiency of reconnection in such an outflow with a random flipping of the magnetic field is uncertain, 
and may be somewhat lower than for a periodic flipping that was mentioned above. This might make it hard to satisfy the 
high gamma-ray emission efficiencies that are commonly inferred from GRB observations \citep{Panaitescu(2002),Granot(2006),Fan(2006)}. However, as was 
recently shown by \cite{Beniamini(2015)} these efficiencies should actually be only moderate ($\sim 0.1$), significantly 
less than what was suggested by some earlier estimates ($\sim 0.9$).

Throughout most of this work we have focused on the properties of the emission that are direct results of anisotropic 
emission in the jet's bulk frame, and that are largely independent of the specific radiation mechanism/s at work. 
However, as mentioned above, our main motivation comes from magnetic reconnection models for the energy dissipation.
In reconnection models for the prompt emission of GRBs, because of the large expected magnetic fields the most natural 
radiation mechanism is synchrotron. We consider the emitting electrons to be 
in the ``slow cooling" regime in order for the spectrum to be consistent with observations in the optical and X-ray bands \citep{Beniamini(2014)}.
This can be expected if the {frame in which the electromagnetic field becomes purely magnetic, is that of the emitters (``the plasmoids"). However, if the latter is in fact the jet's frame}, then electrons producing the observed sub-MeV peak will isotropize
much faster than they can radiate their energy and the resulting emission would no longer be anisotropic, except at frequencies very close to $\nu_{\rm syn,max}$ (see \S \ref{sec:corrGammap} for details).
In ``one zone" magnetically dominated emission regions, the emitting electrons would not be in the ``slow cooling" regime, 
unless $\Gamma\gtrsim 600$ \citep{Beniamini(2014)}; and even then, slow cooling would only occur for the largest possible 
radii, and mildly relativistic electrons. However, in the anisotropic model, suggested in this paper,
there are two important differences compared to the ``one zone" case. First, as mentioned in \S \ref{motiv} the radius implied 
by the variability time-scale is larger by a factor $\Gamma'$ as compared to the non-boosted case (leading to a weaker magnetic 
field for the same overall luminosity). Second, the local magnetic energy density within the reconnection sheet,
where the particles are emitting, can be up to an order of magnitude lower than the average field. 
Combining these two facts implies that in these models, even for a smaller bulk Lorentz
factor: $\Gamma\approx 300$, electrons with $\gamma_e' \lesssim 10$ will be in the `slow cooling" regime, 
as required by both theory and observations. In addition the upper limit on $\gamma_e'$ will increase significantly 
($\gamma_e'\propto \Gamma^5$) for larger values of $\Gamma$.

We briefly summarize our main results and their comparison to the observations:
\begin{enumerate}
\item \uline{\bf{Light-curve Variability}}: GRB light-curves show highly variable time profiles with the 
times between pulses roughly equal to the pulses durations \citep{Nakar(2002)}. We show that isotropic reconnection 
emission models are in conflict with this observation since they generally imply pulse widths significantly broader 
than the time between pulse peaks. However, we find that the anisotropic emission model presented in this work can 
produce a highly variable emission at large distances from the central source, as required by observations.

\item  \uline{\bf{Pulse Asymmetry}}: observed GRB pulses tend to be asymmetric with a typical 
rise to decay time ratios $t_{\rm rise}/t_{\rm decay}\sim 0.3-0.5$
\citep{Nemiroff(1994),Fishman(1995),Norris(1996),Quilligan(2002),Hakkila(2011)}.
Asymmetric pulses with $t_{\rm rise}<t_{\rm decay}$ naturally occur  in our model (see \S \ref{paramspace} and Table \ref{tbl:cases}) 
with $t_{\rm rise}/t_{\rm decay}$ depending on the radial dependence of the bulk Lorentz factor ($\Gamma\propto R^{-m/2}$)
and co-moving luminosity (power-law $\propto R^a$ or log-normal $f(R/R_0)$), on $\Gamma'$, and on the radial extent 
of the emission producing the pulse, $\Delta R/R_0$. In particular, they naturally occur for 
$\Delta R\gtrsim R_0$ (e.g. when $f(R/R_0)$ is asymmetric, see for instance the log-normal or power-law functions with $a<0$, considered in this paper).
Since the shape of a pulse in this case is determined by $f(R/R_0)$, observations of the pulse structure could be used to deduce the evolution of
the emissivity powered by magnetic reconnection at a given  site in the jet, in a way that may be largely 
independent of the micro-physical properties of the reconnection and instead depend mainly on its gross properties.
We note however, that since we effectively see radiation from only a small angular portion of the jet, it would be hard to use this in order
to learn about the global structure of the jet.
The pulse-shape gradually becomes more symmetric as $\Gamma'$ increases, so long as $1/\Gamma' > \Delta R/R_0$. 
At the same limit, the pulse both becomes narrower and peaks at higher fluxes (for a given total radiated energy) 
for larger $\Gamma'$. 
Finally, we note that for $\Gamma'\gtrsim 2$ and $\Delta R/R_0 \lesssim 1$, the degree of asymmetry depends mainly on $\Delta R/R_0$.
Here, once more, observations of the pulse asymmetry could be used to deduce the radial profile of the emission in the jet as mentioned above.

\item  \uline{\bf{Luminosity-Variability Correlation}}: previous studies have found a correlation between the variability 
of GRB light-curves and their peak luminosities \citep{Stern(1999),Fenimore(2000),Reichart(2001),Guidorzi(2005a),Guidorzi(2005b)}.
For $\Delta R /R \lesssim 1/\Gamma'$ and $\Gamma'\gtrsim 2$ our model predicts that as $\Gamma'$ increases, the peak luminosities of pulses increase while their widths decrease (leading to larger variabilities).
For shells of width $\Delta R/R_0 \lesssim 0.01$ or even (assuming a natural assumption that
the luminosities are correlated with $\Gamma'$) $\Delta R/R_0 \lesssim 0.1$, a correlation between maximal luminosity 
and the lightcurve variability is indeed obtained in our model. Although in the latter case, the power law index of the correlation is significantly smaller, it is still consistent with the values reported by \citep{Guidorzi(2005a),Guidorzi(2005b)}.

\item \uline{\bf{$L_p$--$\nu_p$ Correlation}}:  many studies have claimed the existence of  a correlation between 
the peak luminosities $L_p$ and peak frequencies $\nu_p$ of GRBs \citep{Yonetoku(2004),Ghirlanda(2005),Yonetoku(2010)} 
such that $L_p \propto \nu_p ^{1.5-2}$. Recently, it was further suggested \citep{Guiriec(2015)} that these results could hold 
also within a single burst, comparing different pulses.
In both the non-boosted ($\Gamma'=1$) and anisotropic  models, both the peak luminosity and peak frequency are Doppler 
boosted from the emitters' frame. This results in a relation between the luminosity and the peak frequency in the observer 
frame that is similar to that in the jet's bulk frame: $\frac{L_p}{\nu_p^2}=\frac{L_p'}{\nu_p'^2}$, independent of the 
bulk Lorentz factor $\Gamma$. This relation implies that if such a correlation between peak frequency and the luminosity 
exists in the jet's bulk frame, then it would naturally lead to a similar correlation in the observer frame,
and would not be wiped out by varying $\Gamma$, which is likely to change significantly from burst to burst 
(and even within a single burst) and is unlikely to be directly related to the properties of the flow in the jet's bulk frame.
In case of a blob-like anisotropic emission with $\Delta R/R_0  \gtrsim 1/\Gamma'$ this result can be extended further 
as it holds also in case the emission properties are set in the emitters' frame.

\item \uline{\bf{Pulse Widths and Spectral Lags}}: observations of GRB light-curves at different frequencies show 
that the pulse widths tend to decrease with frequency approximately as $W(\nu)\propto \nu^{-0.4}$
\citep{Fenimore(1995),Norris(1996),Norris(2005),Bhat(2012)}. A related observation is that at larger frequencies, 
pulses tend to peak earlier -- this is usually referred to as spectral lags \citep{Norris(1996),Band(1997)}. 
We note that in case the pulse shape at different frequencies is the same, and the emission starts at 
the same time, this would follow directly from the anti-correlation between pulse width and frequency mentioned above.
In our model, in case there is a correlation between the bulk Lorentz factors of the electrons,$\gamma_e'$, and $\Gamma'$, 
then it is possible to obtain a negative correlation between the pulse width and the observed frequency, similar to the 
observed one. This is because higher $\gamma_e'$ typically correspond to both a larger $\nu_{\rm syn}$ and a larger 
$\Gamma'$ that leads a narrower spike. The observed relations are reproduced for $ \Gamma'\propto \gamma_e'^{0.8-1}$. However, we stress that due to isotropization of the electrons' orbit due to Larmour gyration,
this effect is expected to be mostly dominant near $\nu_{\rm syn,max}\approx few\times$GeV, whereas in observations, the pulse widening can be seen at much lower frequencies, down to X-rays.

\item \uline{\bf{Single Pulse $L-\nu_p$ Correlation}}: observations of GRB pulses have shown two typical behaviours for 
their evolution in time: {\it intensity tracking}, where the peak frequency follows the overall flux or luminosity, 
and {\it hard to soft} evolution, where the peak frequency is constantly decreasing in time 
\citep{Ford(1995),Preece(2000),Kaneko(2006),Lu(2012)}.
Interestingly, both behaviours are possible in our model, depending on $\Gamma'$.
For small anisotropy ($\Gamma'\lesssim 2$), the the peak frequency always monotonically decreases, while the flux first increases and later  decreases. Therefore, one may expect a 
``hard to soft" evolution in this case.  However, for larger values of $\Gamma'$ ($>2$), the peak frequency also increases 
at first as the flux rises towards the peak of the pulse, and only later decreases during the tail of the pulse, so in this case 
one expects an ``intensity tracking" behaviour.

\item \uline{\bf{Rapid Decay Phase}}: Finally, observations of GRBs in the early afterglow phase exhibit a so called 
``rapid decay" phase \citep{Tagliaferri(2005)}. This early phase of rapid flux decay at the end of the prompt emission
was unexpected since the flux often falls faster than would be expected from high-latitude emission (which is the shortest
time-scale on which the flux can decay, even if the emission shuts off instantly). A possible solution in many cases is that the 
initial fast decay is dominated by the last pulse so that the zero-point for the power-law decay should be taken as the onset 
of that last pulse rather than of the whole GRB, i.e. the onset of the first pulse. In the anisotropic model the initial flux decay
of a single pulse is significantly more rapid than in the non-boosted case (and increases with $\Gamma'$), and is thus able to 
account even for the most extreme cases of such rapid decay phases. A similar solution, producing rapid decays due to anisotropic
emission was considered by \cite{Beloborodov(2011)} who suggested an anisotropic emission model for GRB afterglows.
\end{enumerate}

We conclude that the simple anisotropic emission model presented in this paper is able to reproduce many of the observational
features seen in the prompt emission of GRBs, and therefore seems very promising.
In a future work we plan to extend this
model to more realistic magnetic reconnection configurations, such as those obtained in relativistic reconnection simulations. 
This would help determine some of the free parameters that still exist in this model and could provide a self-consistent model 
for energy dissipation and emission in the prompt phase of GRBs.

\acknowledgements
We thank Tsvi Piran, Daniel Kagan, Pawan Kumar and Wenbin Lu for helpful discussions. We also thank the anonymous referee for
helpful suggestions and comments.
This research was supported in part by the ISF grant 719/14 (JG).

\appendix
\numberwithin{equation}{section}

\section[A] {approximate expression for the flux}
\label{approx}
Here we derive the result in Eq.~(\ref{approxFnu2}), by solving the double-integral in Eq.~(\ref{flux2}) in the limit $\Gamma'\gg 1$. 
First, the inner integral, may be written as
\begin{equation}
I_1 = \int_0^{2\pi}d\phi(1-\beta'\sin\theta'\cos\phi)^{k-\alpha-3}\ . 
\end{equation}
Since the power $k-\alpha-3$ is typically negative, the integrand becomes maximal when $(1-\beta'\sin\theta'\cos\phi)$ is minimal, 
i.e. for $\phi=0, \theta'=\pi/2$, and it rapidly drops for different values of $\phi$, $\theta'$. Therefore, in order to evaluate
the inner integral we can use the approximation $\phi\ll1$ and $\cos\phi\approx 1-\phi^2/2$, and take the limits of integration
to be form $-\infty$ to $\infty$ (as the dominant contribution is anyway from a small region where $|\phi|\lesssim 1/\Gamma'\ll 1$),
\begin{equation}
I_1 \approx \int_{-\infty}^{\infty}d\phi\left[1-\beta'\sin\theta'\left(1-\frac{\phi^2}{2}\right)\right]^{k-\alpha-3}
= \frac{\sqrt{2\pi}\,\Gamma(2.5+\alpha-k)}{\Gamma(3+\alpha-k)}\frac{(1-\beta'\sin\theta')^{k-\alpha-2.5}}{\sqrt{\beta'\sin\theta'}}\ ,
\end{equation}
where here $\Gamma$ is the gamma function. As for the outer integration, over $y$, since most of the contribution is
from $\theta'\approx\pi/2$, this corresponds to $y\approx y_0 = (m+2)^{-1/(m+1)}$, so that
\begin{equation}
\left(\frac{m\!+\!1}{m\!+\!y^{-\!m\!-\!1}}\right)^{2+\alpha} y^{a-1-\frac{m}{2}(1+\alpha)}
\approx 2^{-2-\alpha}(m+2)^\frac{m(1+\alpha)+2(1-a)}{2(m+1)}\ .
\end{equation}
Also, the integral does not vanish only if the corresponding radius, $R = yR_L\approx y_0 R_L = y_0 R_0(T/T_0)^{1/(m+1)}$ is in the
emission region, i.e. between $R_0$ and $R_f$, or
\begin{equation}\label{eq:int_y}
1<\left[\frac{T}{(m+2)T_0}\right]^\frac{1}{m+1}<\frac{R_f}{R_0}\ .
\end{equation}
Moreover, $\sqrt{\beta'\sin\theta'}\approx 1$, so that the only term in the integrand that significantly varies in the region
where the dominant contribution comes from is $(1-\beta'\sin\theta')^{k-\alpha-2.5}$. Now we can define $\tilde{y}=y-y_0\ll 1$
and approximate
\begin{equation}
\sin\theta' = 2\frac{\sqrt{(m+1)(y^{-m-1}-1)}}{m+y^{-m-1}} \approx 1-\frac{\tilde{y}^2}{8y_0^{2(m+2)}}\ .
\end{equation}
Taking also the limit where $\Gamma'\gg 1$ so that $\beta'\approx1-1/2\Gamma'^2$, we have
\begin{equation}
(1-\beta'\sin\theta')^{k-\alpha-2.5} \approx \left(\frac{1+\frac{1}{4}\Gamma'^2\tilde{y}^2y_0^{-2(m+2)}}{2\Gamma'^2}\right)^{k-\alpha-2.5}\ .
\end{equation}
Changing variables to $\bar{y} = \tilde{y}\Gamma'/2y_0^{m+2}$, we have $dy = d\tilde{y} = (2y_0^{m+2}/\Gamma')d\bar{y}$.
Since the main contribution to the integral is from $\bar{y}$ of order unity, then for $\Gamma'\gg 1$ we can take the limits of
integration over $\bar{y}$ form $-\infty$ to $\infty$, and when the condition in Eq.~(\ref{eq:int_y}) is satisfied one obtains
\begin{equation}
\begin{split}
\int_{y_{\rm min}}^{y_{\rm max}}\! \! dy(1-\beta'\sin\theta')^{k-\alpha-2.5} 
&\approx \Gamma'^{\,4+2\alpha-2k}2^{3.5+\alpha-k}y_0^{m+2}\int_{-\infty}^\infty d\bar{y}(1+\bar{y}^2)^{k-\alpha-2.5}
\\
&= \Gamma'^{\,4+2\alpha-2k}2^{3.5+\alpha-k}y_0^{m+2}\frac{\sqrt{\pi}\,\Gamma(2+\alpha-k)}{\Gamma(2.5+\alpha-k)}\ .
\end{split}
\end{equation}
Putting all the terms together we obtain
\begin{equation}
F_{\nu}(T) \approx \,\frac{\Gamma_0 \Gamma'^{\,1-2k} L''_{\nu''_0}}{2^{k}\pi D^2 }\!\fracb{\nu}{\nu_0}^{-\alpha}
\fracb{T}{T_0}^{2a-m(1+\alpha) \over 2(m+1)}
\frac{(a+1)(m+2)^\frac{m(\alpha-1)-2(a+1)}{2(m+1)}}{\left(1+\frac{\Delta R}{R_0}\right)^{a+1}-1}
\frac{\Gamma(2+\alpha-k)}{\Gamma(3+\alpha-k)} \ .
\end{equation}



\section[C]{Angular time-scale in the anisotropic emission model}
\label{sec:angtimescale}

We aim to find here the angular time-scale for the model with anisotropic emission presented in this paper and show that
it is $T_\theta=(1+z)R/c\Gamma^2 \Gamma'$. We define $\xi=(\Gamma\theta)^2$. In the relativistic limit, $\Gamma\gg 1$,
the main contribution to the radiation comes from  $\theta\ll 1$ for which $\cos\theta\approx 1-\theta^2/2$, and
$\beta = v/c\approx 1-1/2\Gamma^2$ so that
\begin{equation}
R = \int_0^t \beta(t') cdt' \approx ct-\int_0^R \frac{dR'}{2\Gamma^2(R')} = ct - \frac{R}{2(m+1)\Gamma^2(R)}
\quad\Longrightarrow\quad  t\approx\frac{R}{c}\left[1+\frac{1}{2(m+1)\Gamma^2}\right]\ .
\end{equation}
Using the definition of the EATS along with the relations $R/R_0 = y(T/T_0)^{1/(m+1)}$ and $R/\Gamma^2 = 
(R/R_0)^{m+1}R_0/\Gamma_0^2 = y^{m+1}(T/T_0)R_0/\Gamma_0^2$ (see Eqs. [\ref{eq:T0}] and [\ref{eq:scalings}])  
one obtains:
\begin{equation}
 T=(1+z)\left[t-\frac{R\cos\theta}{c}\right]\approx \frac{(1+z)R}{2(m+1)c\Gamma^2}\left[1+(m+1)\xi^2\right]
 =T y^{m+1}\left[1+(m+1)\xi^2\right]\ .
\end{equation}
This can be inverted to obtain an expression for $\xi$ (see for instance \citep{Genet(2009)}):
\begin{equation}
 \xi=(\Gamma\theta)^2=\frac{y^{-m-1}-1}{m+1}\quad\Longleftrightarrow\quad 
 y^{-m-1}=1+(m+1)\xi\ .
\end{equation}
Plugging this into the expression for $\sin\theta'$ (Eq. \ref{trigonometric}) we obtain:
\begin{equation}
 \sin^2\theta'=\frac{4(m+1) (y^{-m-1}-1)}{(m+y^{-m-1})^2}=\frac{4 \xi}{(1+\xi)^2}\ .
\end{equation}

Let us also define the angle between $\hat{n'}$ and $\hat{\beta '}$ as $\chi'$. By definition, 
and using the fact that $cos^2\phi=cos^2\phi'$, we have:
\begin{equation}
\cos^2\chi'=(\hat{n'} \cdot \hat{\beta '})^2= \sin^2\theta' \cos^2\phi' = \frac{4\xi}{(1+\xi)^2}\cos^2 \phi\ .
\end{equation}
Defining the parameter
\begin{equation}\label{eq:f}
f = \frac{\cos^2\phi}{\cos^2\chi'}\ ,
\end{equation}
we thus obtain a quadratic equation for $\xi$,
\begin{equation}
 4f\xi=(1+\xi)^2 \quad\Longrightarrow\quad \xi^2 + \xi (2-4f) + 1=0\ ,
\end{equation}
for which the solutions are:
\begin{equation}\label{eq:xi_pm}
 \xi_\pm=2f-1 \pm \sqrt{(2f-1)^2-1}\ .
\end{equation}
The dominant contribution to the emission arises from within the beaming cone, i.e. from $\chi' \leq 1/\Gamma'$.
Therefore, in order to find the boundaries of this region, we simply substitute $\chi'=1/\Gamma'$ in the expression
for the parameter $f$ in Eq.~(\ref{eq:f}). Hence, each value of $f$ corresponds to a particular value of $\cos^2\phi$,
which in turn correspond to 4 different values of $\phi = \pm\phi_0$ and $\pi\pm\phi_0$. For each of these $\phi$ 
values, there are two values of $\xi = \xi_\pm$ according to Eq.~(\ref{eq:xi_pm}), which each correspond to a value 
of $\theta$. As shown in Fig.~\ref{fig:emission_region}, for a given value of $\Gamma'$  this implies two approximately
circular regions of angular radius $\approx 1/(\Gamma'\Gamma)$ around the two points of intersection of the $x'$-axis
($\phi = 0,\,\pi$) with the circle of angular radius $\theta =1/\Gamma$ around the line of sight (i.e. $\xi=1$), i.e.
$(\xi,\,\phi) = (1,0)$ and $(1,\,\pi)$ or $(\theta,\,\phi) = (1/\Gamma,0)$ and $(1/\Gamma,\,\pi)$.
For a given value of $\chi'=1/\Gamma'$, there are solutions for $\xi$ only for $f$ in the range $1\leq f\leq 1/\cos^2\chi'$. 
The outer edges of this regions in the $\phi$ direction correspond to $f = 1$ and $\phi_0 = \chi' = 1/\Gamma'$.
The corresponding angular separation or distance from the center of the relevant region is 
$\approx\phi_0/\Gamma = 1/(\Gamma'\Gamma)$.
Along the $x'$-axis $\phi_0=0$,  $f=1/cos^2\chi' = 1/cos^2(1/\Gamma') \approx 1+1/\Gamma'^2$ or $2f-1\approx 1+2/\Gamma'^2$, where the last approximation holds for $\Gamma'\gg 1$, for which in this case one obtains 
\begin{equation}
 \xi_\pm\approx1+\frac{2}{\Gamma'^2} \pm \sqrt {(1+\frac{2}{\Gamma'^2})^2-1} \approx 1 \pm \frac{2}{\Gamma'}\ ,
\end{equation}
for which the corresponding values of $\theta$ are:
\begin{equation}
 \Gamma\theta_\pm =\sqrt{1 \pm \frac{2}{\Gamma'} } \approx 1 \pm \frac{1}{\Gamma'} 
 \quad\Longrightarrow\quad \theta_\pm \approx \frac{1}{\Gamma} \pm \frac{1}{\Gamma \Gamma'}\ ,
\end{equation}
which again correspond to an angular separation or distance from the center of the relevant region (that corresponds to 
$\theta=1/\Gamma$ with the same value of $\phi = 0$ or $\pi$) of $\approx\phi_0/\Gamma = 1/(\Gamma'\Gamma)$.

We see that the typical range of $\theta$ from which a dominant contribution to the emission arrives is indeed $1/(\Gamma \Gamma')$.
We can now estimate the angular time-scale, which is the difference in arrival times of photons from $\theta_1=1/\Gamma$ and $\theta_2=1/\Gamma+1/(\Gamma \Gamma')$:
\begin{equation}
 \frac{T_\theta}{1+z}=\frac{R}{c} (\cos\theta_{-}-\cos\theta_{+})\approx
 \frac{R}{c} (\frac{\theta_{+}^2}{2}-\frac{\theta_{-}^2}{2})
 =\frac{R}{2c} (\theta_{+}+\theta_{-})(\theta_{+}-\theta_{-})\approx \frac{2R}{c\Gamma^2\Gamma'}\ .
\end{equation}
This is taking the diameter of this approximately circular region. A better estimate is obtained by using its angular radius,
which results a a time smaller by a factor of $2$,
\begin{equation}
 T_\theta\approx \frac{(1+z)R}{c\Gamma^2\Gamma'}\ ,
\end{equation}
that is shorter by a factor  $\Gamma'/2$ than the angular time for the case of non-boosted emission in the jet's bulk frame.

\section[D]{The typical electrons' energy in the emitters' frame}
\label{sec:zeromomentum}
Consider a distribution of electrons, such that for a given $\gamma_e'$, electrons velocities are distributed uniformly within 
a cone of half-opening angle $\theta_0(\gamma_e')$ around the $x'$-axis in the jet's bulk frame.
The four velocity of an electron with a random Lorentz factor $\gamma_e'$ travelling at some angle $\tilde{\theta}$ to the $x'$-axis 
in the jet's bulk frame, is given by:
\begin{equation}
 u^{\mu}=(\gamma_e',\gamma_e'\beta_e\tilde{\mu},\gamma_e'\beta_e\sqrt{1-\tilde{\mu}^2},0)
\end{equation}
where $\beta_e=\sqrt{1-\gamma_e'^{-2}}$ and $\tilde{\mu}=\cos\tilde{\theta}$.
Transforming this expression to the emitters' frame, moving at $\Gamma'=(1-\beta'^{\,2})^{-1/2}$
in the $x'$-direction in the jet's bulk frame (notice that $\Gamma'$ can be a function of $\gamma_e'$), we obtain:
\begin{equation}
 u'^{\mu}=(\Gamma' \gamma_e'(1-\beta' \beta_e\tilde{\mu}),\Gamma'\gamma_e'(\beta_e\tilde{\mu}-\beta'),\gamma_e'\beta_e\sqrt{1-\tilde{\mu}^2},0)
\end{equation}
The average momentum in the $x'$ direction should be zero in the emitters' frame, as it may be defined as the particles' 
local center of momentum frame. This implies that:
\begin{equation}
\label{eq.beta}
 0=\langle u'^{1} \rangle=\langle \Gamma' \gamma_e'(\beta_e\tilde{\mu}-\beta') \rangle=
 \Gamma' \gamma_e'(\beta_e\langle\tilde{\mu}\rangle-\beta') \quad\Longrightarrow\quad 
 \beta'=\beta_e\langle \tilde{\mu} \rangle\ .
\end{equation}
The average $\langle \tilde{\mu} \rangle$ can be written in terms of $\theta_0$ or 
$\mu_0=\cos\theta_0\approx 1-\frac{1}{2}\theta_0^2$:
\begin{equation}
 \langle \tilde{\mu} \rangle=\frac{1}{1-\mu_0} \int_{\mu_0}^1 \tilde{\mu} d \tilde{\mu}=
 \frac{1+\mu_0}{2}\approx 1-\frac{\theta_0^2}{4}\ .
\end{equation}
Note that $\tilde{\theta}_{\langle\tilde{\mu}\rangle}\equiv\arccos\langle\tilde{\mu}\rangle\approx \theta_0/\sqrt{2}$ 
differs from $\langle\tilde{\theta}\rangle = (\theta_0\cos\theta_0-\sin\theta_0)/(1-\cos\theta_0)\approx\frac{2}{3}\theta_0$.
Given $\beta'$ from Eq. \ref{eq.beta} in terms of the average properties, we can also obtain $\Gamma'$:
\begin{equation}
\label{theta_0}
 \Gamma'=\frac{1}{\sqrt{1-\beta_e^2 \langle \tilde{\mu}\rangle^2}}=
 \frac{1}{\sqrt{1-\beta_e^2 (\frac{1+\mu_0}{2})^2}}\approx \frac{\gamma_e'}{\sqrt{1+\frac{(\gamma_e' \theta_0)^2}{2}}}=
 \left\{ \begin{array}{ll} \gamma_e' & \gamma_e' \theta_0 \ll 1,\\
\frac{\sqrt{2}}{\theta_0} & \gamma_e' \theta_0 \gg 1\ .
\end{array} \right.
\end{equation}
Plugging this back into the zeroth coordinate of the four velocity in the emitters' frame we get that the average energy of particles in this frame is given by:
\begin{equation}
 \gamma_e''\equiv \langle u'^0 \rangle=\Gamma' \gamma_e'(1-\beta_e^2\langle \tilde{\mu} \rangle^2 )=\frac{\gamma_e'}{\Gamma'}\ .
\end{equation}
as given by Eq. \ref{Kdef}.

\end{document}